\documentclass[nofootinbib,aps,prd,superscriptaddress,preprintnumbers,%
 showpacs,showkeys,floatfix,amssymb,amsfonts]{revtex4-1}  

\usepackage[footskip=0.15in, margin=0.7in]{geometry}

\usepackage{graphicx}
\usepackage{bm}
\usepackage{amsmath}
\usepackage{amssymb}
\usepackage{amsfonts}
\usepackage{float}
\usepackage{dsfont}  
\usepackage{slashed}  
\usepackage{booktabs}
\usepackage{multirow}
\usepackage{subfigure}
\usepackage[sort&compress]{natbib}
\usepackage{ulem}
\usepackage{empheq}
\usepackage{color}

\newcommand{\be}{\begin{equation}}  
\newcommand{\ee}{\end{equation}}  
\newcommand{\beq}{\begin{eqnarray}} 
\newcommand{\eeq}{\end{eqnarray}}

\newcommand{\csw}{c_\mathrm{sw}}

\newcommand{\bea}{\begin{eqnarray}}
\newcommand{\eea}{\end{eqnarray}}

\usepackage{xr-hyper}

\usepackage[svgnames,x11names,table]{xcolor}
\usepackage[colorlinks=true,linkcolor=MediumBlue,citecolor=MediumBlue,urlcolor=MediumBlue]{hyperref}
\usepackage{hyperref}

\begin{document}
\preprint{LA-UR-25-24946}
\title{Generalized Parton Distributions from Lattice QCD \\[1ex] with Asymmetric Momentum Transfer: Tensor case}
\author{Shohini Bhattacharya}
\email{shohinib@uconn.edu}
\affiliation{Department of Physics, University of Connecticut, Storrs, CT 06269, USA}
\affiliation{Theoretical Division, Los Alamos National Laboratory, Los Alamos, NM 87545, USA}
\author{Krzysztof Cichy}
\affiliation{Faculty of Physics and Astronomy, Adam Mickiewicz University, ul.\ Uniwersytetu Pozna\'nskiego 2, 61-614 Pozna\'{n}, Poland}
\author{Martha Constantinou}
\email{marthac@temple.edu}
\affiliation{Department of Physics,  Temple University,  Philadelphia,  PA 19122 - 1801,  USA}
\author{Andreas Metz}
\affiliation{Department of Physics,  Temple University,  Philadelphia,  PA 19122 - 1801,  USA}
\author{Joshua Miller}
\email{joshua.miller0007@temple.edu}
\affiliation{Department of Physics,  Temple University,  Philadelphia,  PA 19122 - 1801,  USA}
\author{Peter Petreczky}
\affiliation{Physics Department, Brookhaven National Laboratory, Upton, New York 11973, USA}
\author{Fernanda Steffens}
\affiliation{Institut f\"ur Strahlen- und Kernphysik, Rheinische Friedrich-Wilhelms-Universit\"at Bonn,\\ Nussallee 14-16, 53115 Bonn}
%
\begin{abstract}

The calculation of generalized parton distributions (GPDs) in lattice QCD was traditionally done by calculating matrix elements in the symmetric frame. Recent advancements have significantly reduced computational costs by calculating these matrix elements in the asymmetric frame, allowing us to choose the momentum transfer to be in either the initial or final states only. The theoretical methodology requires a new parametrization of the matrix element to obtain Lorentz-invariant amplitudes, which are then related to the GPDs. The formulation and implementation of this approach have already been established for the unpolarized and helicity GPDs. Building upon this idea, we extend this formulation to the four leading-twist quark transversity GPDs  ($H_T$, $E_T$, $\widetilde{H}_T$, $\widetilde{E}_T$). We also present numerical results for zero skewness using an $N_f=2+1+1$ ensemble of twisted mass fermions with a clover improvement. The light quark masses employed in these calculations correspond to a pion mass of about 260 MeV. Furthermore, we include a comparison between the symmetric and asymmetric frame calculations to demonstrate frame independence of the Lorentz-invariant amplitudes. Analysis of the matrix elements in the asymmetric frame is performed at several values of the momentum transfer squared, $-t$, ranging from 0.17 GeV$^2$ to 2.29 GeV$^2$.
\end{abstract}
\maketitle

\section{Introduction}
\label{s:Intro}
Parton distribution functions (PDFs) are fundamental tools for understanding the internal quark and gluon composition of strongly interacting systems~\cite{Collins:1981uw}. These functions can be probed in processes such as inclusive deep-inelastic lepton-nucleon scattering, revealing information on how partons are distributed within hadrons as a function of their momentum fraction, represented by $x$. PDFs are characterized by matrix elements of operators constructed from parton fields, where the fields are separated by a light-like distance and the operators are evaluated between the same initial and final hadron states. Generalized parton distributions (GPDs) build upon the concept of PDFs by considering light-like parton operators for different initial and final states~\cite{Mueller:1998fv, Ji:1996ek, Radyushkin:1996nd}. GPDs introduce additional dependencies on the longitudinal momentum transfer, $\xi$, and the invariant momentum transfer to the target, $t$, in addition to the parton momentum fraction $x$. Although this makes GPDs inherently more complex, they provide a much richer source of information than PDFs. Specifically, GPDs allow us to construct three-dimensional representations of hadrons~\cite{Burkardt:2000za, Ralston:2001xs, Diehl:2002he, Burkardt:2002hr}, give access to parton angular momentum~\cite{Ji:1996ek}, and offer insight into the internal pressure and shear forces in hadrons~\cite{Polyakov:2002wz, Polyakov:2002yz, Polyakov:2018zvc}. Recent studies have also highlighted that GPDs contain chiral and trace anomaly poles, providing new perspectives on phenomena such as mass generation in QCD, chiral symmetry breaking, and confinement~\cite{Tarasov:2020cwl,Tarasov:2021yll,Bhattacharya:2022xxw,Bhattacharya:2023wvy,Bhattacharya:2023ksc,Bhattacharya:2024geo,Tarasov:2025mvn}. Investigating these anomalies can enhance our understanding of the underlying principles of QCD. For comprehensive discussions on the physics of GPDs, we refer readers to various review articles~\cite{Goeke:2001tz, Diehl:2003ny, Ji:2004gf, Belitsky:2005qn, Boffi:2007yc, Guidal:2013rya, Mueller:2014hsa, Kumericki:2016ehc}.

Experimental access to (chiral-even) GPDs is primarily achieved through hard exclusive scattering processes, such as deep virtual Compton scattering (DVCS)~\cite{Mueller:1998fv, Ji:1996ek, Radyushkin:1996nd, Ji:1996nm, Collins:1998be} and exclusive meson production~\cite{Radyushkin:1996ru, Collins:1996fb, Mankiewicz:1997uy}. However, extracting GPDs from these processes is notoriously difficult without relying on model assumptions, due to the integration over the parton momentum fraction $x$ inherent in quantities such as Compton form factors~\cite{Bertone:2021yyz, Moffat:2023svr}.  
Significant efforts have been made to model and fit GPDs using global experimental data, as demonstrated in studies such as the ones discussed in Refs.~\cite{Polyakov:2002wz, Guidal:2004nd, Goloskokov:2005sd, Mueller:2005ed, Kumericki:2009uq, Goldstein:2010gu, Gonzalez-hernandez:2012xap, Kriesten:2021sqc, Hashamipour:2021kes, Guo:2022upw, Guo:2023ahv, Hashamipour:2022noy}. 
In the present work we concentrate on the four leading-twist chiral-odd quark GPDs.~\cite{Diehl:2001pm}. 
Generally, spatial images of hadrons are provided through impact-parameter distributions, which are Fourier-transforms of GPDs at $\xi = 0$. 
For a spin-$\frac{1}{2}$ hadron, a total of six physically distinct such distributions exist~\cite{Diehl:2005jf}, half of which are related to the twist-2 chiral-odd GPDs.
This point alone provides a strong motivation for studying chiral-odd GPDs.
Furthermore, it was proposed in Ref.~\cite{Burkardt:2005hp} that chiral-odd GPDs allow for a decomposition of quark angular momentum with respect to quarks of definite transversity.    
Also, a suitable combination of chiral-odd GPDs can be used to calculate the correlation between quark spin and quark angular momentum in an unpolarized nucleon~\cite{Diehl:2005jf, Burkardt:2005hp, Bhoonah:2017olu}. 
Of particular interest is the quantity \( \kappa_T = \int dx \big(E_T(x, 0, 0) + 2\widetilde{H}_T(x, 0, 0) \big) \) representing a measure for the dipole-type distortion of the distribution of transversely polarized quarks in an unpolarized nucleon, which involves the two chiral-odd GPDs \( E_T \) and \( \widetilde{H}_T \) evaluated at $\xi = t = 0$.
These examples underscore the significance of chiral-odd GPDs. From this point forward, we shall refer to them as transversity GPDs. Due to their chiral-odd nature, they can neither be accessed through DVCS nor, at leading twist, in hard exclusive meson production~\cite{Collins:1999un}.  One has to resort to (model-dependent) sub-leading twist observables~\cite{Ahmad:2008hp, Goloskokov:2009ia} or more complicated processes~\cite{Ivanov:2002jj}.
As a result, transversity GPDs have largely remained elusive, highlighting the need for direct insights from lattice QCD calculations---although a few model calculations exist, such as those based on the spectator model~\cite{Bhattacharya:2019cme}, the bag model~\cite{Tezgin:2024tfh}, and large-$N_c$ analyses~\cite{Kim:2024ibz}. 

Obtaining GPDs directly from lattice QCD is challenging due to their light-cone definition, which prevents straightforward calculations on a Euclidean lattice. Instead, studies have primarily focused on extracting limited information from Mellin moments of GPDs (see, e.g., Refs.~\cite{Hagler:2003jd, QCDSF-UKQCD:2007gdl, Alexandrou:2011nr, Alexandrou:2013joa, Constantinou:2014tga}). Recent advancements have enabled simulations closer or at the physical point~\cite{Green:2014xba, Alexandrou:2017ypw, Alexandrou:2017hac, Hasan:2017wwt, Gupta:2017dwj, Capitani:2017qpc, Alexandrou:2018sjm, Shintani:2018ozy, Bali:2018qus, Bali:2018zgl, Alexandrou:2019ali, Jang:2019jkn, Constantinou:2020hdm, Alexandrou:2022dtc, Jang:2023zts}. Despite these developments, capturing the full $x$-dependence of GPDs remains a significant challenge.
In recent years, alternative methods for accessing GPDs in momentum space have fueled a promising new direction in lattice QCD research. Our study employs the quasi-distribution approach~\cite{Ji:2013dva}, which involves calculating matrix elements using momentum-boosted hadrons and non-local operators. To relate these quasi-distributions to light-cone GPDs, we utilize the framework of Large-Momentum Effective Theory (LaMET)~\cite{Ji:2014gla,Ji:2020ect}. For comprehensive discussions on approaches to extracting $x$-dependent distribution functions, readers may consult the reviews in Refs.~\cite{Cichy:2018mum,Ji:2020ect,Constantinou:2020pek,Cichy:2021lih,Cichy:2021ewm}. The literature is rich with studies on PDFs but much less extensive when it comes to GPDs~\cite{Ji:2015qla,Xiong:2015nua,Bhattacharya:2018zxi, Liu:2019urm, Bhattacharya:2019cme, Chen:2019lcm, Radyushkin:2019owq, Ma:2019agv, Luo:2020yqj, Alexandrou:2020zbe, Alexandrou:2021bbo,CSSMQCDSFUKQCD:2021lkf, Dodson:2021rdq, Ma:2022ggj,Bhattacharya:2022aob, Bhattacharya:2023jsc, Bhattacharya:2024qpp,Cichy:2024afd}.

In our recent work, detailed in Refs.~\cite{Bhattacharya:2022aob, Bhattacharya:2023jsc, Bhattacharya:2024qpp}\footnote{Additional results are provided in Refs.~\cite{Bhattacharya:2023tik, Constantinou:2022fqt, Cichy:2023dgk,Bhattacharya:2025cia}.}, we have made substantial progress in improving the computational efficiency of lattice QCD calculations for off-forward matrix elements. Results from these calculations have also been used to compute Mellin moments of unpolarized and helicity distributions in the transverse plane utilizing short-distance factorization~\cite{Bhattacharya:2023ays,Bhattacharya:2024wtg}.
Our approach involves asymmetric frames, rather than the conventional symmetric frame, where the entire momentum transfer $\Delta$ is applied to the initial nucleon state. This strategy not only lowers computational costs but also allows us to explore a dense range of $t \equiv \Delta^2$ at such reduced cost, facilitating a more comprehensive mapping of GPDs across larger $t$ values. Previously, our focus was on unpolarized quark GPDs ($H$ and $E$) and axial-vector GPD ($\widetilde{H}$), where we introduced a novel Lorentz-covariant parameterization of the matrix element using Lorentz-invariant amplitudes. This parameterization enabled us to relate matrix elements from different frames. We also proposed a frame-independent definition of quasi-GPDs, which could potentially minimize power corrections in matching relations to light-cone GPDs. In our current work, we build upon this amplitude-based framework to compute the four leading-twist quark transversity GPDs (denoted by $H_T$, $E_T$, $\widetilde{H}_T$, $\widetilde{E}_T$ in Ref.~\cite{Diehl:2001pm}), and present numerical results for zero skewness, $\xi = 0$. 

The paper is organized as follows. In Sec.~\ref{sec:theory}, we introduce the definitions of the transversity light-cone and quasi-GPDs. We then discuss the decomposition of tensor matrix elements in terms of Lorentz-invariant amplitudes and establish their relationships to the transversity GPDs. Based on these amplitudes, we propose potential candidates for a frame-independent definition of quasi-GPDs. In Sec.~\ref{sec:lattice_setup}, we present the Euclidean decomposition of lattice-calculable matrix elements in terms of these amplitudes and outline our lattice setup for position-space calculations. Sec.~\ref{sec:lattice_results_main} presents our numerical results, comparing the outcomes for symmetric and asymmetric frames at various stages in both coordinate and momentum space. Specifically, we provide numerical results for the invariant amplitudes and all transversity GPDs at $\xi=0$. Finally, in Sec.~\ref{sec:summary}, we summarize our findings and discuss future research directions.

\section{Strategy of Frame Transformation}
\label{sec:theory}
Computing GPDs in the symmetric frame poses significant challenges in lattice QCD, as each value of the momentum transfer, \(\Delta\), requires a separate calculation, severely limiting the density of accessible momentum transfer values. This limitation has led to a growing interest by the authors of this manuscript to explore asymmetric frames, which offer computational advantages. One method, developed in our previous work for vector GPDs~\cite{Bhattacharya:2022aob} and axial-vector GPDs~\cite{Bhattacharya:2023jsc}, uses transverse Lorentz transformations (or transverse boosts) to connect symmetric and asymmetric frames. Another approach we have introduced involves a Lorentz-covariant formalism that enables calculations in \textit{any} chosen frame by parameterizing the relevant matrix elements in terms of Lorentz-invariant, frame-independent amplitudes. This construction establishes direct relations between different frames and will be further developed in the following sections in the context of computing the tensor GPDs from matrix elements in asymmetric frames. All expressions presented in this section are given in Minkowski space.

\subsection{Definitions of GPDs}
To begin with, we revisit the concept of light-cone quark GPDs for a spin-1/2 hadron. In position space, GPDs are defined through non-local matrix elements of quark fields, and are expressed as follows:
\begin{equation}
F^{[\Gamma]}(z^-, \Delta, P) = \langle p_f, \lambda' | \bar{\psi} \left( -\frac{z}{2} \right) \Gamma \mathcal{W} \left( -\frac{z}{2}, \frac{z}{2} \right) \psi \left( \frac{z}{2} \right) | p_i, \lambda \rangle \Big|_{z^{+}=0, \vec{z}_{\perp} = \vec{0}_{\perp}} \, .
\label{e:corr_standard_GPD}
\end{equation}
where $\Gamma$ is a general gamma matrix. The Wilson line maintains the gauge invariance of the above expression
\begin{equation}
\mathcal{W} \left( -\frac{z}{2}, \frac{z}{2} \right) \Big|_{z^{+}=0, \vec{z}_{\perp} = \vec{0}_{\perp}} = \mathcal{P} \, \exp \left( -i g \int_{-\frac{z^{-}}{2}}^{\frac{z^{-}}{2}} \, dy^{-} A^{+}(0^{+}, y^{-}, \vec{0}_{\perp}) \right) \, .
\label{e:wilson_line_standard_GPD}
\end{equation}
In Eq.~(\ref{e:wilson_line_standard_GPD}), the symbol $g$ stands for the strong coupling constant, and $A^{+}$ refers to the plus component of the gluon field on the light-cone. The initial and final hadronic states in Eq.~(\ref{e:corr_standard_GPD}) are characterized by their four-momenta $p_i$ and $p_f$, and helicities $\lambda$ and $\lambda'$, respectively.
We introduce the following kinematic variables: the average four-momentum $P$ of the hadrons, the four-momentum transfer $\Delta$, the skewness $\xi$ (which is used to describe the longitudinal momentum transfer to the hadron having a large light-cone plus momentum), and the invariant squared momentum transfer $t$:
\begin{equation}
P = \frac{1}{2} (p_i + p_f), \qquad
\Delta = p_f - p_i, \qquad 
\xi = \frac{p^{+}_i - p^{+}_f}{p^{+}_i + p^{+}_f}, \qquad
t = \Delta^2 \, .
\label{e:kinematics}
\end{equation}
These kinematic definitions, given in Eq.~\eqref{e:kinematics}, are used in both symmetric and asymmetric frames.

At twist-2, the correlator with $\Gamma = i \sigma^{j+} \gamma_5$ ($j$ is a transverse index) in Eq.~(\ref{e:corr_standard_GPD}) can be characterized by four distinct tensor/transversity GPDs: $H_T$, $E_T$, $\widetilde{H}_T$, and $\widetilde{E}_T$. In position space, the expression is given by~\cite{Diehl:2001pm}:
\begin{align}
F^{[i\sigma^{j+}\gamma_5]} (z^-, \Delta, P) & = -i \epsilon^{-+ij} \bar{u}(p_f, \lambda') \bigg[ i \sigma^{+i} H_T (z^-, \xi, t) + \frac{\gamma^+ \Delta^i_{\perp} - \Delta^+ \gamma^i_{\perp}}{2m} E_T (z^-, \xi, t) \nonumber \\[0.2cm]
& + \frac{P^+ \Delta^i_{\perp} - P^i_\perp \Delta^+}{m^2} \widetilde{H}_T (z^-, \xi, t) + \frac{\gamma^+ P^i_\perp - P^+ \gamma^i_{\perp}}{m} \widetilde{E}_T (z^-, \xi, t) \bigg] u(p_i, \lambda) \, .
\label{e:GPD_def_pos}
\end{align}
For the expression corresponding to Eq.~(\ref{e:corr_standard_GPD}) in momentum space, the Fourier transform is performed with respect to $P \cdot z$, while keeping $P^+$ fixed, resulting in:
\begin{align}
F^{[i\sigma^{j+}\gamma_5]} (x, \Delta)&  =  \dfrac{1}{2P^+} \int \dfrac{d(P \cdot z)}{2\pi} e^{i x P \cdot z}  F^{[i\sigma^{j+}\gamma_5]} (z^-, \Delta, P) \nonumber \\
& = \dfrac{-i \epsilon^{-+ij}}{2P^+} \bar{u}(p_f, \lambda') \bigg[ i \sigma^{+i} H_T (x, \xi, t) + \frac{\gamma^+ \Delta^i_{\perp} - \Delta^+ \gamma^i_{\perp}}{2m} E_T (x, \xi, t) \nonumber \\[0.2cm]
& + \frac{P^+ \Delta^i_{\perp} - P^i_\perp \Delta^+}{m^2} \widetilde{H}_T (x, \xi, t) + \frac{\gamma^+ P^i_\perp - P^+ \gamma^i_{\perp}}{m} \widetilde{E}_T (x, \xi, t) \bigg] u(p_i, \lambda) \, .
\label{e:GPD_def}
\end{align}

Now, let us shift our focus to quasi-GPDs, which are defined in position space through the equal-time correlator~\cite{Ji:2013dva}:
\begin{equation}
F^{[\Gamma]}(z^3, \Delta, P) =  \langle p_f,\lambda'| \bar{\psi}(-\tfrac{z}{2}) \, \Gamma \, {\cal W}(-\tfrac{z}{2}, \tfrac{z}{2}) \psi (\tfrac{z}{2})|p_i, \lambda \rangle \Big |_{z^{0}=0, \vec{z}_{\perp}=\vec{0}_{\perp}} \, .
\label{e: corr_quasi_GPD_1}
\end{equation}
Here, the Wilson line is given by
\begin{equation}
{\cal W}(-\tfrac{z}{2}, \tfrac{z}{2}) \Big|_{z^{0}=0, \vec{z}_{\perp}=\vec{0}_{\perp}} = {\cal P} \, \exp \bigg( -ig \int_{-\tfrac{z^{3}}{2}}^{\tfrac{z^{3}}{2}} \, dy^{3} A^{3}(0, \vec{0}_{\perp}, y^{3}) \bigg )\, .
\end{equation}
For $\Gamma = i \sigma^{j0} \gamma_5$, one finds 
\begin{align}
F^{[i\sigma^{j0}\gamma_{5}]} (z^3, \Delta, P) & = -i\epsilon^{03ij} \bar{u}(p_f, \lambda ') \bigg [i\sigma^{3i}{\cal H}_{T} (z^3, \xi, t; P^3) + \frac{\gamma^{3}\Delta^{i}_{\perp}-\Delta^{3}\gamma^{i}_{\perp}}{2m} {\cal E}_{T} (z^3, \xi, t; P^3) \nonumber \\
& + \frac{P^{3}\Delta^{i}_{\perp} - P^{i}_\perp\Delta^{3}}{m^{2}}{\cal \widetilde{H}}_{T} (z^3, \xi, t; P^3)  + \frac{P^{i}_\perp \gamma^{3} - P^{3}\gamma^{i}_{\perp}}{m} {\cal \widetilde{E}}_{T} (z^3, \xi, t; P^3) \bigg ] u(p_i, \lambda) \, ,
\label{e:qGPD_def_pos}
\end{align}
with the quasi-GPDs ${\cal H}_{T} (z^3, \xi, t; P^3)$, and so on. Eq.~\eqref{e:qGPD_def_pos} is the quasi-GPD counterpart of Eq.~\eqref{e:GPD_def_pos}. For the expression corresponding to Eq.~(\ref{e:qGPD_def_pos}) in momentum space, we perform a Fourier transform with respect to $P \cdot z$, while keeping $P^3$ fixed, yielding
\begin{align}
F^{[i\sigma^{j0}\gamma_{5}]}(x, \Delta; P^3) & =  \dfrac{1}{2P^0}\int \frac{d(P \cdot z)}{2\pi}  e^{i x P \cdot z} F^{[i\sigma^{j0}\gamma_{5}]} (z^3, \Delta, P) \nonumber \\
& = \dfrac{-i\epsilon^{03ij} }{2P^0} \bar{u}(p_f, \lambda ') \bigg [i\sigma^{3i}{\cal H}_{T} (x, \xi, t; P^3) + \frac{\gamma^{3}\Delta^{i}_{\perp}-\Delta^{3}\gamma^{i}_{\perp}}{2m} {\cal E}_{T} (x, \xi, t; P^3) \nonumber \\
& + \frac{P^{3}\Delta^{i}_{\perp} - P^{i}_\perp\Delta^{3}}{m^{2}}{\cal \widetilde{H}}_{T} (x, \xi, t; P^3)  + \frac{P^{i}_\perp \gamma^{3} - P^{3}\gamma^{i}_{\perp}}{m} {\cal \widetilde{E}}_{T} (x, \xi, t; P^3) \bigg ] u(p_i, \lambda) \, .
\label{e: corr_quasi_GPD}
\end{align}
In Ref.~\cite{Constantinou:2017sej}, it was proposed to define the quasi-counterpart of the light-cone GPDs using $\Gamma = i \sigma^{j0} \gamma_5$, as presented in Eq.~\eqref{e:qGPD_def_pos}. The rationale for selecting $i \sigma^{j0} \gamma_5$ instead of, for instance, $i \sigma^{j3} \gamma_5$, is the absence of mixing with other operators under renormalization. This mixing is considered a lattice artifact resulting from chiral symmetry breaking~\cite{Constantinou:2017sej}. Furthermore, Ref.~\cite{Bhattacharya:2019cme} argues that, for this definition, it is necessary to substitute $i \sigma^{j+} \gamma_5 / P^+$ with $i \sigma^{j0} \gamma_5 / P^0$ to ensure consistency with the forward limit. The definition in Eq.~(\ref{e: corr_quasi_GPD}) also yields the correct local limit when integrated with respect to $x$ (see Appendix~\ref{sec:local}).

\subsection{Parameterization of the tensor matrix elements}
Now, let us discuss the Lorentz-covariant decomposition of the tensor matrix elements, that is, Eq.~\eqref{e:corr_standard_GPD} with $\Gamma = i \sigma^{\mu\nu} \gamma_5$ for spin-1/2 particles in position space. By incorporating parity constraints, we establish that the tensor matrix element can be expressed as a combination of twelve distinct Dirac structures, each multiplied by a corresponding Lorentz-invariant amplitude. The choice of basis is not unique, and here we employ the following decomposition:
\begin{align}
& F^{[i\sigma^{\mu\nu}\gamma_5]} (z, P, \Delta)
\equiv \langle p_f;\lambda'| \bar{\psi} (-\tfrac{z}{2})\, i\sigma^{\mu\nu} \gamma_5 \, {\cal W}(-\tfrac{z}{2},\tfrac{z}{2}) \psi (\tfrac{z}{2})|p_i;\lambda \rangle \nonumber \\[0.3cm]
& = \bar{u}(p_f,\lambda') \bigg [P^{[\mu}z^{\nu]}\gamma_5A_{T1} + \frac{P^{[\mu}\Delta^{\nu]}}{m^2}\gamma_5A_{T2} + z^{[\mu}\Delta^{\nu]}\gamma_5A_{T3} + \gamma^{[\mu}\bigg(\frac{P^{\nu]}}{m}A_{T4} + m z^{\nu]}A_{T5} + \frac{\Delta^{\nu]}}{m}A_{T6}\bigg)\gamma_5 \nonumber \\[0.1cm]
&  + m\slashed{z}\gamma_5\bigg(P^{[\mu}z^{\nu]}A_{T7} + \frac{P^{[\mu}\Delta^{\nu]}}{m^2}A_{T8} + z^{[\mu}\Delta^{\nu]}A_{T9}\bigg)+i\sigma^{\mu\nu}\gamma_5A_{T10}  +i\epsilon^{\mu\nu Pz}A_{T11} + i\epsilon^{\mu\nu z\Delta}A_{T12} \bigg ] u(p_i,\lambda) \, ,
\label{eq:tensor_para}
\end{align}
where $A^{[\mu}B^{\nu]} = A^{\mu}B^{\nu}-A^{\nu}B^{\mu}$, $\epsilon^{\mu\nu Pz} = \epsilon^{\mu\nu\alpha\beta}P_{\alpha}z_{\beta}$, and $\epsilon^{\mu\nu z\Delta} = \epsilon^{\mu\nu\alpha\beta}z_{\alpha}\Delta_{\beta}$. We note that the above equation holds for a general value of $z$ and has a smooth $z \to 0$ limit (see Appendix~\ref{sec:local} for the consistency of this decomposition with the corresponding local current).
For brevity, we adopt the concise notation $A_{Ti} \equiv A_{Ti} (z \cdot P, z \cdot \Delta, \Delta^2, z^2)$ for the Lorentz-invariant amplitudes. The procedure for deriving these results closely follows the steps outlined in Ref.~\cite{Meissner:2009ww}. Twelve independent lattice matrix elements are required to disentangle all the amplitudes.

We now briefly discuss the consistency of this decomposition in Eq.~(\ref{eq:tensor_para}) in the forward limit. By setting the momentum transfer $\Delta \to 0$, we simplify the expression and identify several terms involving different spin and momentum structures. Notably, some of these structures are basically identical due to relations between spinor terms, such as the connection between $\bar{u}(p) (\gamma^\mu \gamma_5) u(p)$ and $u(p)(i \sigma^{\mu \nu} \gamma_5) u(p)$. 
Eliminating redundant terms in this way leaves us with four structures and the corresponding amplitudes, for which we choose to be $A_{T5}$, $A_{T7}$, $A_{T10}$, and $A_{T11}$.
Furthermore, hermiticity and time-reversal symmetry (see Appendix~\ref{sec:symmetry}) imply that the amplitude $A_{T11}$ is odd in $\xi$ and therefore vanishes in the forward limit. 
This leaves us with only three amplitudes, consistent with the number of corresponding PDFs in the forward limit~\cite{Jaffe:1991kp}. This completes the consistency check of our decomposition in this limit.

We now establish connections between the light-cone GPDs and the amplitudes. 
After substituting $(\mu, \nu) = (j, +)$ in Eq.~(\ref{eq:tensor_para}), we apply a basis transformation to express the amplitudes $A_{Ti}$ in terms of the GPDs defined in Eq.~(\ref{e:GPD_def_pos}):
\begin{align}
H_T(z\cdot P, z \cdot \Delta, \Delta^2) & = - 2 A_{T2} \bigg ( 1-\dfrac{P^{2}}{m^2}\bigg ) + A_{T4} - \Delta \cdot z A_{T8} +A_{T10} \, , \label{e:LC1}\\[0.2cm]
E_T (z\cdot P, z \cdot \Delta, \Delta^2)& = 2 A_{T2} - A_{T4} + \Delta \cdot z A_{T8} \, , \label{e:LC2}\\[0.2cm]
\widetilde{H}_T (z\cdot P, z \cdot \Delta, \Delta^2)& = - A_{T2} \, , \label{e:LC3}\\[0.2cm]
\widetilde{E}_T(z\cdot P, z \cdot \Delta, \Delta^2) & = - 2A_{T6} - 2 P \cdot z A_{T8} \, ,
\label{e:LC4}
\end{align}
where the $A_{Ti}$'s are evaluated at \( z^2 = 0 \). In Eqs.~(\ref{e:LC1}) - (\ref{e:LC4}), the arguments of the light-cone GPDs are expressed differently from those in Eq.~(\ref{e:GPD_def_pos}) to highlight their Lorentz-invariant structure; this change is purely notational. In the above expressions, one needs to choose the kinematical variables from the relevant frame, keeping in mind that in the symmetric frame, the momentum transfer is equally distributed between the initial and final states. In contrast, in an asymmetric frame, it is considered non-symmetric.  
However, as is evident, these equations exhibit Lorentz invariance, ensuring their validity and applicability across all reference frames. This means that the numerical values of these GPDs remain the same in all frames.  
In Appendix~\ref{sec:symmetry}, we present a comprehensive discussion on the symmetry properties of the amplitudes, as dictated by hermiticity and time-reversal transformations. In particular, the relations combining these symmetries determine how the amplitudes transform under skewness reversal, $\xi \rightarrow -\xi$. We find that these relations imply that the amplitudes $A_{T1}$, $A_{T6}$, $A_{T8}$, $A_{T9}$, and $A_{T11}$ are odd in $\xi$. This symmetry behavior, together with the requirement for a well-defined forward limit of the matrix element, leads us to conclude that these amplitudes vanish at $\xi = 0$. Notably, the vanishing of $A_{T6}$ and $A_{T8}$ implies that $\widetilde{E}_T$ also vanishes, consistent with the known result that this GPD is odd in $\xi$~\cite{Diehl:2001pm, Meissner:2007rx, Bhattacharya:2019cme}. In our lattice data analysis, we found these amplitudes to be numerically consistent with zero, which serves as a crucial consistency check for our results. Further discussion, along with confirmation of the vanishing of the GPD \( \widetilde{E}_T \) at zero skewness, can be found in Sec.~\ref{sec:lattice_results_main}.

Next, we turn our attention to the quasi-GPDs. As emphasized in Refs.~\cite{Bhattacharya:2022aob, Bhattacharya:2023jsc}, a natural approach to defining quasi-GPDs is to extend the Lorentz-invariant light-cone definitions in Eqs.~(\ref{e:LC1}) - (\ref{e:LC4}) to include $z^2 \neq 0$. Throughout this discussion, we refer to these objects as the Lorentz-invariant (LI) quasi-GPDs, 

\begin{align}
{\cal H} _T(z\cdot P, z \cdot \Delta, \Delta^2, z^2) & = - 2 A_{T2} \bigg ( 1-\dfrac{P^{2}}{m^2}\bigg ) + A_{T4} - \Delta \cdot z A_{T8} +A_{T10} \, , \label{e:LI1}\\[0.2cm]
{\cal E}_T (z\cdot P, z \cdot \Delta, \Delta^2, z^2)& = 2 A_{T2} - A_{T4} + \Delta \cdot z A_{T8} \, ,  \label{e:LI2}\\[0.2cm]
\widetilde{\cal H}_T (z\cdot P, z \cdot \Delta, \Delta^2, z^2)& = - A_{T2} \, ,  \label{e:LI3}\\[0.2cm]
\widetilde{\cal E}_T(z\cdot P, z \cdot \Delta, \Delta^2, z^2) & = - 2A_{T6} - 2 P \cdot z A_{T8} \, ,
\label{e:LI4}
\end{align}
where now the $A_{Ti}$ are evaluated at $z^2 \neq 0$. This definition of the quasi-GPDs retains the same functional form in terms of the \( A_{Ti} \) as the light-cone GPDs. Specifically, it is constructed using an operator that, for zero skewness, combines \( (i \sigma^{j0}\gamma_5, i \sigma^{j3}\gamma_5) \), rather than the standard operator \( i \sigma^{j0}\gamma_5 \) alone (see below). It is known that the two operators \( i \sigma^{j0}\gamma_5 \) and \( i \sigma^{j3}\gamma_5 \) have different matching coefficients, even for zero skewness, which is relevant for our analysis. Thus, this implies that the LI definition should have a different matching coefficient compared to the standard definition corresponding to the operator \( i \sigma^{j0}\gamma_5 \) alone. However, the systematics introduced by using the same matching for both the definitions is negligible compared to other, larger systematics in these calculations, particularly the mixing that the \( i \sigma^{j3}\gamma_5 \) operator undergoes under renormalization or the lattice formulation~\cite{Constantinou:2017sej}. (It is interesting to note that the mixing under renormalization appears to be suppressed when a clover term is included in the fermionic action, as used in this work. As shown in Ref.~\cite{Constantinou:2017sej}, at one-loop level, the mixing is of the form $a - b\,\csw$, and the combination becomes very small for $\csw$ around 1-1.5.) We note that in our earlier works~\cite{Bhattacharya:2022aob, Bhattacharya:2023jsc, Bhattacharya:2024qpp}, where we employed the LI definition, modifications to the operator were required to account for contributions from twist-3 operators. However, this is not the case here. Thus, the analysis is simpler in terms of the matching for the transversity case.

We now consider the quasi-GPDs introduced in Eq.~\eqref{e:qGPD_def_pos}. By setting $(\mu, \nu) = (j, 0)$ in Eq.~(\ref{eq:tensor_para}), we perform a basis transformation to rewrite the resulting expression, thereby establishing a mapping between the $A_{Ti}$ amplitudes and those quasi-GPDs. The explicit relations in the symmetric frame are given by:
\begin{eqnarray}
    \mathcal{H}^s_{T} (z, P^s, \Delta^s)  &=& -2A_{T2}\bigg(1-\frac{(P^s)^2}{m^2}\bigg) + A_{T4} + z^3A_{T8}\bigg(\frac{(P^{0,s})^2\Delta^{0,s}}{P^{3,s}P^{0,s}}\bigg) + A_{T10},   \label{e:standard1}\hspace*{2cm}\\[1.5ex]
    \mathcal{E}^s_{T} (z, P^s, \Delta^s) &=& 2A_{T2} - A_{T4} - z^3A_{T8}\bigg(\frac{(P^{0,s})^2\Delta^{0,s}}{P^{3,s}P^{0,s}} \bigg) \label{e:standard2},  \hspace*{2cm}\\[1.5ex]
    \widetilde{\mathcal{H}}^s_{T} (z, P^s, \Delta^s) &=& -A_{T2} + z^3A_{T12}\frac{m^2}{P^{3,s}}, \label{e:standard3} \hspace*{2cm}\\[1.5ex]
    \widetilde{\mathcal{E}}^s_{T} (z, P^s, \Delta^s) &=& -2A_{T6} + z^3A_{T8}\frac{2(P^{0,s})^2}{P^{3,s}} \label{e:standard4}.
\end{eqnarray}

\noindent
We emphasize that the relations between quasi-GPDs and amplitudes, as presented above, are defined exclusively in the symmetric frame. While similar relations can be derived in the asymmetric frame leading to ${\cal H}^a_T, {\cal E}^a_T, \widetilde{{\cal H}}^a_T, \widetilde{{\cal E}}^a_T$, the functional form of these relations—both in terms of the kinematic coefficients multiplying the amplitudes and the specific amplitudes that contribute—will differ from those presented above.
We have previously done this for the unpolarized GPDs~\cite{Bhattacharya:2022aob} and established that the standard operator definitions of quasi-GPDs (in contrast to the LI definitions) exhibit frame dependence. For simplicity, this work focuses on (standard) quasi-GPDs defined in the symmetric frame, where the kinematic factors are specific to this frame. Defining quasi-GPDs in different frames was also addressed in Ref.~\cite{Braun:2023alc}. As discussed in some detail in Ref.~\cite{Bhattacharya:2022aob}, these objects can also be accessed using matrix elements from the asymmetric frame via appropriate Lorentz transformations.

We reiterate that the two sets of quasi-GPDs discussed above—the LI definition (Eqs.~(\ref{e:LI1}) - (\ref{e:LI4})) and the standard operator definition (Eqs.~(\ref{e:standard1}) - (\ref{e:standard4}))—are not equivalent, as they differ in the contributing amplitudes as well as in both explicit and implicit power corrections.\footnote{Throughout this discussion, power corrections specifically refer to terms proportional to $z^2$.} For instance, the additional amplitudes present in the quasi-GPD $\widetilde{{\cal H}}^s_{T}$ (see Eq.~(\ref{e:standard3}) versus Eq.~(\ref{e:LI3})) can be interpreted as contamination from explicit power corrections. This suggests that the LI definition $\widetilde{{\cal H}}_{T}$ could potentially exhibit faster convergence compared to the standard definition. However, it is essential to recognize that the amplitudes themselves also contain implicit power corrections, meaning the rate of convergence must be assessed on a case-by-case basis as there might be numerical cancellations between different power corrections~\cite{Bhattacharya:2025cia}. Ultimately, the convergence behavior of different quasi-GPD definitions is dictated by the underlying non-perturbative dynamics. Thus, numerical comparisons are essential for assessing their convergence and quantifying the relative magnitude of power corrections in each case.

\section{Lattice Calculation}
\label{sec:lattice_setup}
\subsection{Methodology}
In this section, we present a description of the lattice QCD methodology for calculating the matrix elements of the tensor operator. All expressions are presented in Euclidean space with lowered indices to avoid confusion with the Minkowski space equations of Sec.~\ref{sec:theory}. 
As mentioned earlier, we restrict our numerical implementation to the case of zero skewness, although the theoretical formalism is developed for general kinematics. This choice is motivated by more than one consideration. First, the zero-skewness limit provides an ideal setting to validate the methodology, including the frame independence of the extracted amplitudes and the expected vanishing of certain components, such as $\widetilde{E}_T$ at $\xi = 0$. These serve as important internal consistency checks and establish a foundation for subsequent studies. Second, the case of zero skewness is physically distinct, with a well-defined interpretation in impact-parameter space and a trivial ERBL region. For these reasons, it is common practice to analyze the $\xi=0$ regime independently.
We define the initial ($p_i$) and final ($p_f$) momenta in the symmetric frame through
\begin{equation}
\vec{p}_i^{~s}= \vec{P} - \frac{\vec{\Delta}}{2} = \bigg(\frac{-\Delta_1}{2}, \frac{-\Delta_2}{2},P_3 - \frac{\Delta_3}{2}\bigg), \qquad
\vec{p}_f^{~s} = \vec{P} + \frac{\vec{\Delta}}{2} = \bigg(\frac{+\Delta_1}{2}, \frac{+\Delta_2}{2},P_3+\frac{\Delta_2}{2}\bigg)\,.
\label{eq:sym_frame}
\end{equation}
The asymmetric frame of choice assigns the momentum transfer in the momentum of the initial state, that is
\begin{equation}
\vec{p}_i^{~a}= \vec{P}-\vec{\Delta} = (-\Delta_1,-\Delta_2,P_3-\Delta_3) , \qquad
\vec{p}_f^{~a} = \vec{P} = (0,0,P_3)\,.
\label{eq:asym_frame}
\end{equation}
Eqs.~\eqref{eq:sym_frame} - \eqref{eq:asym_frame} are presented in units of $\tfrac{2\pi}{L}$ where $L$ is the spatial extent of the lattice. We note that all numerical calculations presented in this paper are done at zero-skewness ($\Delta_3=0$), as in the case of the unpolarized~\cite{Bhattacharya:2023ays} and helicity~\cite{Bhattacharya:2024wtg} GPDs.  In either of the frames, we parameterize the lattice matrix elements using the trace
\begin{equation}
    K\mathrm{Tr}\bigg[\Gamma_{\kappa}\bigg(\frac{-i\slashed{p}_f+m}{2m}\bigg)F^{[i\sigma^{\mu\nu}\gamma_5]}_{\lambda,\lambda'}\bigg(\frac{-i\slashed{p}_i+m}{2m}\bigg)\bigg], \qquad \mu,\kappa = 0,1,2,3,
    \label{eq:Trace}
\end{equation}
where $F^{[i\sigma^{\mu\nu}\gamma_5]}_{\lambda,\lambda'}$ is given in Eq.~\eqref{eq:tensor_para}, and $K$ is a kinematic factor that has been obtained based on the normalization of the proton state,
\begin{equation}
    K = \frac{2m^2}{\sqrt{E_fE_i(E_f+m)(E_i+m)}}.
\end{equation}  
We use the four parity projectors: the unpolarized, $\Gamma_0$, and the three polarized, $\Gamma_{\kappa}$, defined as
\begin{eqnarray}
    \Gamma_0 &=& \frac{1}{4} \left(1 + \gamma_0\right)\,, \\
    \Gamma_k &=& \frac{1}{4} \left(1 + \gamma_0\right) i \gamma_5 \gamma_k\,, \quad k=1,2,3\,.
\end{eqnarray}
With these four projectors as well as the six independent values of $\mu$ and $\nu$ for the tensor operator, we can disentangle all twelve amplitudes $A_{Ti}$ for any kinematic setup. 
Below, we provide the expressions for the various matrix elements and their decomposition into the $A_{Ti}$. 
These expressions depend on the kinematic frame, the projector, and the operator indices, so we label them as $\Pi^{a/s}_{\mu\nu}(\Gamma_j)$. 
Note that, for consistency with the work of Refs.~\cite{Alexandrou:2021bbo,Bhattacharya:2021moj} we use an alternative basis for the operator, $i\sigma^{\mu\nu}$, which can be related to the one with a $\gamma_5$ via the identity $i\gamma_5\sigma^{\mu\nu} = -\tfrac{1}{2}\epsilon^{\mu\nu\alpha\rho}\sigma_{\alpha\rho}$. In this basis, the standard twist-2 operator employed is $\sigma^{3j}$ ($j=1,2$.)
The general expressions for Eq.~\eqref{eq:Trace} in the symmetric frame at zero skewness are given below. Any combinations of operators and projectors that result in vanishing matrix elements are not reported.

\begin{eqnarray}
   \label{eq:Pi01G0_s}
   \Pi^s_{01}(\Gamma_0) &=& 
 i\,  K\,\left(-A_{T4}\frac{  \Delta_1  P_3^2}{4 m^3}+A_{T5}\frac{  \Delta_1 z
    P_3}{4 m}+A_{T10} \frac{( E+m)  \Delta_1}{4 m^2} \right) \hspace*{2cm}\\[3ex]
\Pi^s_{01}(\Gamma_1) &=&  
   K\,\left(A_{T2}\frac{ ( E+m)  P_3  \Delta_1  \Delta_2}{4 m^4}-A_{T3} \frac{( E+m)  \Delta_1 z  \Delta_2}{4 m^2}-A_{T4}\frac{  P_3  \Delta_1  \Delta_2}{8 m^3}+A_{T5}\frac{  \Delta_1 z  \Delta_2}{8 m} \right) \hspace*{2cm}\\[3ex]
   \Pi^s_{01}(\Gamma_2) &=& 
   K\,\left(A_{T2} \frac{( E+m)  P_3  \Delta_2^2}{4 m^4}-A_{T3} \frac{( E+m) z  \Delta_2^2}{4 m^2}+A_{T4}\frac{  P_3 \left( \Delta_1^2+4 m
   ( E+m)\right)}{8 m^3}\right. \nonumber \\[1.25ex] 
   && \left. \qquad -A_{T5}\frac{ \left( \Delta_1^2+4 m ( E+m)\right) z}{8 m}+A_{T10} \frac{( E+m)  P_3}{2 m^2} \right) \hspace*{2cm}\\[3ex]
    \Pi^s_{01}(\Gamma_3) &=& 
   K\,\left(-A_{T6} \frac{ E
   ( E+m)  \Delta_2}{2 m^3} -A_{T8} \frac{ E ( E+m)  P_3  \Delta_2 z}{2 m^3} + A_{T9}  \frac{E ( E+m)  \Delta_2 z^2}{2 m} \right) \hspace*{2cm} \\[3ex]
      \Pi^s_{02}(\Gamma_0) &=&
 i\,  K\,\left(-A_{T4} \frac{ \Delta_2  P_3^2}{4 m^3}+A_{T5}\frac{  \Delta_2 z  P_3}{4 m}+A_{T10} \frac{( E+m)  \Delta_2}{4 m^2} \right) \hspace*{2cm}\\[3ex]
    \Pi^s_{02}(\Gamma_1) &=& 
   K\,\left(-A_{T2}\frac{ ( E+m)  P_3  \Delta_1^2}{4 m^4}+A_{T3} \frac{( E+m) z
    \Delta_1^2}{4 m^2}-A_{T4}\frac{  P_3 \left( \Delta_2^2+4 m ( E+m)\right)}{8 m^3}\right. \nonumber \\[1.25ex] 
   && \left. \qquad+A_{T5}\frac{ \left( \Delta_2^2+4 m ( E+m)\right) z}{8 m}-A_{T10} \frac{( E+m)  P_3}{2 m^2} \right) \\[3ex]
    \Pi^s_{02}(\Gamma_2) &=&  
   K\,\left(-A_{T2}\frac{ ( E+m)  P_3  \Delta_1  \Delta_2}{4
   m^4}+A_{T3} \frac{( E+m)  \Delta_1 z  \Delta_2}{4 m^2}+A_{T4}  \frac{P_3  \Delta_1  \Delta_2}{8 m^3}-A_{T5}\frac{  \Delta_1 z  \Delta_2}{8 m} \right) \hspace*{2cm}\\[3ex]
    \Pi^s_{02}(\Gamma_3) &=& 
   K\,\left(A_{T6} \frac{ E ( E+m)  \Delta_1}{2 m^3}+A_{T8} \frac{ E ( E+m)  P_3  \Delta_1 z}{2 m^3}-A_{T9}  \frac{E ( E+m)  \Delta_1
   z^2}{2 m} \right) \hspace*{2cm}\\[3ex]
    \Pi^s_{03}(\Gamma_0) &=& 
 i\,  K\,\left(A_{T6} \frac{ P_3 \left( \Delta_1^2+ \Delta_2^2\right)}{4 m^3}-A_{T11}\frac{  E \left( E ( E+m)- P_3^2\right) z}{2 m^2} \right) \hspace*{2cm}\\[3ex]
   \Pi^s_{03}(\Gamma_1) &=& 
   K\,\left(A_{T6}\frac{ \left( E ( E+m)- P_3^2\right)  \Delta_2}{2 m^3}+A_{T11}\frac{  E  P_3 z  \Delta_2}{4 m^2} \right) \hspace*{2cm}\\[3ex]
   \Pi^s_{03}(\Gamma_2) &=& 
   K\,\left(A_{T6}\frac{ \left( P_3^2- E
   ( E+m)\right)  \Delta_1}{2 m^3}-A_{T11}\frac{  E  P_3  \Delta_1 z}{4 m^2} \right) \hspace*{2cm}\\[3ex]
    \Pi^s_{12}(\Gamma_1) &=&
  i\, K\,\left(A_{T1}\frac{ E ( E+m)   \Delta_1 z}{4 m^2} \right) \hspace*{2cm}\\[3ex]
    \Pi^s_{12}(\Gamma_2) &=& 
  i\,  K\,\left(A_{T1} \frac{ E ( E+m)   \Delta_2 z}{4 m^2} \right) \hspace*{2cm}\\[3ex]
    \Pi^s_{12}(\Gamma_3) &=& 
  i\,  K\,\left(-A_{T4} \frac{(E+m) ( E^2 - P_3^2)}{2 m^3} -A_{T5} \frac{( E+m)  P_3 z}{2 m} -A_{T7} \frac{ E^2
   ( E+m) z^2}{2 m}-A_{T10}\frac{ ( E+m)}{2 m} \right) \\[3ex]
   \label{Pi_31_s}
   \Pi^s_{31}(\Gamma_0) &=& 
   K\,\left(-A_{T4} \frac{ E  P_3  \Delta_1}{4 m^3}+A_{T10}\frac{  P_3  \Delta_1}{4 m^2} +A_{T12}\frac{ \left( P_3^2- E ( E+m)\right) z  \Delta_1}{2 m^2} \right) \hspace*{2cm}
\end{eqnarray}   
\begin{eqnarray} 
   \Pi^s_{31}(\Gamma_1) &=& 
 i\,  K\,\left(-A_{T2}\frac{  E ( E+m)  \Delta_1  \Delta_2}{4 m^4}+A_{T4} \frac{ E  \Delta_1  \Delta_2}{8 m^3} -A_{T10}\frac{  \Delta_1  \Delta_2}{8
   m^2}-A_{T12}\frac{  P_3  \Delta_1 z  \Delta_2}{4 m^2} \right) \hspace*{2cm}\\[2.5ex]
   \Pi^s_{31}(\Gamma_2) &=& 
 i\,    K\,\left(-A_{T2} \frac{ E ( E+m)  \Delta_2^2}{4 m^4}-A_{T4} \frac{ E \left( \Delta_1^2+4 m ( E+m)\right)}{8 m^3} +A_{T10}\frac{ \left( \Delta_1^2-4  E ( E+m)\right)}{8 m^2}+A_{T12}\frac{  P_3 z  \Delta_1^2}{4
   m^2} \right) \hspace*{0.5cm}\\[2.5ex]
   \Pi^s_{31}(\Gamma_3) &=&  
 i\,    K\,\left(A_{T6}\frac{ ( E+m)  P_3  \Delta_2}{2 m^3}+A_{T8}\frac{
   ( E+m)  \Delta_2 z  E^2}{2 m^3} \right) \hspace*{2cm}\\[2.5ex]
   \Pi^s_{32}(\Gamma_0) &=&  
   K\,\left(-A_{T4}  \frac{E  P_3  \Delta_2}{4 m^3}+A_{T10} \frac{ P_3  \Delta_2}{4 m^2}+A_{T12} \frac{\left( P_3^2- E ( E+m)\right) z  \Delta_2}{2 m^2} \right) \hspace*{2cm}  \\[2.5ex]
   \Pi^s_{32}(\Gamma_1) &=&  
 i\,    K\,\left(A_{T2} \frac{ E ( E+m)  \Delta_1^2}{4 m^4}+A_{T4} \frac{ E \left( \Delta_2^2+4 m ( E+m)\right)}{8 m^3}+A_{T10} \frac{\left(4  E ( E+m)- \Delta_2^2\right)}{8 m^2} -A_{T12}\frac{
    P_3  \Delta_2^2 z}{4 m^2} \right) \\[2.5ex] 
   \Pi^s_{32}(\Gamma_2) &=& 
 i\,    K\,\left(A_{T2}  \frac{E ( E+m)  \Delta_1  \Delta_2}{4 m^4}-A_{T4} \frac{ E  \Delta_1  \Delta_2}{8 m^3}+A_{T10} \frac{ \Delta_1  \Delta_2}{8 m^2}+A_{T12}\frac{  P_3  \Delta_1 z  \Delta_2}{4 m^2} \right)  \\[2.5ex]
   \label{eq:Pi32G3_s}
   \Pi^s_{32}(\Gamma_3) &=&  
  i\,   K\,\left(-A_{T6}\frac{
   ( E+m)  P_3  \Delta_1}{2 m^3} -A_{T8}\frac{ ( E+m)  \Delta_1 z  E^2}{2 m^3} \right)  
\end{eqnarray}
where $K = \frac{2m^2}{E(E+m)}$ since $E_f=E_i\equiv E$ in the symmetric frame when $\xi=0$. Using Eq.~\eqref{eq:Trace} in the asymmetric frame, the general expressions for zero skewness are given below. Once again, vanishing matrix elements are not shown.
\begin{eqnarray}
   \label{eq:Pi01G0_a}
   \Pi^a_{01}(\Gamma_0) &=& 
   K\,\left( A_{T4} \frac{  i  (m^2- E_f^2)  \Delta_1}{4 m^3}+ A_{T5} \frac{
    i   P_3 z  \Delta_1}{4 m} + A_{T10} \frac{  i  ( E_f+m)  \Delta_1}{4 m^2} \right) \\[3ex]
\Pi^a_{01}(\Gamma_1) &=&  
   K\,\left( - A_{T1} \frac{ ( E_f+m)  \Delta_1  \Delta_2 z}{8 m^2}+ A_{T2} \frac{ ( E_f+m)  P_3  \Delta_1  \Delta_2}{4
   m^4} - A_{T3} \frac{ ( E_f+m)  \Delta_1  \Delta_2 z}{4 m^2} - A_{T4} \frac{  P_3  \Delta_1  \Delta_2}{8 m^3}\right. \nonumber \\[1.25ex]  && \left. \qquad + A_{T6} \frac{  P_3  \Delta_1  \Delta_2}{4 m^3} - A_{T7} \frac{  P_3  \Delta_1  \Delta_2 z^2}{8 m}+ A_{T8} \frac{ ( E_f^2-m^2) \Delta_1  \Delta_2 z}{4 m^3}- A_{T9} \frac{  P_3
    \Delta_1  \Delta_2 z^2}{4 m}\right) \\[3ex]
   \Pi^a_{01}(\Gamma_2) &=& 
   K\,\left( - A_{T1} \frac{ ( E_f+m) z  \Delta_2^2}{8 m^2} + A_{T2} \frac{ ( E_f+m)  P_3
    \Delta_2^2}{4 m^4} - A_{T3} \frac{ ( E_f+m) z  \Delta_2^2}{4 m^2} \right. \nonumber \\[1.25ex]  && \left. \qquad + A_{T4} \frac{  P_3 \left(- \Delta_2^2-2 ( E_f- E_i-2 m) ( E_f+m)\right)}{8 m^3}+ A_{T5} \frac{ ( E_f- E_i-2 m) ( E_f+m) z}{4 m} + A_{T6} \frac{  P_3  \Delta_2^2}{4 m^3} \right. \nonumber \\[1.25ex]  && \left. \qquad- A_{T7} \frac{  P_3 z^2  \Delta_2^2}{8 m} + A_{T8} \frac{ ( E_f^2-m^2) z  \Delta_2^2}{4 m^3} - A_{T9} \frac{  P_3 z^2  \Delta_2^2}{4 m}+ A_{T10} \frac{ ( E_f+ E_i+2 m) P_3}{4 m^2} \right) \\[3ex]
    \Pi^a_{01}(\Gamma_3) &=& 
   K\,\left(A_{T1} \frac{ ( E_f{-} E_i)  P_3  \Delta_2 z}{8 m^2} + A_{T2} \frac{ ( E_i{-} E_f) ( E_f^2{-}m^2) \Delta_2}{4 m^4}+ A_{T3} \frac{ ( E_f{-} E_i)  P_3  \Delta_2 z}{4 m^2} + A_{T4} \frac{ ( E_f{+}m) (E_i{-} E_f{+}2 m)  \Delta_2}{8 m^3} \right. \nonumber \\[1.25ex]  && \left. \qquad + A_{T5} \frac{  P_3  \Delta_2 z}{4 m} - A_{T6} \frac{ ( E_f{+} E_i) ( E_f{+}m)
    \Delta_2}{4 m^3} + A_{T7} \frac{ ( E_f{+} E_i) ( E_f{+}m)  \Delta_2 z^2}{8 m}- A_{T8} \frac{ ( E_f{+} E_i) ( E_f{+}m)  P_3  \Delta_2 z}{4
   m^3}\right. \nonumber \\[1.25ex]  
   && \left. \qquad + A_{T9} \frac{ ( E_f+ E_i) ( E_f+m)  \Delta_2 z^2}{4 m}+ A_{T10} \frac{ ( E_f+m)  \Delta_2}{4 m^2} \right) 
\end{eqnarray}
\begin{eqnarray}     
   \Pi^a_{02}(\Gamma_0) &=&
   K\,\left( A_{T4} \frac{  i  (m^2- E_f^2)  \Delta_2}{4 m^3}+ A_{T5} \frac{  i   P_3 z  \Delta_2}{4 m} + A_{T10} \frac{  i  ( E_f+m)  \Delta_2}{4 m^2} \right) \\[3ex]
    \Pi^a_{02}(\Gamma_1) &=& 
   K\,\left(A_{T1} \frac{ ( E_f{+}m) z  \Delta_1^2}{8 m^2} - A_{T2} \frac{ ( E_f{+}m)  P_3  \Delta_1^2}{4 m^4} + A_{T3} \frac{ ( E_f{+}m) z  \Delta_1^2}{4 m^2}  + A_{T4} \frac{  P_3
   \left( \Delta_1^2{+}2 ( E_f{-} E_i{-}2 m) ( E_f{+}m)\right)}{8 m^3} \right. \nonumber \\[1.25ex]  
   && \left. \qquad + A_{T5} \frac{ ( E_f+m) (- E_f+ E_i+2 m) z}{4 m}  - A_{T6} \frac{  P_3
    \Delta_1^2}{4 m^3} + A_{T7} \frac{  P_3 z^2  \Delta_1^2}{8 m}\right. \nonumber \\[1.25ex]  
   && \left. \qquad + A_{T8} \frac{ (m^2- E_f^2)  z  \Delta_1^2}{4 m^3}+ A_{T9} \frac{  P_3 z^2  \Delta_1^2}{4 m} - A_{T10} \frac{ ( E_f+ E_i+2 m)  P_3}{4 m^2} \right) \\[3ex]
    \Pi^a_{02}(\Gamma_2) &=&  
   K\,\left( A_{T1} \frac{ ( E_f+m)  \Delta_1  \Delta_2 z}{8 m^2} - A_{T2} \frac{ ( E_f+m)  P_3  \Delta_1  \Delta_2}{4 m^4} + A_{T3} \frac{ ( E_f+m)  \Delta_1  \Delta_2 z}{4 m^2} \right. \nonumber \\[1.25ex]  
   && \left. \qquad  + A_{T4} \frac{
    P_3  \Delta_1  \Delta_2}{8 m^3}- A_{T6} \frac{  P_3  \Delta_1  \Delta_2}{4 m^3} + A_{T7} \frac{  P_3  \Delta_1  \Delta_2 z^2}{8 m} + A_{T8} \frac{ (m^2- E_f^2)  \Delta_1  \Delta_2 z}{4 m^3} + A_{T9} \frac{  P_3  \Delta_1  \Delta_2 z^2}{4
   m} \right)\\[3ex]
    \Pi^a_{02}(\Gamma_3) &=& 
   K\,\left(A_{T1} \frac{ ( E_i- E_f)  P_3  \Delta_1 z}{8
   m^2}+ A_{T2} \frac{
   ( E_f- E_i) ( E_f^2-m^2)  \Delta_1}{4 m^4} + A_{T3} \frac{ ( E_i- E_f)  P_3  \Delta_1 z}{4 m^2}\right. \nonumber \\[1.25ex]  
   && \left. \qquad   + A_{T4} \frac{ ( E_f{-} E_i{-}2 m) (E_f{+}m)  \Delta_1}{8 m^3} - A_{T5} \frac{  P_3  \Delta_1 z}{4 m} + A_{T6} \frac{ (E_f{+} E_i) ( E_f{+}m)  \Delta_1}{4 m^3}  - A_{T7} \frac{ ( E_f{+} E_i) ( E_f{+}m)
    \Delta_1 z^2}{8 m}\right. \nonumber \\[1.25ex]  
   && \left. \qquad  + A_{T8} \frac{ ( E_f+ E_i) ( E_f+m)  P_3  \Delta_1 z}{4 m^3} - A_{T9} \frac{ ( E_f+ E_i) ( E_f+m)  \Delta_1 z^2}{4 m} - A_{T10} \frac{ ( E_f+m)  \Delta_1}{4 m^2} \right) \\[3ex]
    \Pi^a_{03}(\Gamma_0) &=& 
   K\,\left(A_{T4} \frac{( E_f^2- E_i^2)  i   P_3}{8 m^3}+ A_{T6} \frac{( E_i^2- E_f^2)  i   P_3}{4 m^3}+  A_{T10} \frac{( E_i- E_f)  i   P_3}{4 m^2}\right. \nonumber \\[1.25ex]  
   && \left. \qquad + A_{T11} \frac{( E_f+ E_i)  i  ( E_f- E_i-2 m) ( E_f+m) z}{8 m^2}- A_{T12}\frac{
   ( E_f- E_i)  i  ( E_f- E_i-2 m) ( E_f+m) z}{4 m^2} \right) \\[3ex]
   \Pi^a_{03}(\Gamma_1) &=& 
   K\,\left(A_{T4} \frac{ ( E_f- E_i-2 m) ( E_f+m)  \Delta_2}{8 m^3}+ A_{T6} \frac{ ( E_f+m) (- E_f+ E_i+2 m)  \Delta_2}{4 m^3}\right. \nonumber \\[1.25ex]  
   && \left. \qquad- A_{T10} \frac{ ( E_f+m)  \Delta_2}{4 m^2}+ A_{T11} \frac{ ( E_f+ E_i)  P_3 z  \Delta_2}{8 m^2}+ A_{T12} \frac{ ( E_i- E_f)  P_3 z
    \Delta_2}{4 m^2} \right)\\[3ex]
   \Pi^a_{03}(\Gamma_2) &=& 
   K\,\left( A_{T4} \frac{ ( E_f+m) (- E_f+ E_i+2 m)  \Delta_1}{8 m^3}+ A_{T6} \frac{
   ( E_f- E_i-2 m) ( E_f+m)  \Delta_1}{4 m^3}\right. \nonumber \\[1.25ex]  
   && \left. \qquad + A_{T10}\frac{ ( E_f+m)  \Delta_1}{4 m^2}- A_{T11} \frac{ ( E_f+ E_i)  P_3 z  \Delta_1}{8
   m^2}+ A_{T12} \frac{ ( E_f- E_i)  P_3 z  \Delta_1}{4 m^2} \right)\\[3ex]
   \Pi^a_{12}(\Gamma_1) &=& 
   K\,\left( A_{T1} \frac{ ( E_f+ E_i)  i  ( E_f+m)  \Delta_1 z}{8 m^2} + A_{T2} \frac{ ( E_f- E_i)  i  ( E_f+m)  P_3  \Delta_1}{4 m^4} - A_{T3} \frac{ ( E_f- E_i)  i  ( E_f+m)  \Delta_1 z}{4 m^2} \right. \nonumber \\[1.25ex]  
   && \left. \qquad - A_{T4} \frac{  i  ( E_f- E_i+2 m)  P_3  \Delta_1}{8 m^3}+ A_{T5} \frac{  i  ( E_f+m)
    \Delta_1 z}{4 m} + A_{T6} \frac{ ( E_f- E_i)  i   P_3
    \Delta_1}{4 m^3} + A_{T7} \frac{ ( E_f+ E_i)  i   P_3  \Delta_1 z^2}{8 m}\right. \nonumber \\[1.25ex]  
   && \left. \qquad  + A_{T8} \frac{ ( E_f- E_i)  i  ( E_f^2-m^2)  \Delta_1 z}{4
   m^3} +  A_{T9} \frac{ ( E_i- E_f)  i   P_3  \Delta_1 z^2}{4 m}- A_{T10} \frac{  i   P_3  \Delta_1}{4 m^2} \right) 
\end{eqnarray}   
\begin{eqnarray}   
     \Pi^a_{12}(\Gamma_2) &=&
   K\,\left(A_{T1} \frac{ ( E_f+ E_i)  i  ( E_f+m)  \Delta_2 z}{8 m^2}+ A_{T2} \frac{ ( E_f- E_i)
    i  ( E_f+m)  P_3  \Delta_2}{4 m^4} - A_{T3} \frac{
   ( E_f- E_i)  i  ( E_f+m)  \Delta_2 z}{4 m^2}\right. \nonumber \\[1.25ex]  
   && \left. \qquad  - A_{T4} \frac{  i  ( E_f- E_i+2 m)  P_3  \Delta_2}{8 m^3} + A_{T5} \frac{  i  ( E_f+m)  \Delta_2 z}{4 m} + A_{T6} \frac{ ( E_f- E_i)  i   P_3  \Delta_2}{4 m^3} + A_{T7} \frac{ ( E_f+ E_i)  i   P_3  \Delta_2 z^2}{8 m} \right. \nonumber \\[1.25ex]  
   && \left. \qquad + A_{T8} \frac{ ( E_f- E_i)  i  ( E_f^2-m^2) \Delta_2 z}{4 m^3}+ A_{T9} \frac{ ( E_i- E_f)  i   P_3  \Delta_2 z^2}{4 m}- A_{T10} \frac{  i   P_3  \Delta_2}{4 m^2} \right)  \\[3ex]
  \Pi^a_{12}(\Gamma_3) &=& 
   K\,\left(A_{T1} \frac{
   ( E_i^2- E_f^2)  i   P_3 z}{8 m^2}- A_{T2} \frac{  i  ( E_f^2-m^2)  ( E_f- E_i)^2}{4 m^4}+ A_{T3} \frac{  i   P_3 z ( E_f- E_i)^2}{4 m^2} \right. \nonumber \\[1.25ex]  
   && \left. \qquad + A_{T4} \frac{  i  ( E_f{+}m) \left( E_f^2{+}2 m  E_f{-} E_i^2{-}2 m ( E_i{+}2 m)\right)}{8 m^3}- A_{T5} \frac{  i  ( E_f{+} E_i{+}2 m)  P_3 z}{4 m} - A_{T6} \frac{  i  ( E_f{+}m) ( E_f^2{-} E_i^2)}{4 m^3}  \right. \nonumber \\[1.25ex]  
   && \left. \qquad - A_{T7} \frac{ ( E_f{+} E_i)^2  i  ( E_f{+}m)
   z^2}{8 m} - A_{T8} \frac{  i  ( E_f{+}m)  P_3 z ( E_f^2{-} E_i^2)}{4 m^3} + A_{T9} \frac{  i  ( E_f{+}m) z^2
   ( E_f^2{-} E_i^2)}{4 m} \right. \nonumber \\[1.25ex]  
   && \left. \qquad + A_{T10} \frac{  i  ( E_f- E_i-2 m) ( E_f+m)}{4 m^2} \right) \\[3ex]
   \label{Pi_31_a}
\Pi^a_{31}(\Gamma_0) &=& 
   K\,\left(- A_{T4} \frac{ ( E_f+ E_i)
    P_3  \Delta_1}{8 m^3} + A_{T6} \frac{ ( E_i- E_f)  P_3  \Delta_1}{4 m^3}+ A_{T10} \frac{  P_3  \Delta_1}{4 m^2}\right. \nonumber \\[1.25ex]  
   && \left. \qquad + A_{T11} \frac{ ( E_f- E_i-2 m) ( E_f+m) z  \Delta_1}{8 m^2}+ A_{T12} \frac{ ( E_f- E_i-2 m) ( E_f+m) z  \Delta_1}{4 m^2} \right)\\[3ex]
   \Pi^a_{31}(\Gamma_1) &=& 
   K\,\left(- A_{T2} \frac{  E_f  i  ( E_f+m)  \Delta_1  \Delta_2}{4 m^4} + A_{T4} \frac{  i  ( E_f+m)  \Delta_1  \Delta_2}{8 m^3}- A_{T6} \frac{  i  ( E_f+m)  \Delta_1  \Delta_2}{4 m^3}\right. \nonumber \\[1.25ex]  
   && \left. \qquad- A_{T8} \frac{  E_f  i   P_3  \Delta_1 z  \Delta_2}{4 m^3} - A_{T11} \frac{  i   P_3
    \Delta_1 z  \Delta_2}{8 m^2}- A_{T12} \frac{  i   P_3  \Delta_1 z  \Delta_2}{4 m^2} \right) \\[3ex]
   \Pi^a_{31}(\Gamma_2) &=& 
   K\,\left(- A_{T2} \frac{  E_f  i  ( E_f+m)  \Delta_2^2}{4 m^4}- A_{T4} \frac{  i  ( E_f+m) \left( \Delta_1^2+2 ( E_f+ E_i) m\right)}{8 m^3} \right. \nonumber \\[1.25ex]  
   && \left. \qquad+ A_{T6} \frac{  i  ( E_f+m) \left( \Delta_1^2+2 ( E_f- E_i) ( E_f-m)\right)}{4
   m^3} - A_{T8} \frac{  E_f  i   P_3  \Delta_2^2 z}{4 m^3}\right. \nonumber \\[1.25ex]  
   && \left. \qquad - A_{T10} \frac{ ( E_f+ E_i)  i  ( E_f+m)}{4 m^2}+  A_{T11} \frac{  i   P_3 z  \Delta_1^2}{8 m^2}+ A_{T12} \frac{  i 
    P_3 z  \Delta_1^2}{4 m^2} \right) \hspace*{2cm}\\[3ex]
   \Pi^a_{31}(\Gamma_3) &=&  
   K\,\left( A_{T2} \frac{  E_f ( E_f- E_i)  i   P_3  \Delta_2}{4 m^4} - A_{T4} \frac{  i   P_3  \Delta_2}{4 m^2} + A_{T6} \frac{  i  ( E_f+m)  P_3  \Delta_2}{2
   m^3}\right. \nonumber \\[1.25ex]  
   && \left. \qquad + A_{T8} \frac{  E_f ( E_f+ E_i)  i  ( E_f+m) z  \Delta_2}{4
   m^3} - A_{T10} \frac{  i   P_3  \Delta_2}{4 m^2} \right) \hspace*{2cm}\\[3ex]
   \Pi^a_{32}(\Gamma_0) &=&  
   K\,\left(- A_{T4} \frac{ ( E_f+ E_i)  P_3  \Delta_2}{8 m^3} + A_{T6} \frac{ ( E_i- E_f)  P_3  \Delta_2}{4 m^3} + A_{T10} \frac{  P_3  \Delta_2}{4 m^2} \right. \nonumber \\[1.25ex]  && \left. \qquad + A_{T11} \frac{ ( E_f- E_i-2 m) ( E_f+m) z  \Delta_2}{8
   m^2}+ A_{T12} \frac{ ( E_f- E_i-2 m) ( E_f+m) z  \Delta_2}{4 m^2} \right) \hspace*{2cm}
\end{eqnarray}
\begin{eqnarray}   
   \Pi^a_{32}(\Gamma_1) &=&  
   K\,\left(A_{T2} \frac{  E_f  i  ( E_f+m)  \Delta_1^2}{4 m^4} + A_{T4} \frac{  i  ( E_f+m) \left( \Delta_2^2+2 ( E_f+ E_i) m\right)}{8 m^3} \right. \nonumber \\[1.25ex]  && \left. \qquad- A_{T6} \frac{  i  ( E_f+m) \left( \Delta_2^2+2
   ( E_f- E_i) ( E_f-m)\right)}{4 m^3} + A_{T8} \frac{  E_f  i   P_3 z  \Delta_1^2}{4 m^3}\right. \nonumber \\[1.25ex]  && \left. \qquad + A_{T10} \frac{ ( E_f+ E_i)  i  ( E_f+m)}{4 m^2}- A_{T11} \frac{  i   P_3  \Delta_2^2 z}{8 m^2}- A_{T12} \frac{  i   P_3  \Delta_2^2 z}{4
   m^2} \right) \hspace*{2cm}\\[3ex]
   \Pi^a_{32}(\Gamma_2) &=& 
   K\,\left(A_{T2} \frac{  E_f  i 
   ( E_f+m)  \Delta_1  \Delta_2}{4 m^4}- A_{T4} \frac{  i  ( E_f+m)  \Delta_1  \Delta_2}{8 m^3}+ A_{T6} \frac{  i  ( E_f+m)  \Delta_1  \Delta_2}{4 m^3} \right. \nonumber \\[1.25ex]  && \left. \qquad + A_{T8} \frac{  E_f  i   P_3  \Delta_1 z  \Delta_2}{4 m^3} + A_{T11} \frac{  i   P_3  \Delta_1 z  \Delta_2}{8 m^2}+ A_{T12} \frac{  i   P_3  \Delta_1 z  \Delta_2}{4 m^2} \right) \hspace*{2cm}\\[3ex]
   \label{eq:Pi32G3_a}
   \Pi^a_{32}(\Gamma_3) &=&  
   K\,\left( A_{T2} \frac{  E_f ( E_i- E_f)  i   P_3  \Delta_1}{4 m^4} + A_{T4} \frac{  i   P_3  \Delta_1}{4 m^2} - A_{T6} \frac{  i  ( E_f+m)  P_3  \Delta_1}{2
   m^3}\right. \nonumber \\[1.25ex]  && \left. \qquad- A_{T8} \frac{  E_f ( E_f+ E_i)  i  ( E_f+m) z  \Delta_1}{4
   m^3}+ A_{T10} \frac{  i   P_3  \Delta_1}{4 m^2}  \right) \hspace*{2cm}
\end{eqnarray}
As discussed in Sec.~\ref{sec:theory}, the amplitudes $zA_{T1},~A_{T6},~zA_{T8},~z^2A_{T9}$ and $zA_{T11}$ vanish at zero skewness. However, we include them in the above equations so that we can test if the amplitudes do in fact vanish numerically, which serves as a non-trivial check. 
For the quasi-GPDs, we can then also check whether their inclusion reduces statistical noise due to cancellation of matrix elements. We can utilize the matrix elements to disentangle all twelve amplitudes $A_{Ti}$ in both the symmetric (Eqs.~\eqref{eq:Pi01G0_s} - \eqref{eq:Pi32G3_s}) and the asymmetric (Eqs.~\eqref{eq:Pi01G0_a} - \eqref{eq:Pi32G3_a}) frames. In practice, we solve 12 linear equations numerically for each value of $P_3$ and $\vec{\Delta}$. The analytic expressions can be very lengthy, so we only provide the example of the symmetric frame for $\vec{\Delta} = (\Delta,0,0)$ for each amplitude presented below. The corresponding expressions for the asymmetric frame are much more complicated and impractical to present here. We find 
\begin{eqnarray}
\label{eq:AT1}
z A_{T1}&=& -\,i\, \frac{2 }{\Delta_1} \Pi^s_{12}(\Gamma_1)\\[3ex]
A_{T2}&=& \frac{P_3  m^3}{(E+m) \left(E^2-P_3^2\right) \Delta_1}\Pi^s_{31}(\Gamma_0)-\,i\, \frac{2 \left(E^2+m E-P_3^2\right) 
   m^3}{(E+m) \left(E^2-P_3^2\right) \Delta_1^2}\Pi^s_{31}(\Gamma_2) \nonumber \\[1.25ex]  && \qquad
   + \,i\,\frac{ \left(E^2+m E-P_3^2\right)  m^2}{(E+m) \left(E^2-P_3^2\right) \Delta_1}\Pi^s_{01}(\Gamma_0)+\frac{P_3  m^2}{2
   (E+m) \left(P_3^2-E^2\right)}\Pi^s_{01}(\Gamma_2)-\,i\, \frac{2  m^2}{\Delta_1^2}\Pi^s_{32}(\Gamma_1)\\[3ex]
 z A_{T3}&=& \frac{m  P_3^2}{(E+m) \left(E^2-P_3^2\right) \Delta_1}\Pi^s_{31}(\Gamma_0)+\,i\, \frac{E  m
    P_3}{(E+m) \left(E^2-P_3^2\right) \Delta_1}\Pi^s_{01}(\Gamma_0)  - \,i\,\frac{2 P_3}{\Delta_1^2}\Pi^s_{32}(\Gamma_1)\nonumber \\[1.25ex]  && \qquad
   +\,i\, \frac{2 m \left(P_3^2-E (E+m)\right)  P_3}{(E+m) \left(E^2-P_3^2\right) \Delta_1^2}\Pi^s_{31}(\Gamma_2)+\frac{2 E m \left(E^2+m E-P_3^2\right) }{(E+m) \left(E^2-P_3^2\right) \Delta_1^2}\Pi^s_{01}(\Gamma_2)+\frac{2 E }{\Delta_1^2}\Pi^s_{02}(\Gamma_1)\\[3ex]
A_{T4}&=& \,i\, \frac{2 \left(E^2+m E-P_3^2\right)  m^2}{(E+m) \left(E^2-P_3^2\right) \Delta_1}\Pi^s_{01}(\Gamma_0)+\frac{P_3  m^2}{(E+m)
   \left(P_3^2-E^2\right)}\Pi^s_{01}(\Gamma_2)+\frac{P_3 \Delta_1  m}{2 (E+m) \left(P_3^2-E^2\right)}\Pi^s_{31}(\Gamma_0)\nonumber \\[1.25ex]  && \qquad
   +\,i\, \frac{ \left(E^2+m E-P_3^2\right)  m}{(E+m)
   \left(E^2-P_3^2\right)}\Pi^s_{31}(\Gamma_2)\\[3ex]
 z A_{T5}&=& \frac{\Delta_1 P_3^2}{2 m (E+m) \left(P_3^2-E^2\right)}\Pi^s_{31}(\Gamma_0) -\,i\, \frac{E  \Delta_1 P_3}{2 m (E+m)
   \left(E^2-P_3^2\right)}\Pi^s_{01}(\Gamma_0)\nonumber \\[1.25ex]  && \qquad
   +\,i\, \frac{ \left(E^2+m E-P_3^2\right)  P_3}{m (E+m) \left(E^2-P_3^2\right)}\Pi^s_{31}(\Gamma_2)+\frac{E \left(P_3^2-E (E+m)\right)}{m (E+m) \left(E^2-P_3^2\right)}\Pi^s_{01}(\Gamma_2)
\end{eqnarray}
\begin{eqnarray}   
A_{T6}&=& \frac{E m \left(P_3^2-E (E+m)\right) }{(E+m) \left(E^2-P_3^2\right) \Delta_1}\Pi^s_{03}(\Gamma_2)-\,i\, \frac{E  m
   P_3 }{2 (E+m) \left(E^2-P_3^2\right)}\Pi^s_{03}(\Gamma_0)\\[3ex]
z^2 A_{T7}&=& \,i\, \frac{ \Delta_1  P_3^2}{2 E m (E+m) \left(E^2-P_3^2\right)}\Pi^s_{01}(\Gamma_0)+\frac{\left(E^2+m
   E-P_3^2\right)  P_3}{E m (E+m) \left(E^2-P_3^2\right)}\Pi^s_{01}(\Gamma_2) +\,i\, \frac{1}{E m}\Pi^s_{12}(\Gamma_3)\nonumber \\[1.25ex]  && \qquad
  +\frac{\Delta_1 P_3}{2 m (E+m) \left(E^2-P_3^2\right)}\Pi^s_{31}(\Gamma_0) +\,i\, \frac{ \left(P_3^2-E (E+m)\right)}{m (E+m) \left(E^2-P_3^2\right)}\Pi^s_{31}(\Gamma_2)\\[3ex]
z A_{T8}&=& \,i\, \frac{ m P_3^2}{2 E (E+m)
   \left(E^2-P_3^2\right)}\Pi^s_{03}(\Gamma_0)+\frac{m \left(E^2+m E-P_3^2\right) P_3}{E (E+m) \left(E^2-P_3^2\right) \Delta_1}\Pi^s_{03}(\Gamma_2)+\,i\, \frac{ m}{E \Delta_1}\Pi^s_{32}(\Gamma_3)\\[3ex]
 z^2 A_{T9}&=& -\frac{1}{m \Delta_1}\Pi^s_{02}(\Gamma_3)-\,i\, \frac{ P_3 }{2 E m (E+m)}\Pi^s_{03}(\Gamma_0)+\frac{\left(P_3^2-E (E+m)\right) }{E m (E+m) \Delta_1}\Pi^s_{03}(\Gamma_2)+\,i\, \frac{ P_3 }{E m \Delta_1}\Pi^s_{32}(\Gamma_3)\\[3ex]
A_{T10}&=& \,i\, \frac{2 \left(P_3^2-E (E+m)\right) }{(E+m) \Delta_1}\Pi^s_{01}(\Gamma_0)+\frac{P_3 }{E+m}\Pi^s_{01}(\Gamma_2)\\[3ex]
 z A_{T11}&=&  
   \,i\, \frac{ \left(E^2+m E-P_3^2\right) }{(E+m) \left(E^2-P_3^2\right)}\Pi^s_{03}(\Gamma_0)+\frac{P_3 \Delta_1 }{2 (E+m) \left(P_3^2-E^2\right)}\Pi^s_{03}(\Gamma_2)\\[3ex]
 z A_{T12}&=&
   \frac{ P_3^2}{2 (E+m) \left(E^2-P_3^2\right)}\Pi^s_{01}(\Gamma_2)+\,i\, \frac{ \left(P_3^2-E (E+m)\right)  P_3}{(E+m) \left(E^2-P_3^2\right) \Delta_1
  }\Pi^s_{01}(\Gamma_0)\nonumber \\[1.25ex]  && \qquad
  -\,i\, \frac{E P_3}{2 (E+m) \left(E^2-P_3^2\right)}\Pi^s_{31}(\Gamma_2) +\frac{E \left(P_3^2-E (E+m)\right) }{(E+m) \left(E^2-P_3^2\right) \Delta_1}    \Pi^s_{31}(\Gamma_0)
  \label{eq:AT12}
\end{eqnarray}

\noindent
We expect that, numerically, the amplitudes $zA_{T1},~A_{T6},~zA_{T8},~z^2A_{T9}$ and $zA_{T11}$ must vanish at zero skewness as indicated by their symmetry properties under the transformation $\xi \to -\xi$. The inclusion or exclusion of these amplitudes in the decomposition does not alter the expressions for the remaining ones. From the extracted amplitudes, we can then compute the quasi-GPDs. As discussed above, the definition of the quasi-GPDs is not unique.
Here we utilize the two definitions presented above, concentrating on the special case $\xi = 0$:
\begin{itemize}
    \item[\textbf{(a)}]  Lorentz-invariant definition for the quasi-GPDs (Eqs.~\eqref{eq:quasi_c_1} - \eqref{eq:quasi_c_2}); \\[-5ex]
    \item[\textbf{(b)}]  standard definition for the quasi-GPDs based on operator $\sigma^{3j}$ in the symmetric frame (Eqs.~\eqref{eq:quasi_a_1} - \eqref{eq:quasi_a_2}).
\end{itemize}   
We reiterate that, for each definition, one may use the amplitudes calculated in any frame due to their Lorentz invariance. 
However, the quasi-GPDs in the standard definition are frame-dependent.
If, in the definition \textbf{(b)}, we use $A_{Ti}$ from the asymmetric frame at a given $-t^a$, the quasi-GPDs must be evaluated at $t = t^a$. As discussed in Ref.~\cite{Bhattacharya:2022aob}, this can be achieved by an appropriate Lorentz transformation, which then ensures a consistent calculation of these quasi-GPDs. 
For zero skewness, the relations between the quasi-GPDs and the amplitudes are given below for: \\
-- the Lorentz invariant definition (\textbf{a})
\begin{eqnarray}
\label{eq:quasi_c_1}
\mathcal{H}_T &=& -2A_{T2}\bigg( 1+ \frac{P^2}{M^2}\bigg) + A_{T4} + A_{T10} \hspace*{4.15cm}\\[2ex]
    \label{eq:quasi_c_3}
\mathcal{E}_T &=& 2A_{T2} -A_{T4} \hspace*{2cm}\\[2ex]
            \label{eq:quasi_c_5}
\widetilde{\mathcal{H}}_T
 &=& -A_{T2} \hspace*{2cm}\\[2ex]
 \widetilde{\mathcal{E}}_T &=& -2A_{T6} - 2P_3\,zA_{T8}
    \label{eq:quasi_c_2}
\end{eqnarray}
-- the standard definitions in symmetric (\textbf{b}) 
\begin{eqnarray}
\label{eq:quasi_a_1}
    \mathcal{H}_T^{s} &=& -2A_{T2}\bigg(1 + \frac{P^2}{M^2}\bigg) + A_{T4} - z A_{T8}\bigg(\frac{E_f^2 - E_i^2}{2 P_3}\bigg) + A_{T10} \hspace*{2cm}\\[2ex]
    \mathcal{E}_T^{s} &=& 2A_{T2} - A_{T4} + z A_{T8}\bigg(\frac{E_f^2 - E_i^2}{2 P_3}\bigg) \hspace*{2cm}\\[2ex]
    \label{eq:quasi_a_3}
    \widetilde{\mathcal{H}}_T^{s} &=& -A_{T2} - z A_{T12}\frac{M^2}{P_3} \hspace*{2cm}\\[2ex]
    \widetilde{\mathcal{E}}_T^{s} &=& -2A_{T6} - z A_{T8}\frac{(E_f+E_i)^2}{2 P_3}
    \label{eq:quasi_a_2}
\end{eqnarray}
where $P =(p_f+p_i)/2$. We note that the combination $P_3\,z$ in Eq.~\eqref{eq:quasi_c_2} originated from the Lorentz invariant combination $P\cdot z$ at zero skewness. The above definitions contain the amplitudes $A_{T6}$ and $A_{T8}$ that vanish for zero skewness. If these amplitudes are removed, we readily see that $\mathcal{H}_T^{s}=\mathcal{H}_T$ and $\mathcal{E}_T^{s}=\mathcal{E}_T$. 
Interestingly, the difference $E_f^2 - E_i^2$ in the coefficient of $A_{T8}$ at $\xi = 0$ is zero in the symmetric frame, which reinforces that the functional forms of these two quasi-GPDs are the same for the two definitions.
In addition, as already discussed above, $\widetilde{\mathcal{E}}_T^{s}= \widetilde{\mathcal{E}}_T=0$.

It is insightful to examine the specific operators entering each definition. For demonstration purposes, we adopt the kinematic setup $\vec{\Delta}=(\Delta_1,0,0)$ and substitute the $A_{Ti}$ derived above into the two definitions of the quasi-GPDs. To ensure clarity in our analysis, we omit the amplitudes that vanish in this scenario, namely $A_{T6}$ and $z A_{T8}$. As demonstrated below, the standard definitions exclusively involve the twist-2 operators $\sigma^{31}$ and $\sigma^{32}$. This property extends to $\mathcal{H}_T$ and $\mathcal{E}_T$ within the Lorentz-invariant definition, which coincide with their standard ones. Notably, the quasi-GPD $\widetilde{\mathcal{H}}_T$ differs by including the other twist-2 counterpart of $\sigma^{+j}$, $\sigma^{01}$. For a more general kinematic setup, $\vec{\Delta}=(\Delta_1,\Delta_2,0)$, the operator $\sigma^{02}$ also emerges within the corresponding matrix elements. 

\begin{eqnarray}
\label{eq:quasiGPDs1}
\mathcal{H}_T &=&  i\, \Pi^s_{32}(\Gamma_1) \\[3ex]
\mathcal{E}_T &=& -\frac{4  i\, m \Pi^s_{31}(\Gamma_2) \left( E ( E+m)- P_3^2\right)}{\Delta_1^2 ( E+m)}+\frac{2 m  P_3 \Pi^s_{31}(\Gamma_0)}{\Delta_1 ( E+m)}-\frac{4  i\, m^2
   \Pi^s_{32}(\Gamma_1)}{\Delta_1^2} \\[3ex]
\widetilde{\mathcal{H}}_T &=& \frac{2  i\, m^3 \Pi^s_{31}(\Gamma_2) \left( E ( E+m)- P_3^2\right)}{\Delta_1^2 ( E+m) \left( E^2- P_3^2\right)}+\frac{2  i\, m^2
   \Pi^s_{32}(\Gamma_1)}{\Delta_1^2}+\frac{m^3  P_3 \Pi^s_{31}(\Gamma_0)}{\Delta_1 ( E+m) \left( P_3^2- E^2\right)} \nonumber \\[1.25ex]  
   && \qquad  -\frac{ i\, m^2 \Pi^s_{01}(\Gamma_0) \left( E ( E+m)- P_3^2\right)}{\Delta_1
   ( E+m) \left( E^2- P_3^2\right)} +\frac{m^2  P_3 \Pi^s_{01}(\Gamma_2)}{2 ( E+m) \left( E^2- P_3^2\right)} 
 \\[3ex]
\widetilde{\mathcal{H}}^s_T &=& \frac{2  i\, m^2 \Pi^s_{31}(\Gamma_2)}{\Delta_1^2}+\frac{2  i\, m^2 \Pi^s_{32}(\Gamma_1)}{\Delta_1^2}+\frac{m^2 \Pi^s_{31}(\Gamma_0)}{ P_3 \Delta_1}
\label{eq:quasiGPDs2}
\end{eqnarray}

\subsection{Computational Setup}

The proton matrix elements required for the calculation are based on a non-local tensor operator, which includes spatially separated quark fields along the $\hat{z}$-direction. The momentum boost and the Wilson line are both along this same direction. The matrix elements involve a momentum transfer between the initial and final proton states. These matrix elements can be expressed as
\begin{equation}
\label{eq:ME}
F^{[i\sigma^{\mu\nu}\gamma_5]}(\Gamma_\kappa,z,p_f,p_i)\equiv \langle N(p_f)|\bar\psi\left(z\right) i\sigma^{\mu\nu} {\cal W}(0,z)\psi\left(0\right)|N(p_i)\rangle\,, \quad \mu, \nu, \kappa: 0,1,2,3\,.
\end{equation}
Here, $|N(p_i)\rangle$ and $|N(p_f)\rangle$ refer to the initial and final proton states, while the operator setup, including momentum transfer and displacements, has been discussed earlier. To enhance the overlap with the proton’s ground state and reduce gauge noise, we employ momentum smearing, as detailed in Ref.~\cite{Bali:2016lva}, which was shown to be critical for calculations of matrix elements with non-local operators~\cite{Alexandrou:2016jqi}. In earlier studies~\cite{Alexandrou:2020zbe}, this method reduced the statistical noise significantly, by a factor of 4-5 for the real part and 2-3 for the imaginary part of the quasi-GPDs. Furthermore, five iterations of stout smearing~\cite{Morningstar:2003gk} with a smearing parameter $\rho=0.15$ are applied to the gauge links, which helps suppress gauge noise, as demonstrated in previous studies~\cite{Alexandrou:2016ekb,Alexandrou:2020sml}.
It should be noted that while stout smearing affects both the matrix elements and the renormalization function, the renormalized matrix elements remain unaffected by the smearing procedure. This has been confirmed in earlier tests where varying the number of smearing steps (ranging from 0 to 20) showed no significant change in the renormalized matrix elements. These tests were performed with zero momentum transfer and on ensembles with physical pion masses~\cite{Alexandrou:2019lfo}. Similarly, studies of gluon PDFs using this ensemble have confirmed the effectiveness of this approach~\cite{Delmar:2023agv}.
The matrix elements are extracted by calculating the ratio
\begin{equation}
\label{eq:ratio}
R_\mu (\Gamma_\kappa, z, p_f, p_i; t_s, \tau) = \frac{C^{\rm 3pt}_\mu (\Gamma_\kappa, z, p_f, p_i; t_s, \tau)}{C^{\rm 2pt}(\Gamma_0, p_f;t_s)} \sqrt{\frac{C^{\rm 2pt}(\Gamma_0, p_i, t_s-\tau)C^{\rm 2pt}(\Gamma_0, p_f, \tau)C^{\rm 2pt}(\Gamma_0, p_f, t_s)}{C^{\rm 2pt}(\Gamma_0, p_f, t_s-\tau)C^{\rm 2pt}(\Gamma_0, p_i, \tau)C^{\rm 2pt}(\Gamma_0, p_i, t_s)}}\,.
 \end{equation}
In this expression, $C^{\rm 2pt}$ and $C^{\rm 3pt}$ are the two-point and three-point correlation functions, respectively, while $\tau$ represents the current insertion time, and $t_s$ is the source-sink time separation. We then extract the ground-state contribution to the matrix elements by fitting $R_\mu$ to a plateau with respect to $\tau$, thus isolating the relevant matrix element  $\Pi_\mu(\Gamma_\kappa)$, where the dependence on $z$, $p_f$, and $p_i$ is not shown explicitly. 

 \begin{table}[h!]
\centering
\renewcommand{\arraystretch}{1.2}
\renewcommand{\tabcolsep}{6pt}
  \begin{tabular}{| l| c | c | c | c | c  | c | c |}
  \hline
    \multicolumn{8}{|c|}{Parameters} \\
    \hline
 Ensemble   & $\beta$ & $a$ [fm] & volume $L^3\times T$ & $N_f$ & $m_\pi$ [MeV] &
$L m_\pi$ & $L$ [fm]\\
    \hline
cA211.32 & 1.726 & 0.093  & $32^3\times 64$  & $u, d, s, c$ & 260
& 4 & 3.0 \\
    \hline
    \end{tabular}
  \caption{\small Parameters of the ensemble used in this work.}
  \label{tab:params}
  \vspace*{0.2cm}
\end{table}
For this study, we use a gauge ensemble with $N_f=2+1+1$ flavors of twisted-mass fermions, with an added clover term~\cite{Alexandrou:2018egz}. The gluon action is Iwasaki-improved, and the lattice volume is $32^3 \times 64$ with a lattice spacing $a=0.093$fm. The ensemble corresponds to a pion mass of 260 MeV. The detailed parameters of the ensemble are provided in Table~\ref{tab:params}.

We focus on the isovector flavor combination $u-d$, which requires only the connected diagram. The matrix elements are computed at a source-sink separation of $t_s=10 a = 0.934$ fm, a choice made to control the need for minimizing excited-state contamination while maintaining statistical accuracy. Although a study of excited-state effects is beyond the scope of this paper, the chosen separation time has been previously found to suppress such effects in similar calculations. 
The statistics of the calculations for both symmetric and asymmetric frames are summarized in Table~\ref{tab:stat}. 
As can be seen, we implement all kinematically equivalent momenta that lead to the same value of $p_i^2$, $p_f^2$, and average the matrix elements leading to the same amplitudes by taking into consideration their symmetry properties as presented in Sec.~\ref{sec:symmetry}.
\begin{table}[h!]
\begin{center}
\renewcommand{\arraystretch}{1.9}
\begin{tabular}{lcccc|cccc}
\hline
frame & $P_3$ [GeV] & $\quad \mathbf{\Delta}$ $[\frac{2\pi}{L}]\quad$ & $-t$ [GeV$^2$] & $\quad \xi \quad $ & $N_{\rm ME}$ & $N_{\rm confs}$ & $N_{\rm src}$ & $N_{\rm tot}$\\
\hline
N/A       & $\pm$1.25 &(0,0,0)  &0   &0   &2   &329  &16  &10528 \\
\hline
symm      & $\pm$0.83 &($\pm$2,0,0), (0,$\pm$2,0)  &0.69   &0   &8   &67 &8  &4288 \\
symm      & $\pm$1.25 &($\pm$2,0,0), (0,$\pm$2,0)  &0.69   &0   &8   &249 &8  &15936 \\
symm      & $\pm$1.67 &($\pm$2,0,0), (0,$\pm$2,0)  &0.69   &0   &8   &294 &32  &75264 \\
symm      & $\pm$1.25 &$(\pm 2,\pm 2,0)$           &1.38   &0   &16   &224 &8  &28672 \\
symm      & $\pm$1.25 &($\pm$4,0,0), (0,$\pm$4,0)  &2.77   &0   &8   &329 &32  &84224 \\
\hline
asymm  & $\pm$1.25 &($\pm$1,0,0), (0,$\pm$1,0)  &0.17   &0   &8   &269 &8  &17216\\
asymm      & $\pm$1.25 &$(\pm 1,\pm 1,0)$       &0.34   &0   &16   &195 &8  &24960 \\
asymm  & $\pm$1.25 &($\pm$2,0,0), (0,$\pm$2,0)  &0.65   &0   &8   &269 &8  &17216\\
asymm      & $\pm$1.25 &($\pm$1,$\pm$2,0), ($\pm$2,$\pm$1,0) &0.81   &0   &16   &195 &8  &24960 \\
asymm  & $\pm$1.25 &($\pm$2,$\pm$2,0)          &1.24    &0   &16  &195 &8   &24960\\
asymm  & $\pm$1.25 &($\pm$3,0,0), (0,$\pm$3,0)  &1.38   &0   &8   &269 &8  &17216\\
asymm      & $\pm$1.25 &($\pm$1,$\pm$3,0), ($\pm$3,$\pm$1,0)  &1.52   &0   &16   &195 &8  &24960 \\
asymm  & $\pm$1.25 &($\pm$4,0,0), (0,$\pm$4,0)  &2.29   &0   &8   &269 &8  &17216\\
\hline
\end{tabular}
\caption{\small Statistics for the symmetric and asymmetric frame data. The momentum unit $2\pi/L$ is 0.417 GeV. $N_{\rm ME}$, $N_{\rm confs}$, $N_{\rm src}$ and $N_{\rm total}$ are the number of matrix elements, configurations, source positions per configuration and total statistics, respectively.}
\label{tab:stat}
\end{center}
\end{table}

For the momentum boost $P_3=1.25$ GeV, we examine three values of momentum transfer $-t$ in the symmetric frame and eight values in the asymmetric frame. Most of the momentum transfer values are in the range $-t \in [0.17 - 1.50]$ GeV$^2$. For $-t=0.69$ GeV$^2$, we analyze three different momentum boosts, that is, 0.83, 1.25, and 1.67 GeV. The asymmetric frame offers distinct computational advantages, as it allows multiple values of $-t$ to be obtained with the same computational setup, at a small overhead cost. In contrast, each value of $-t$ in the symmetric frame requires a separate calculation. The production of the data for the asymmetric frame momenta is grouped into two categories based on different momentum configurations, which provides further efficiency.

Another component of this work is the renormalization of the non-local operator. 
This is done using a variant of the RI scheme, defined at a renormalization scale $\mu_R$. To minimize statistical fluctuations, we compute the vertex functions of the non-local tensor operator using the momentum source method~\cite{Gockeler:1998ye,Alexandrou:2015sea}. The renormalization condition, applied independently at each value of $z$, is expressed as
\begin{equation}
\label{renorm}
{\cal Z}_q^{-1} {\cal Z}_T(z) \, \mathrm{Tr}\left[ {\cal V}_T(p, z) \slashed{p} \right] \Big|_{p^2 = \mu_R^2} = 
\mathrm{Tr}\left[ {\cal P}_T {\cal V}_T^{\text{Born}}(p, z) \right] \Big|_{p^2 = \mu_R^2}\,,
\end{equation}
where ${\cal Z}_T$ represents the renormalization factor for a general tensor structure, and ${\cal V}_T(p, z)$ is the amputated vertex function. Here, ${\cal V}_T^{\text{Born}}(p, z)$ corresponds to the tree-level vertex function, while ${\cal P}_T$ denotes the minimal projector designed to isolate the tree-level contribution~\cite{Alexandrou:2021bbo}.

The quark field renormalization factor, ${\cal Z}_q$, is determined through the quark propagator, using the condition
\begin{equation}
{\cal Z}_q = \frac{1}{12} \mathrm{Tr}\left[ S(p)^{-1} S^{\text{Born}}(p) \right] \Big|_{p^2 = \mu_R^2}\,,
\end{equation}
where $S(p)$ and $S^{\text{Born}}(p)$ are the full and tree-level propagators, respectively.

The prescription in Eq.~\eqref{renorm} is both mass-independent and compatible with the matching formalism employed in this work~\cite{Liu:2019urm}. Residual cutoff effects proportional to $am_q$ may arise, particularly due to lattice artifacts. To address these, we perform the renormalization using five degenerate-quark-mass ensembles with $N_f = 4$ and identical lattice spacings to the main ensemble. These ensembles correspond to pion masses in the range of 350–520 MeV, and the resulting renormalization factors are extrapolated to the chiral limit following
\begin{equation}
\label{eq:Zchiral_fit}
{\cal Z}^{\text{RI}}_T(z, \mu_R, m_\pi) = Z^{\text{RI}}_T(z, \mu_R) + m_\pi^2 \, \bar{Z}^{\text{RI}}_T(z, \mu_R)\,.
\end{equation}
Thus, the extracted ${Z}^{\text{RI}}_T(z, \mu_R)$ is free from mass-dependent effects~\cite{Alexandrou:2019lfo}.

The scale and scheme dependence of ${\cal Z}_T$ requires its definition at a specific RI scale, $\mu_R$. To reduce discretization uncertainties, we choose values of $\mu_R$ that are isotropic in spatial directions and ensure $\frac{p^4}{(p^2)^2} < 0.35$~\cite{Constantinou:2010gr,Constantinou:2022aij}. A range of $\mu_R$ values, $(a\mu_R)^2 \in [1, 5]$, is used to test the robustness of the matching formalism. The final results rely on a single renormalization scale, $(a\mu_R)^2 \approx 2.57$, which is consistent with the matching equations connecting the quasi-GPDs in the RI scheme at $\mu_R$ to the GPDs in the $\overline{\rm MS}$ scheme at 2 GeV.
In our analysis, variations in $\mu_R$ produce negligible differences in the extracted GPDs.

\section{Lattice Results}
\label{sec:lattice_results_main}
In this section, we present a detailed analysis of the lattice QCD results for $H_T,~E_T,~\widetilde{H}_T,~\mathrm{and}~\widetilde{E}_T$, with a focus on the extracted Lorentz-invariant amplitudes and the behavior of these quantities for various values of the momentum transfer. The results are divided into several subsections, each corresponding to different aspects of the calculations with its associated plots. We remind the reader that all results are for the isovector flavor combination.

\subsection{Bare matrix elements: symmetric vs. asymmetric frames}

\begin{figure}[h!]
    \centering
    \includegraphics[scale=0.32]{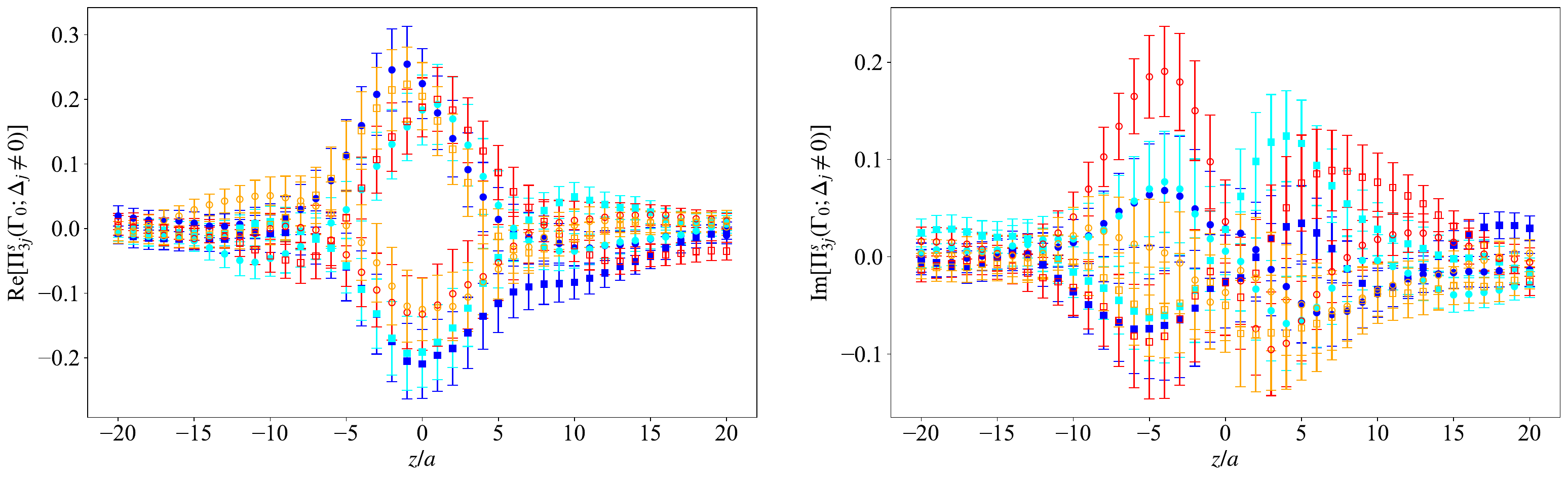}
    \includegraphics[scale=0.32]{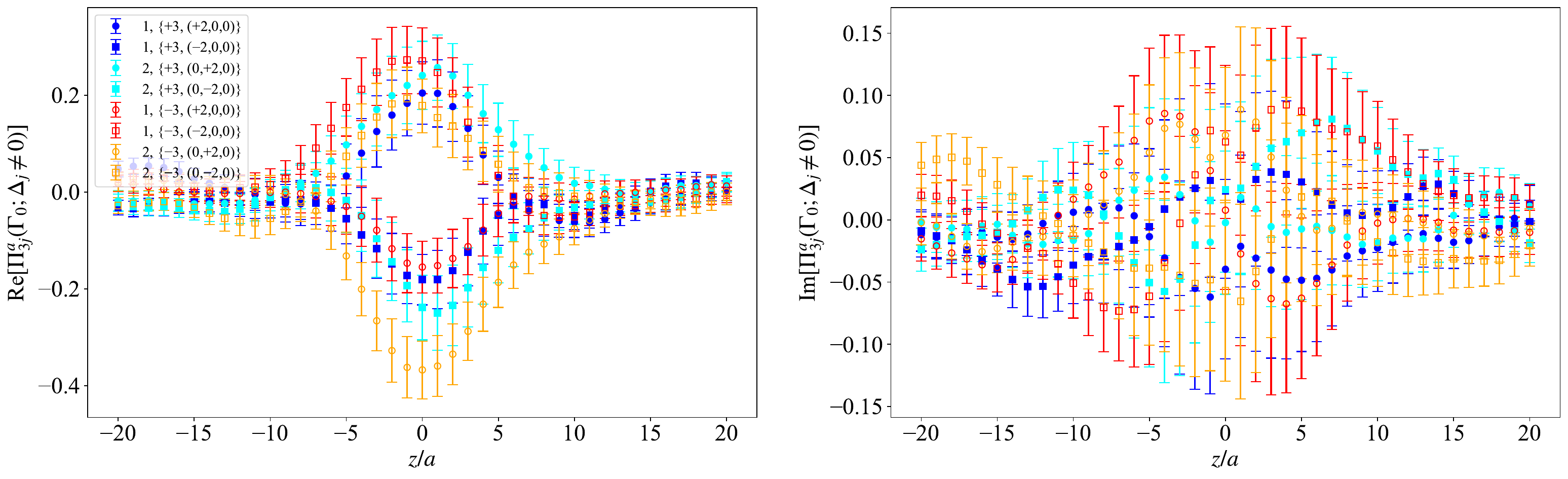}
    \vspace*{-0.4cm}
    \caption{\small{Bare matrix elements $\Pi_{3j}(\Gamma_0)$ in the symmetric frame (top) and in the asymmetric frame (bottom), for $|P_3|=1.25$ GeV and $-t=0.69$ GeV$^2$ ($-t=0.65$ GeV$^2$) for the symmetric (asymmetric) frame. The left (right) panel corresponds to the real (imaginary) part. The notation in the legend is $j,~\{P_3,\vec{\Delta}\}$ in units of $2\pi/L$.}}
    \label{fig:Pi3jG0}
\end{figure}

\begin{figure}[h!]
    \centering
    \includegraphics[scale=0.32]{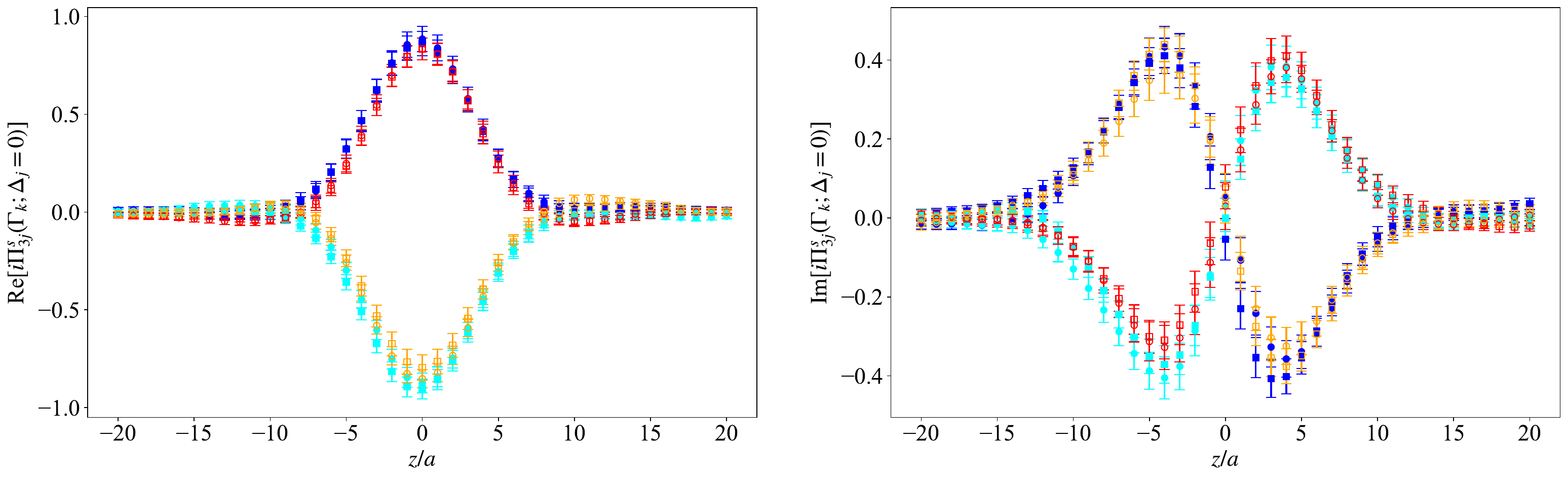}
    \includegraphics[scale=0.32]{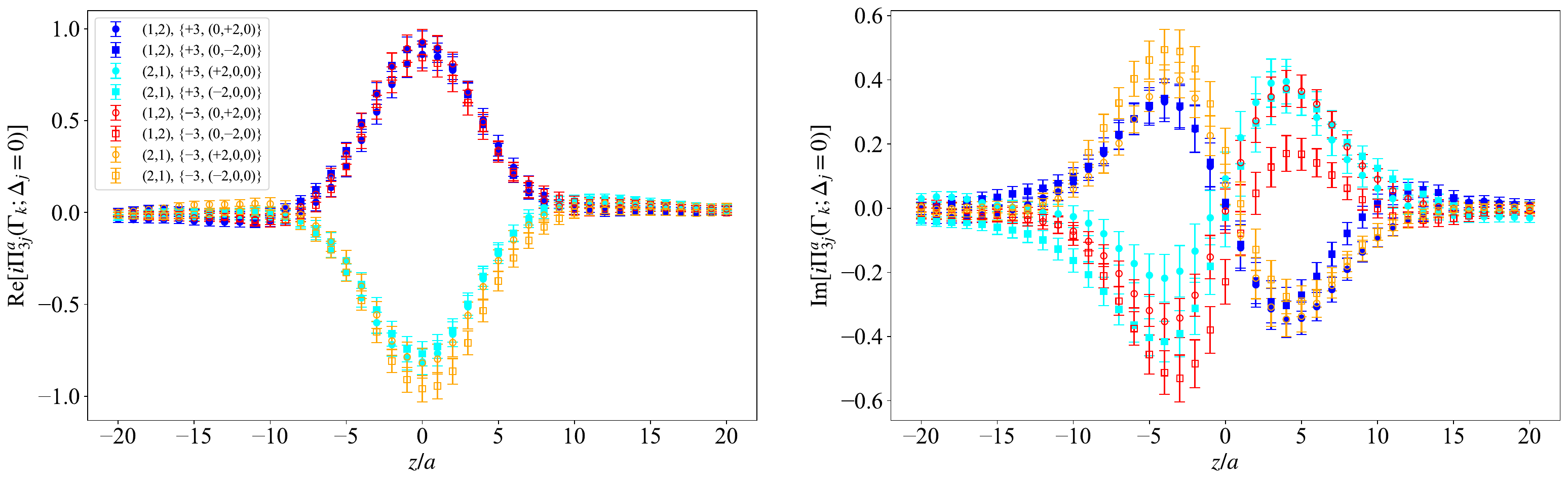}
    \vspace*{-0.4cm}
    \caption{\small{Bare matrix elements $\Pi_{3j}(\Gamma_k; \Delta_j=0)$ in the symmetric frame (top) and in the asymmetric frame (bottom), for $|P_3|=1.25$ GeV and $-t=0.69$ GeV$^2$ ($-t=0.65$ GeV$^2$) for the symmetric (asymmetric) frame. The left (right) panel corresponds to the real (imaginary) part. The notation in the legend is $(j,k),~\{P_3,\vec{\Delta}\}$ in units of $2\pi/L$.}}
    \label{fig:Pi3jGk_D0}
\end{figure}

\begin{figure}[h!]
    \centering
    \includegraphics[scale=0.32]{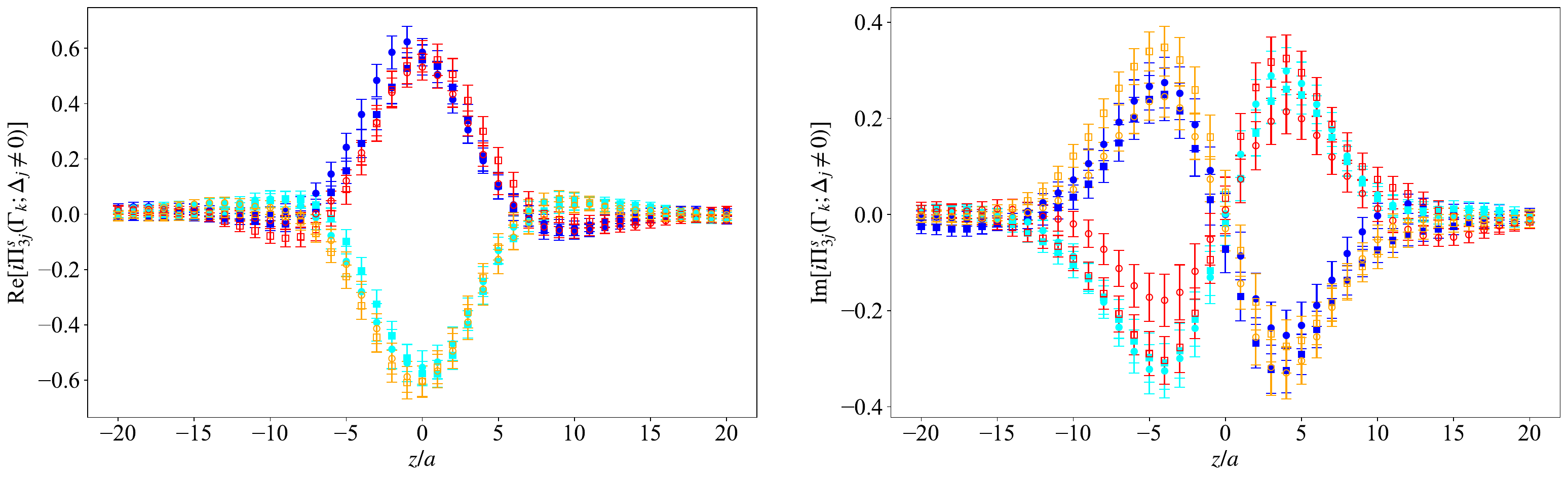}
    \includegraphics[scale=0.32]{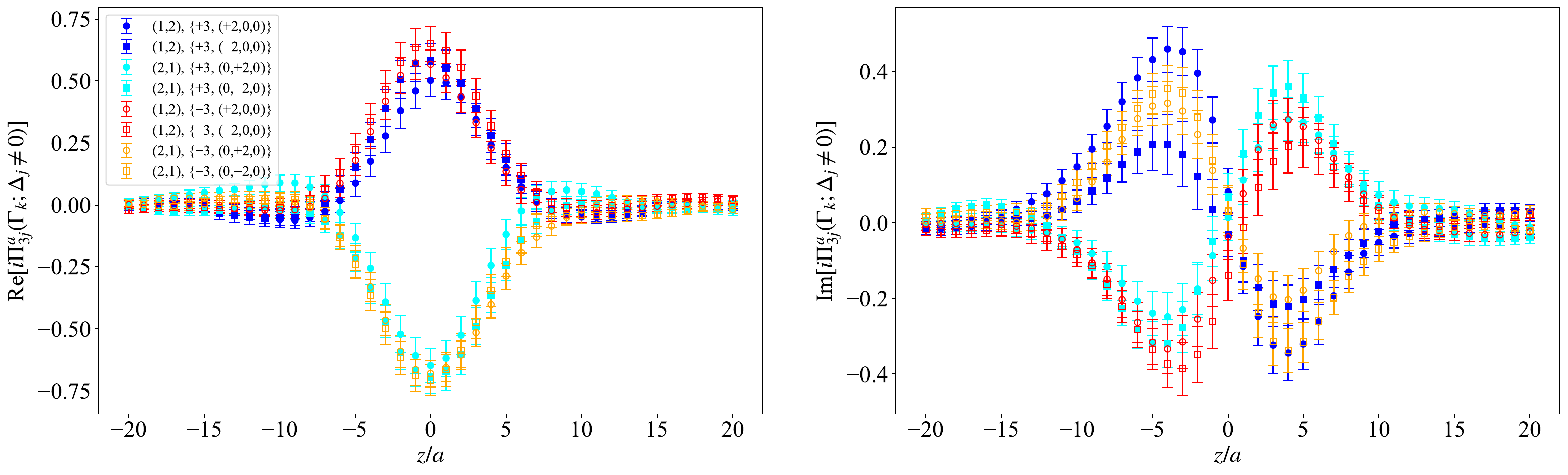}
    \vspace*{-0.4cm}
    \caption{\small{Bare matrix elements $\Pi_{3j}(\Gamma_k; \Delta_j\neq 0)$ in the symmetric frame (top) and in the asymmetric frame (bottom), for $|P_3|=1.25$ GeV and $-t=0.69$ GeV$^2$ ($-t=0.65$ GeV$^2$) for the symmetric (asymmetric) frame. The left (right) panel corresponds to the real (imaginary) part. The notation in the legend is $(j,k),~\{P_3,\vec{\Delta}\}$ in units of $2\pi/L$.}}
    \label{fig:Pi3jGk_D2}
\end{figure}

The presentation of results in this subsection is to numerically validate the use of the asymmetric frame by comparing the values of $A_{Ti}$ to those obtained in the symmetric frame. 
We begin by examining the bare matrix elements $\Pi_{3j}(\Gamma_0)$ and $\Pi_{3j}(\Gamma_\kappa)$ extracted from both frames. 
Fig.~\ref{fig:Pi3jG0}, Fig.~\ref{fig:Pi3jGk_D0}, and Fig.~\ref{fig:Pi3jGk_D2}, shows the real and imaginary components of $\Pi_{3j}(\Gamma_0)$, $\Pi_{3j}(\Gamma_k; \Delta_j=0)$, and $\Pi_{3j}(\Gamma_k; \Delta_j\neq0)$, respectively, for kinematic setups at $-t^s=-0.69$ GeV$^2$ and $-t^a=-0.65$ GeV$^2$. 
Both calculations extract these from the same values of $P_3 = 1.25$ GeV and $\vec{\Delta} = \frac{2\pi}{L}\{(\pm 2,0,0),(0,\pm 2,0)\}$, which result in slightly different $-t$ coming from the different energies. 
The data demonstrate similarities in magnitude even though the matrix elements contain different amplitudes in the two frames (see, e.g., Eq.~\eqref{Pi_31_s} and Eq.~\eqref{Pi_31_a}). 
We find that the signal-to-noise ratio of the imaginary part decreases in the asymmetric frame, a feature also observed in the unpolarized and helicity cases. 
This noise is likely due to the fact that all momentum transfer is placed in the initial state, which amplifies uncertainties in the boosted nucleon propagators. 
Despite these fluctuations, the overall trend in the imaginary parts remains consistent between the two frames. 
Furthermore, a number of matrix elements enter the definition of quasi-GPDs (see Eqs.~\eqref{eq:quasi_a_1} - \eqref{eq:quasi_c_2}), which has the potential of noise cancellation in some cases. 
One such cancellation can be observed by replacing the $A_{Ti}$ of Eqs.~\eqref{eq:AT1} - \eqref{eq:AT12} into Eqs.~\eqref{eq:quasi_c_1} - \eqref{eq:quasi_a_2}, which leads to only two types of amplitudes in Eqs.~\eqref{eq:quasiGPDs1} - \eqref{eq:quasiGPDs2}.
We also remind the reader that the matrix elements in the symmetric frame have definite symmetries, and the various kinematic setups at the same value of $-t^s$ can be combined. 
This symmetry property at the matrix element level does not hold for the asymmetric frame. Therefore, we calculate the average for equivalent kinematic setups at the level of the amplitudes. 
We note that the complexity of the analysis remains the same, regardless of whether the data are averaged at the stage of matrix elements or amplitudes.

\begin{figure}[h!]
    \centering
    \includegraphics[scale=0.315]{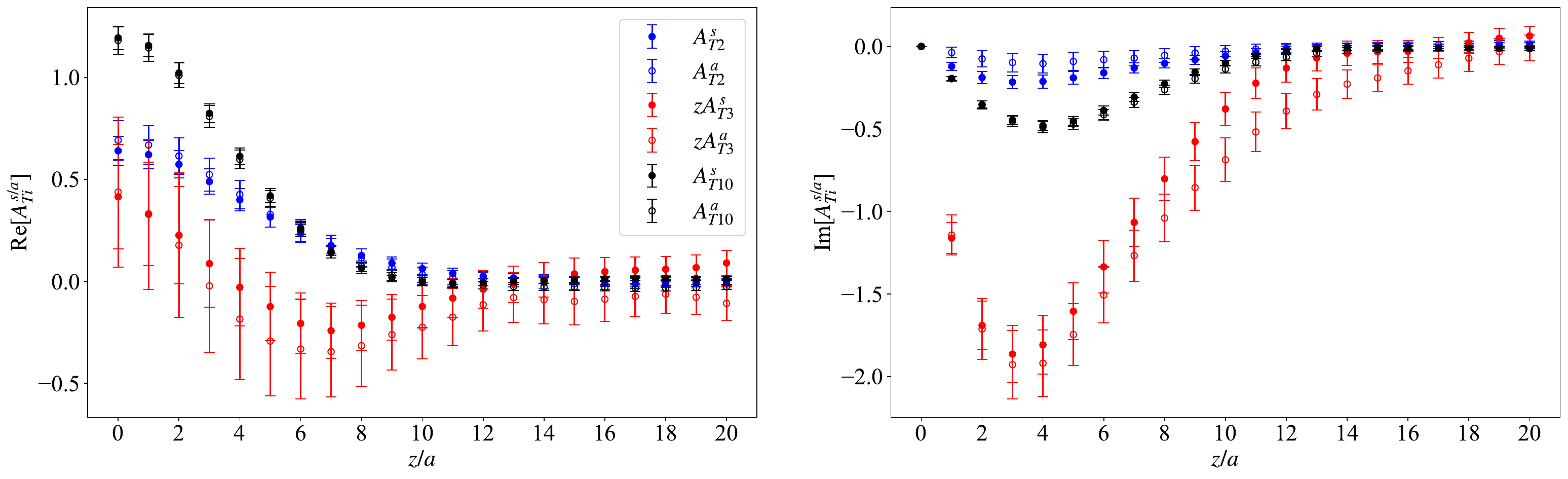}
    \vspace*{-0.4cm}
    \caption{\small{Comparison of amplitudes $A_{T2}$ (blue), $zA_{T3}$ (red), and $A_{T10}$ (black) with the real (imaginary) part on the left (right). $A^s_{Ti}$  (filled symbols) correspond to $-t^a=0.69~\mathrm{GeV}^2$, while $A^a_{Ti}$ (open symbols) to $-t^s=0.65~\mathrm{GeV}^2$. }}
    \label{fig:A2_zA3_A10_frame}
\end{figure}

\begin{figure}[h!]
    \centering
    \includegraphics[scale=0.315]{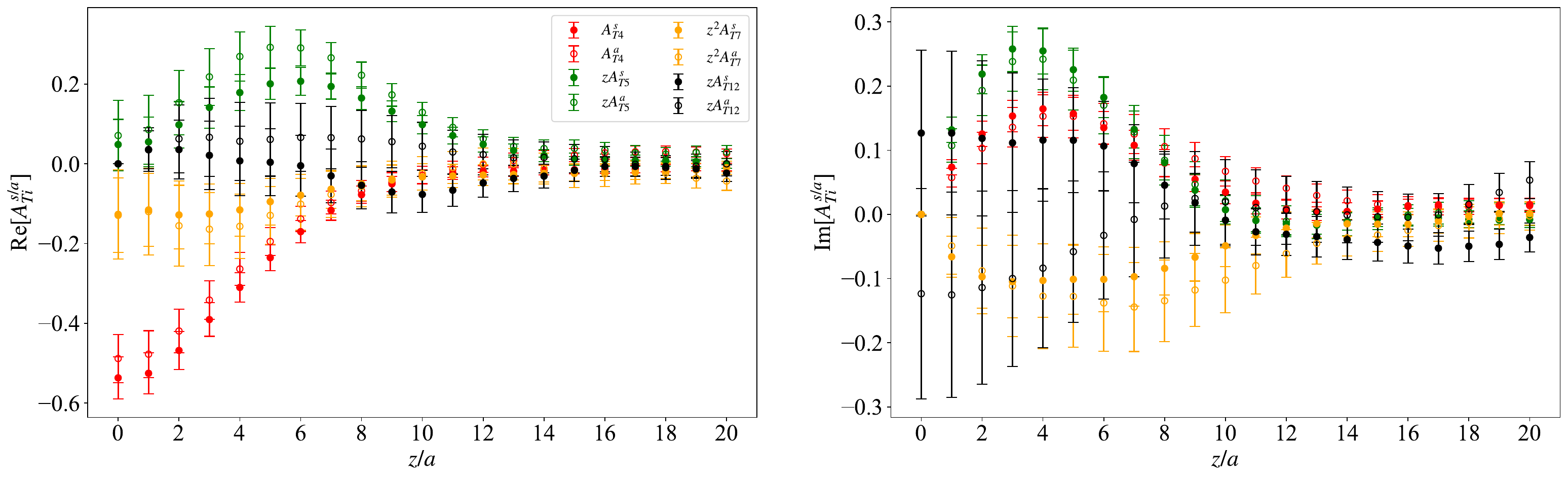}
    \vspace*{-0.4cm}
    \caption{\small{Frame comparison of amplitudes $A_{T4}$, $z A_{T5}$, $z^2 A_{T7}$, $z A_{T12}$ for the real (imaginary) parts on the left (right). $A^s_{Ti}$  (filled symbols) correspond to $-t^a=0.69~\mathrm{GeV}^2$, while $A^a_{Ti}$ (open symbols) to $-t^s=0.65~\mathrm{GeV}^2$.}}
    \label{fig:zA4_A5_A7_zA12_frame}
\end{figure}

\begin{figure}[h!]
    \centering
    \includegraphics[scale=0.315]{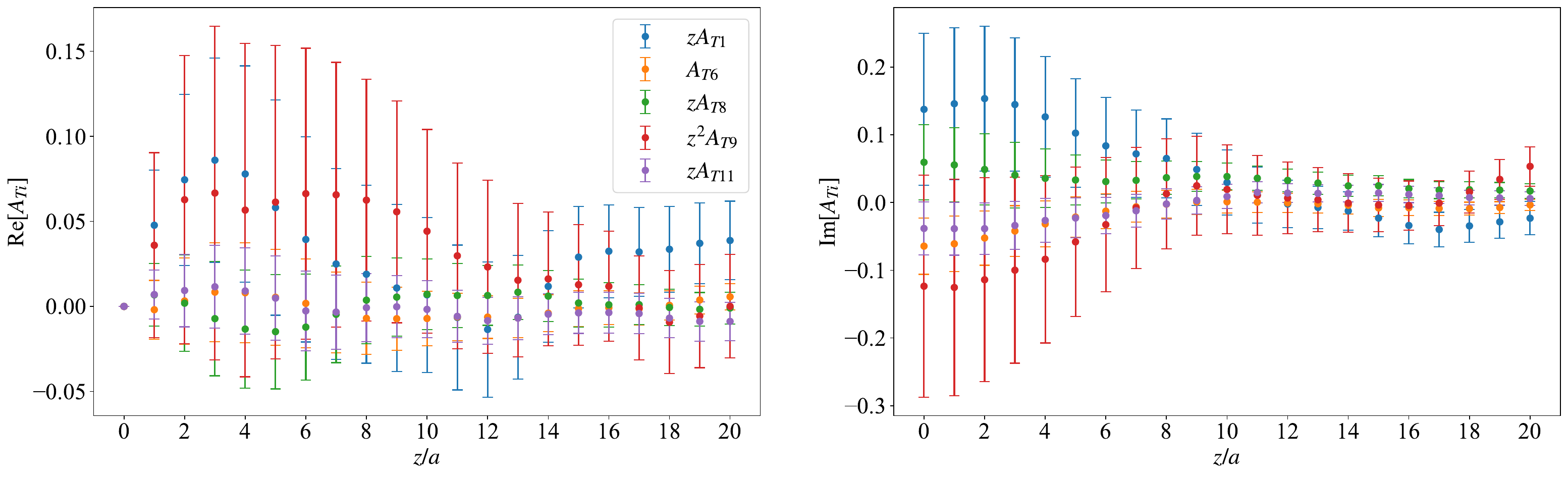}
    \vspace*{-0.4cm}
    \caption{\small{Amplitudes $zA_{T1},~A_{T6},~zA_{T8},~z^2A_{T9}$ and $zA_{T11}$ which are expected to be zero in zero-skewness. Results are shown in the asymmetric frame at $-t^a=0.65~\mathrm{GeV}^2$.}}
    \label{fig:zero_amplitudes}
\end{figure}

\newpage
From the 12 independent matrix elements, we compute the $A_{Ti}$ by solving a system of linear equations.
These amplitudes form the foundation for extracting GPDs, and it is crucial to verify that the results are consistent across different frame choices. 
Figs.~\ref{fig:A2_zA3_A10_frame} - \ref{fig:zero_amplitudes} present the real and imaginary parts of the amplitudes at $-t^s=0.69~\mathrm{GeV}^2$ and $-t^a=0.65~\mathrm{GeV}^2$.
In particular, Fig.~\ref{fig:A2_zA3_A10_frame} shows the three largest amplitudes, $A_{T2}$, $z A_{T3}$, and $A_{T10}$. 
Overall, we find good agreement between the two frames for these amplitudes, but there is a small difference in the imaginary part of $A_{T2}$ around $z/a \in [1-5]$. 
This is most likely a statistical fluctuation inherent in the lattice data. 
Also, we remind the reader that the $-t$ value between the two frames differs by about $6\%$.
Fig.~\ref{fig:zA4_A5_A7_zA12_frame} shows the smaller-magnitude amplitudes $A_{T4}$, $z A_{T5}$, $z^2 A_{T7}$, and $z A_{T12}$. 
As can be seen, the estimates from the two frames agree within errors for all these amplitudes.
Finally, for completeness, Fig.~\ref{fig:zero_amplitudes} demonstrates numerically, for the asymmetric frame at $-t^a=0.65$ GeV$^2$, that $z A_{T1}$, $A_{T6}$, $z A_{T8}$, $z^2 A_{T9}$, $z A_{T11}$ are compatible with zero, as anticipated theoretically. The same conclusion is found for the data in the symmetric frame.
In summary, our analysis confirms the theoretical expectation that the frame choice does not affect the overall extraction of amplitudes and, therefore, the GPDs. This is a non-trivial check of the linear equations relating the matrix elements and the amplitudes, as well as the numerical implementation.
\vspace*{-0.25cm}
\vspace*{-0.25cm}

\begin{figure}[h!]
    \centering
    \includegraphics[scale=0.34]{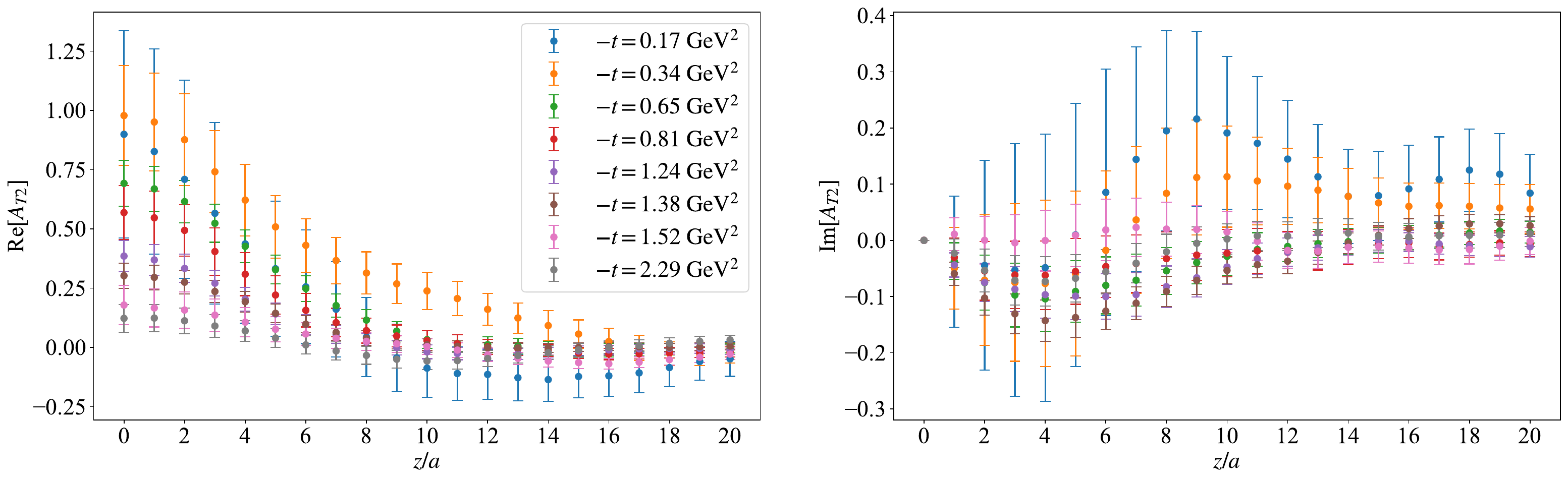}
    \vspace*{-0.4cm}
    \caption{\small{The amplitude $A_{T2}$ for all available values of $-t^a$ shown in Table~\ref{tab:stat}.}}
    \label{fig:A2_t}
\end{figure}

\begin{figure}[h!]
    \centering
    \includegraphics[scale=0.33]{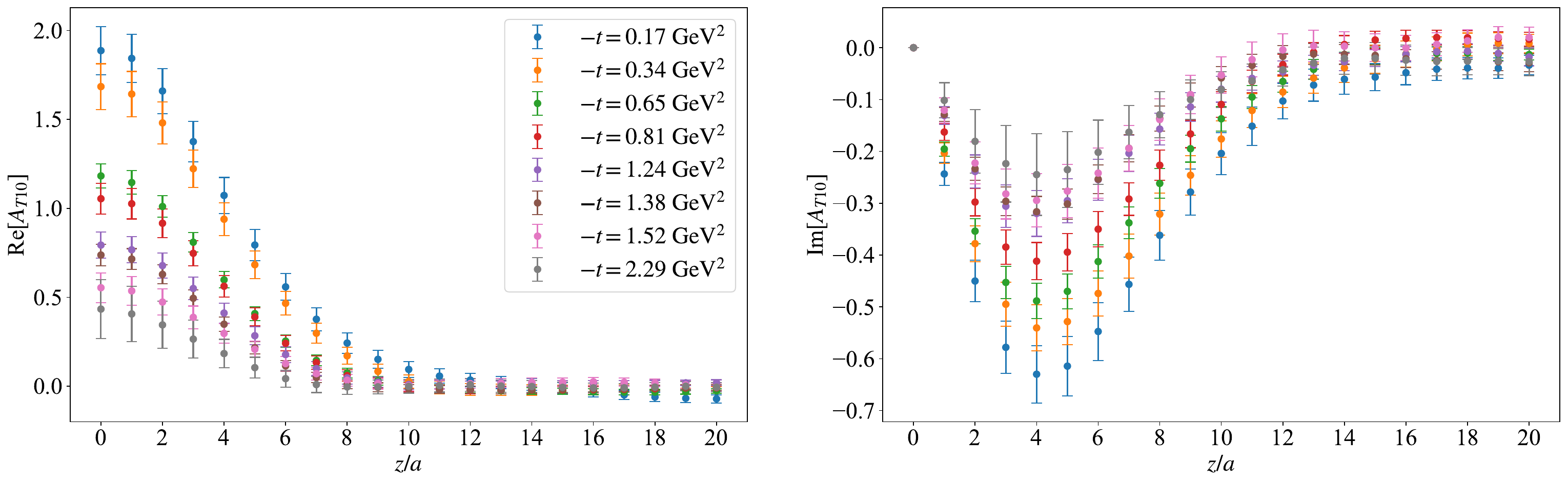}
    \vspace*{-0.4cm}
    \caption{\small{The amplitude $A_{T10}$ for all available values of $-t^a$ shown in Table~\ref{tab:stat}.}}
    \label{fig:A10_t}
\end{figure}

We now explore the behavior of the Lorentz-invariant amplitudes as a function of the momentum transfer $-t$.
As can be seen in Figs.~\ref{fig:A2_t} - \ref{fig:A10_t}, $A_{T2}$ and $A_{T10}$ decrease in magnitude smoothly as $-t$ increases. For better clarity, we only show the results in the asymmetric frame. 
This behavior is consistent with the known properties of GPDs, where the amplitudes are expected to decay at higher $-t$ values. 
An outlier in that regard is the imaginary part of $A_{T2}$, which may be due to large statistical uncertainties.

\begin{figure}[h!]
    \centering
    \includegraphics[scale=0.33]{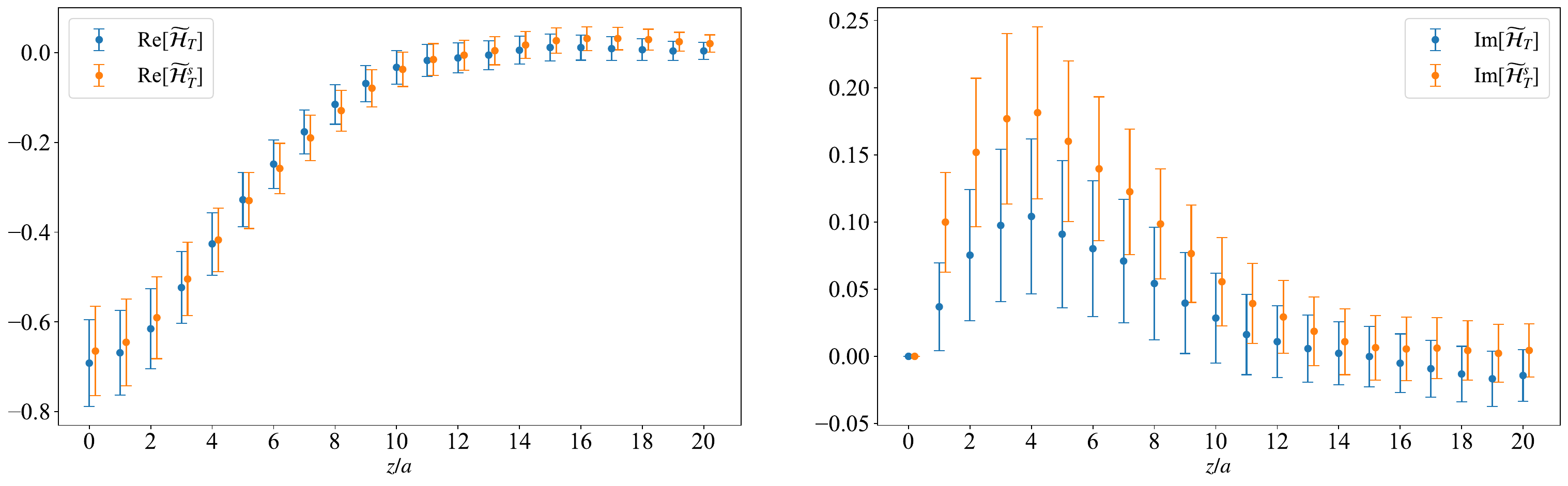}
    \vspace*{-0.4cm}
    \caption{\small{The quasi-GPD $\mathcal{\widetilde{H}}_T$ in coordinate space for the two definitions at $P_3=1.25$ GeV and $-t^a = 0.65 \mathrm{~GeV}^2$.}}
    \label{fig:HTtilde_definitions}
\end{figure}

\begin{figure}[h!]
    \centering
    \includegraphics[scale=0.33]{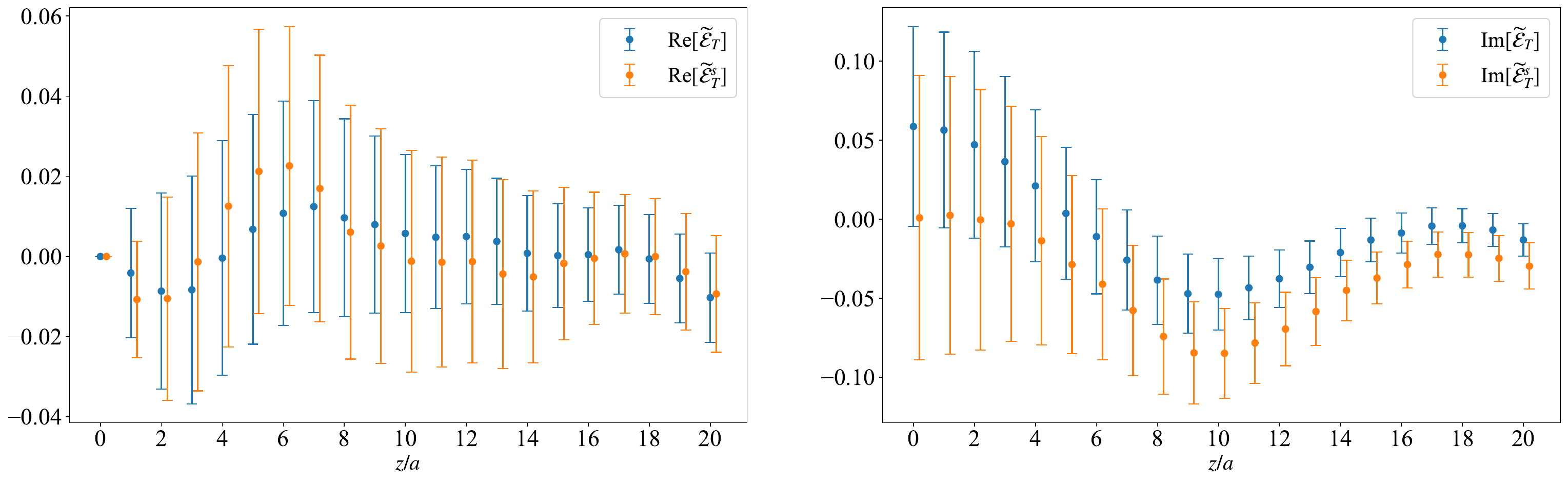}
    \vspace*{-0.4cm}
    \caption{\small{The quasi-GPD $\mathcal{\widetilde{E}}_T$ in coordinate space for the two definitions at $P_3=1.25$ GeV and $-t^a = 0.65 \mathrm{~GeV}^2$.} }
    \label{fig:ETtilde_definitions}
\end{figure}

Before discussing the quasi-GPDs in coordinate space, we restate the similarities of the two definitions. As can be seen in Eq.~\eqref{eq:quasi_c_1} - \eqref{eq:quasi_a_2}, the standard symmetric and LI definitions of the quasi-GPDs $\mathcal{H}_T$ and $\mathcal{E}_T$ coincide at zero skewness. 
For $\widetilde{\mathcal{H}}_T$, the standard definition (Eq.~\eqref{eq:quasi_a_3}) contains a term proportional to $z A_{T12}$, which is absent in the LI definition (Eq.~\eqref{eq:quasi_c_5}). 
$\widetilde{\mathcal{E}}_T$ is theoretically zero at $\xi=0$. This is confirmed in the two definitions (Eq.~\eqref{eq:quasi_a_2}, Eq.~\eqref{eq:quasi_c_2}), as they contain only the amplitudes $A_{T6}$ and $z A_{T8}$, which vanish for zero skewness.
This holds numerically for all values of $-t$, with one of them shown in Fig.~\ref{fig:zero_amplitudes}. 
To further demonstrate the above, we show in Figs.~\ref{fig:HTtilde_definitions} - \ref{fig:ETtilde_definitions} the two definitions for $\mathcal{\widetilde{H}}_T$ and $\mathcal{\widetilde{E}}_T$ both using asymmetric frame $A_{Ti}$ at $t^a=0.65$ GeV$^2$.
For the symmetric frame definition, we apply a Lorentz transformation on the kinematic factors as discussed in Sec.~\ref{sec:theory}. To summarize our findings, we have confirmed numerically that
$\widetilde{\mathcal{E}}_T$ is zero within errors, with the differences between the two definitions arising from the different kinematic coefficient of $A_{T8}$. Also, the two definitions of $\widetilde{\mathcal{H}}_T$ exhibit minor differences in the imaginary part, while the real part is very similar.
Any differences in the case of $\widetilde{\mathcal{H}}_T$ are due to the amplitude $A_{T12}$ entering the standard definition. $A_{T12}$ has a small real part and a nonzero imaginary part at $-t^a=0.65$ GeV$^2$.

\begin{figure}[h!]
    \centering
    \includegraphics[scale=0.34]{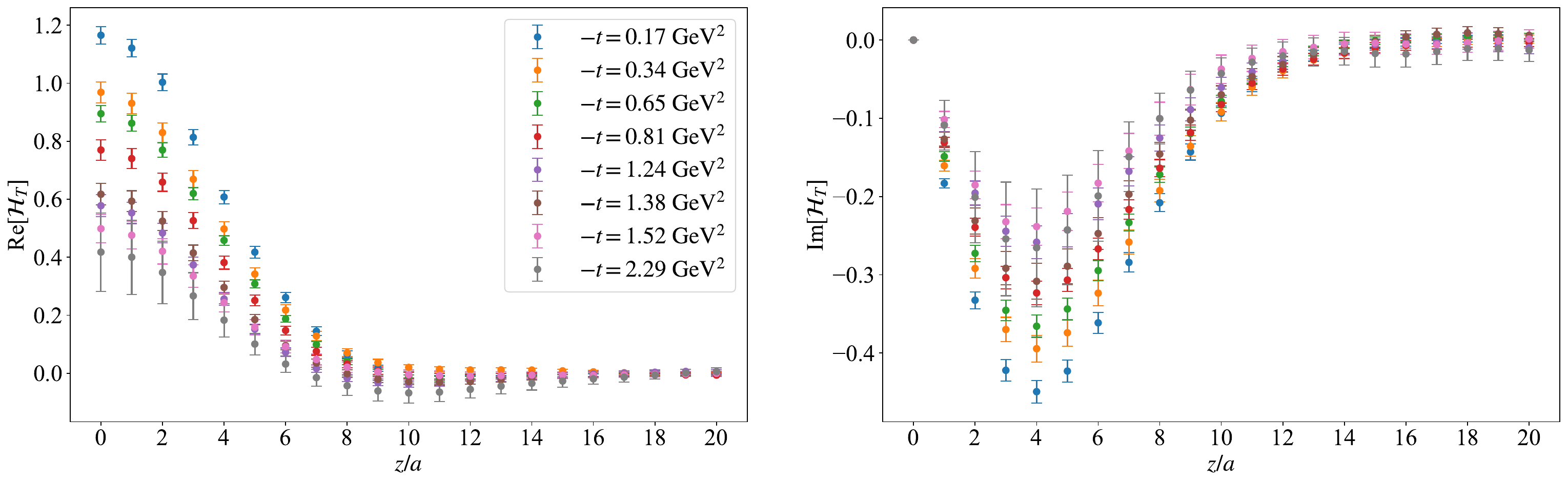}
    \vspace*{-0.4cm}
    \caption{\small{The LI quasi-GPD $\mathcal{H}_T$ in coordinate space for all values of $-t^a$ and at $P_3=1.25$ GeV.}}
    \label{fig:HT_t}
\end{figure}

\begin{figure}[h!]
    \centering
    \includegraphics[scale=0.34]{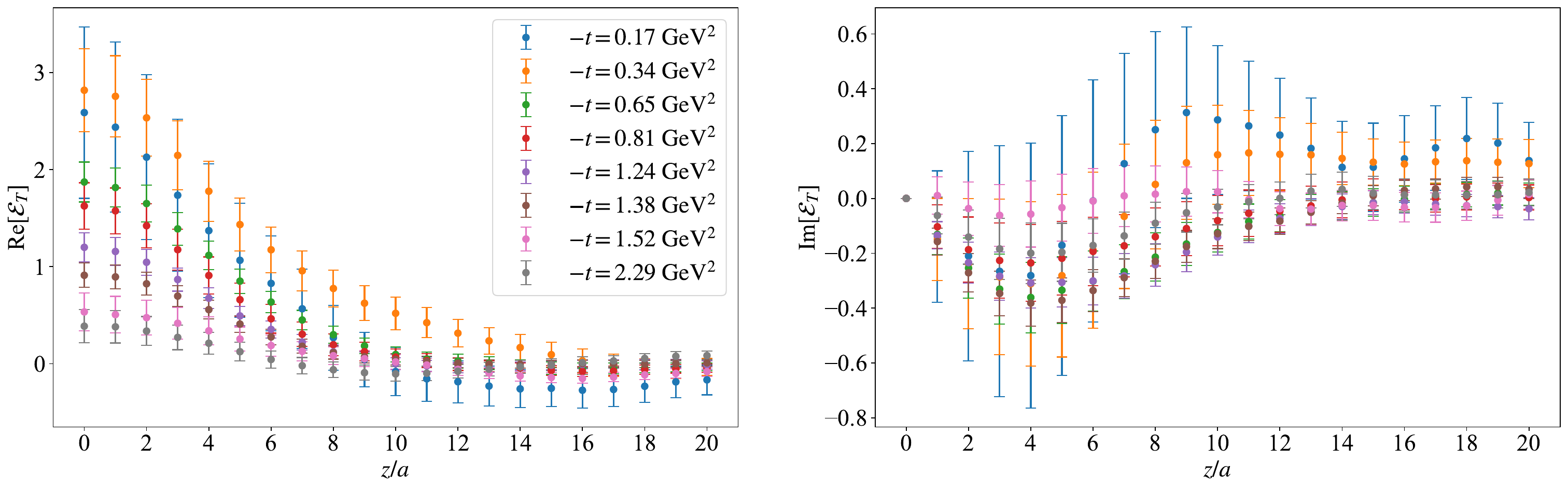}
    \vspace*{-0.4cm}
    \caption{\small{The LI quasi-GPD $\mathcal{E}_T$ in coordinate space for all values of $-t^a$ at $P_3=1.25$ GeV.}}
    \label{fig:ET_t}
\end{figure}

\begin{figure}[h!]
    \centering
\includegraphics[scale=0.34]{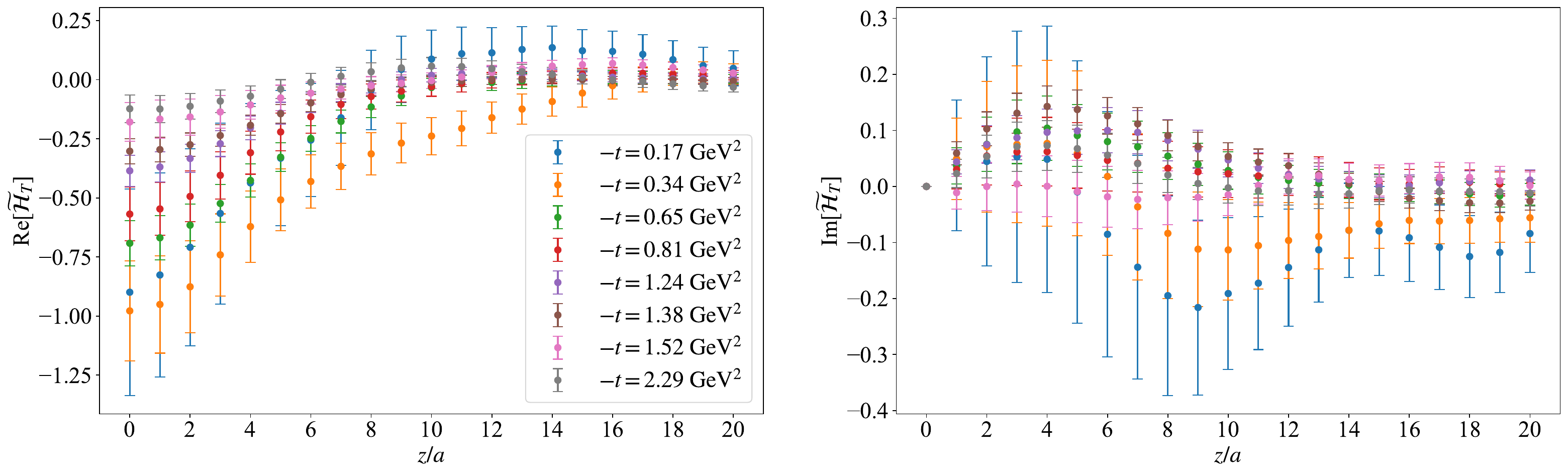}
    \vspace*{-0.4cm}
    \caption{\small{The LI quasi-GPD $\widetilde{\mathcal{H}}_T$ in coordinate space for all values of $-t^a$ at $P_3=1.25$ GeV.}}
    \label{fig:HTtilde_LI_t}
\end{figure}

In the following, we focus on the LI definitions for the quasi-GPDs and provide their $-t$ dependence in Figs.~\ref{fig:HT_t} - \ref{fig:HTtilde_LI_t}. 
For $\mathcal{H}_T$, $\mathcal{E}_T$, and $\mathcal{\widetilde{H}}_T$, we observe smooth behavior with increasing the value of $-t$. 
We exclude $\mathcal{\widetilde{E}}_T$ from this comparison, as it is found to be zero within errors for all data we study in this work; see Fig.~\ref{fig:ETtilde_definitions}.

\begin{figure}[h!]
    \centering
    \includegraphics[scale=0.29]{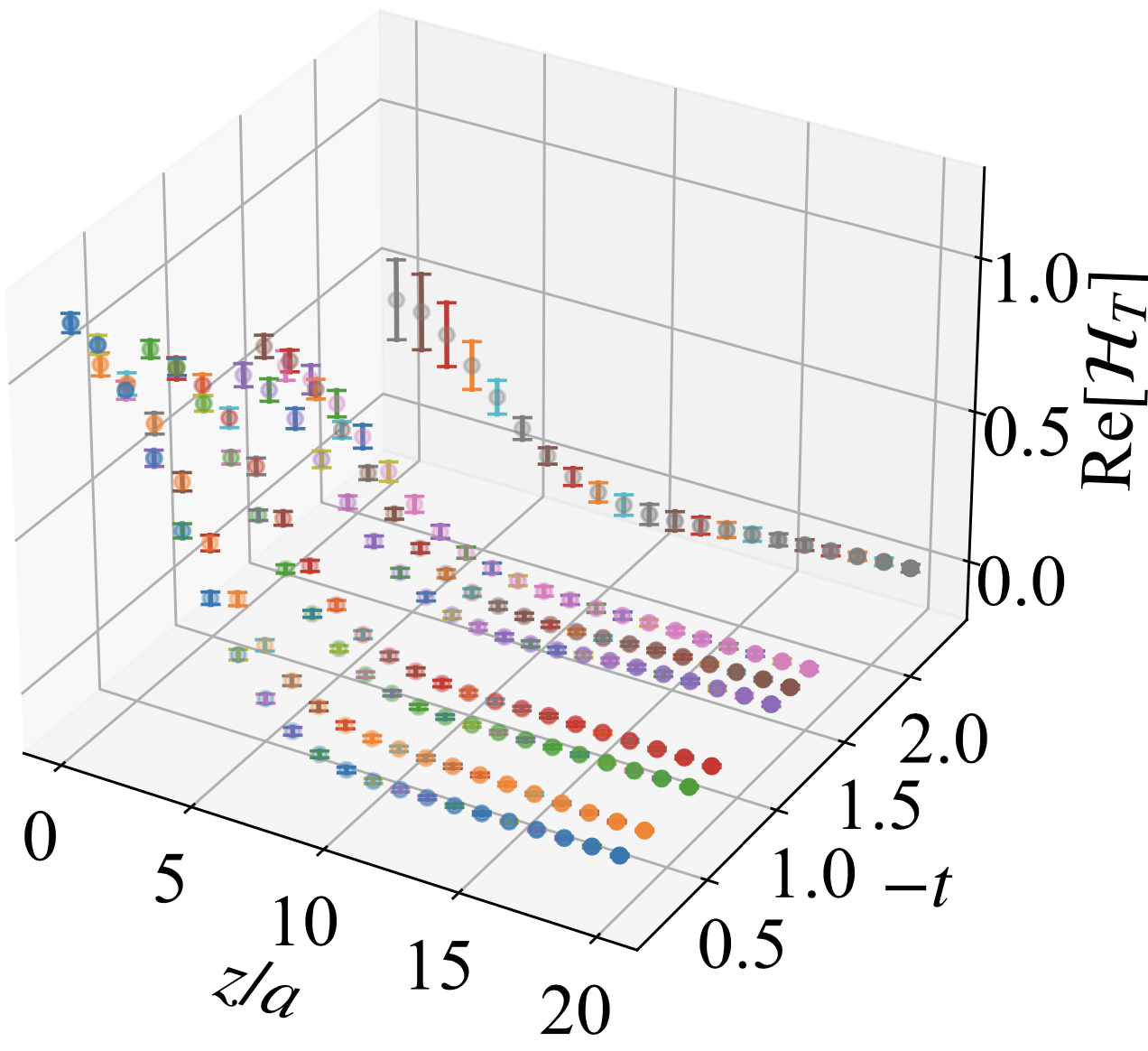} \hspace*{0.65cm}
    \includegraphics[scale=0.29]{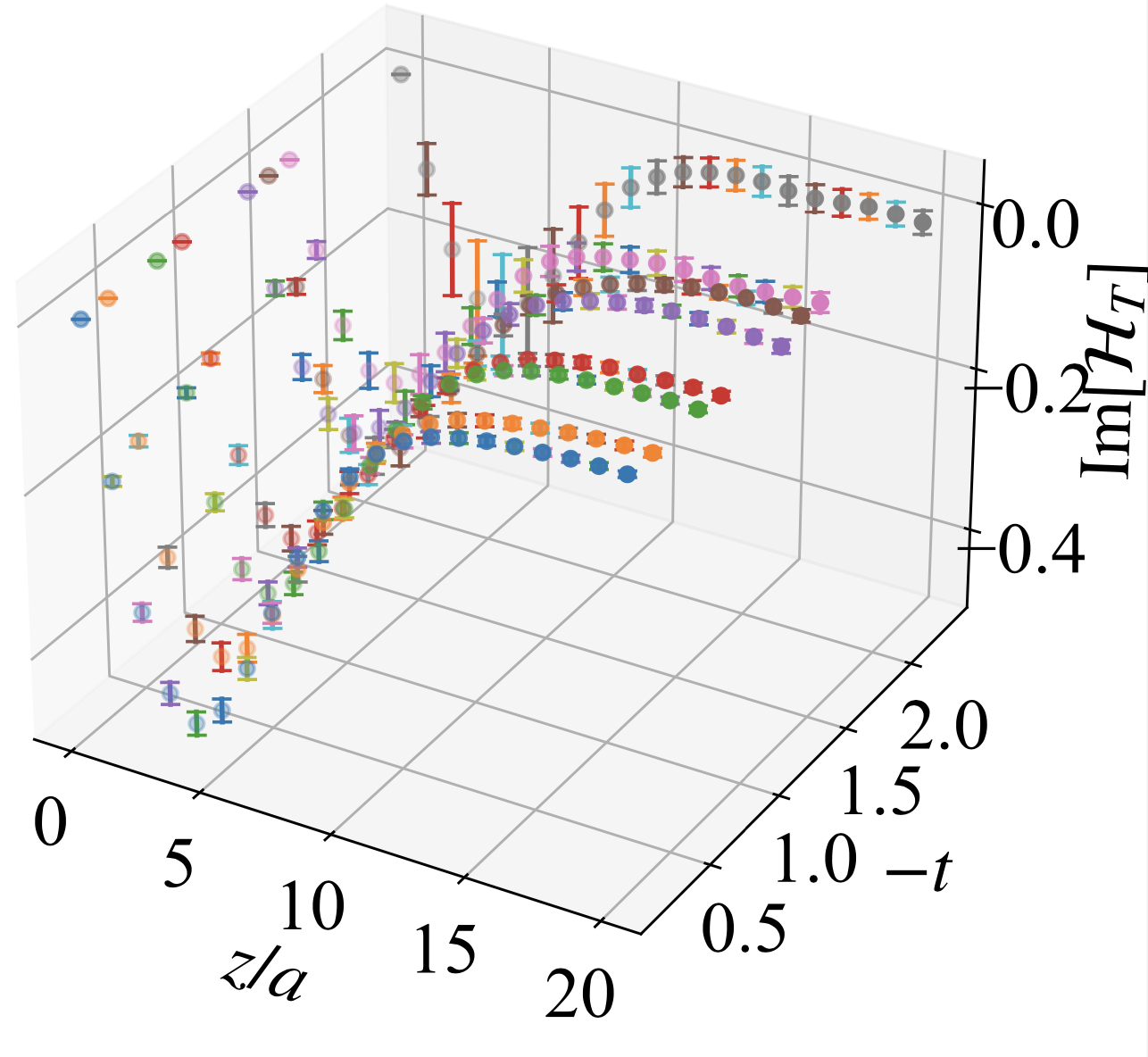}
    \vspace*{-0.1cm}
    \caption{\small{3-D plot of the LI quasi-GPD $\mathcal{H}_T$ as a function of $z/a$ and $-t$. The color coding for $-t$ follows that of Fig.~\ref{fig:HT_t}.}}
    \label{fig:3D_HT_position}
\end{figure}

\begin{figure}[h!]
    \centering
    \includegraphics[scale=0.29]{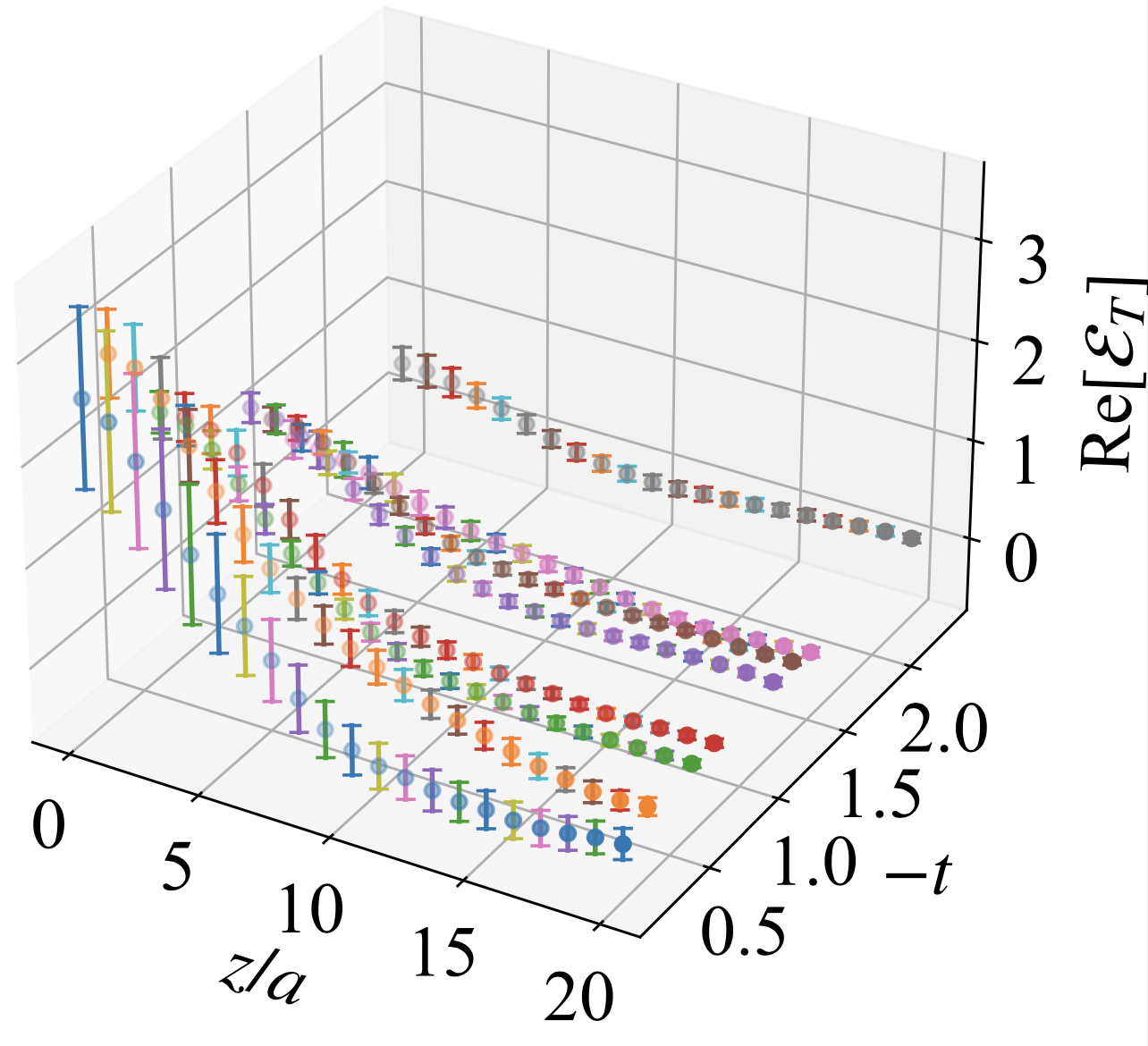} \hspace*{0.65cm}
    \includegraphics[scale=0.29]{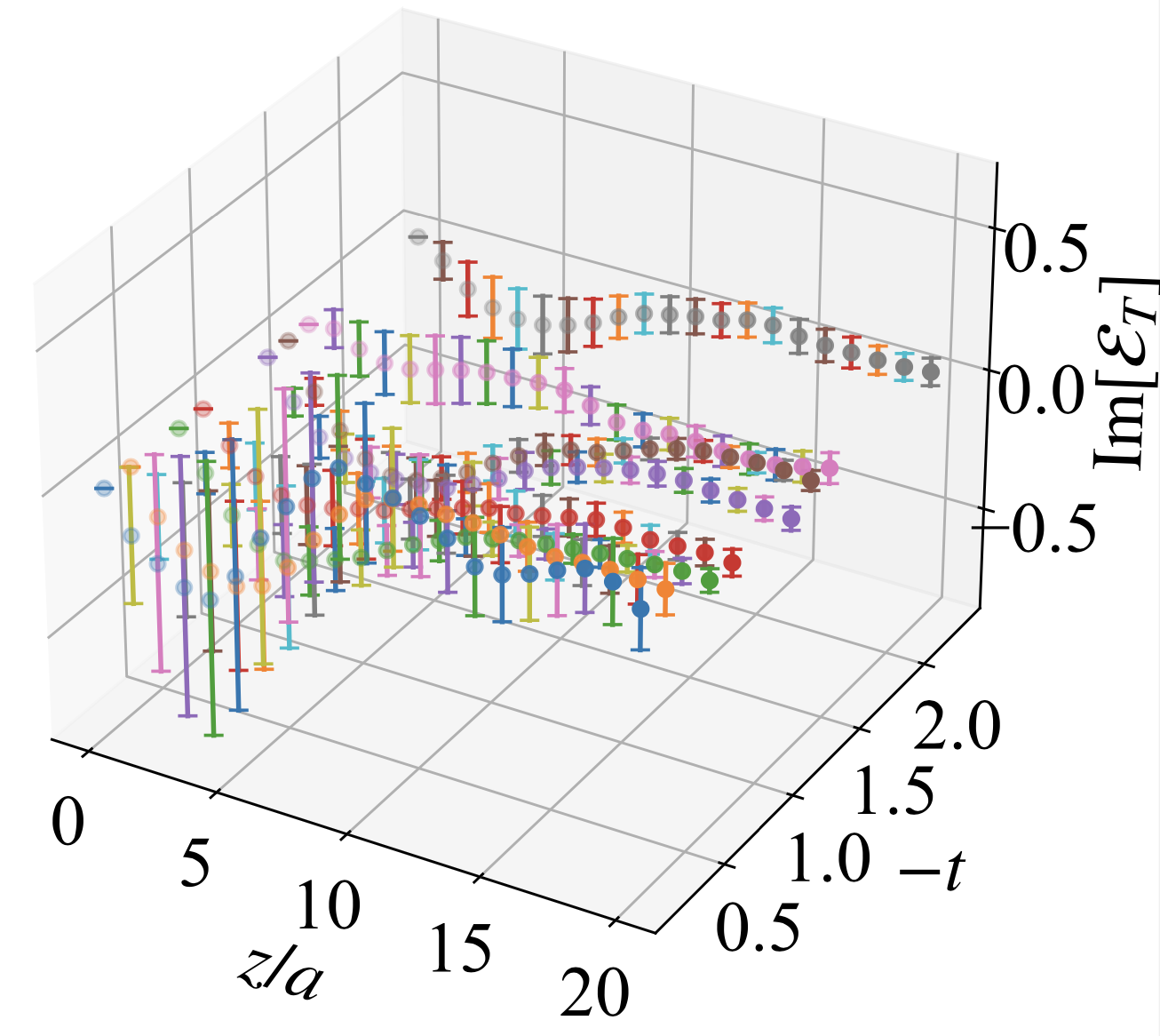}
    \vspace*{-0.1cm}
    \caption{\small{3-D plot of the LI quasi-GPD $\mathcal{E}_T$ as a function of $z/a$ and $-t$. The color coding for $-t$ follows that of Fig.~\ref{fig:HT_t}.}}
    \label{fig:3D_ET_position}
\end{figure}

\begin{figure}[h!]
    \centering
    \includegraphics[scale=0.29]{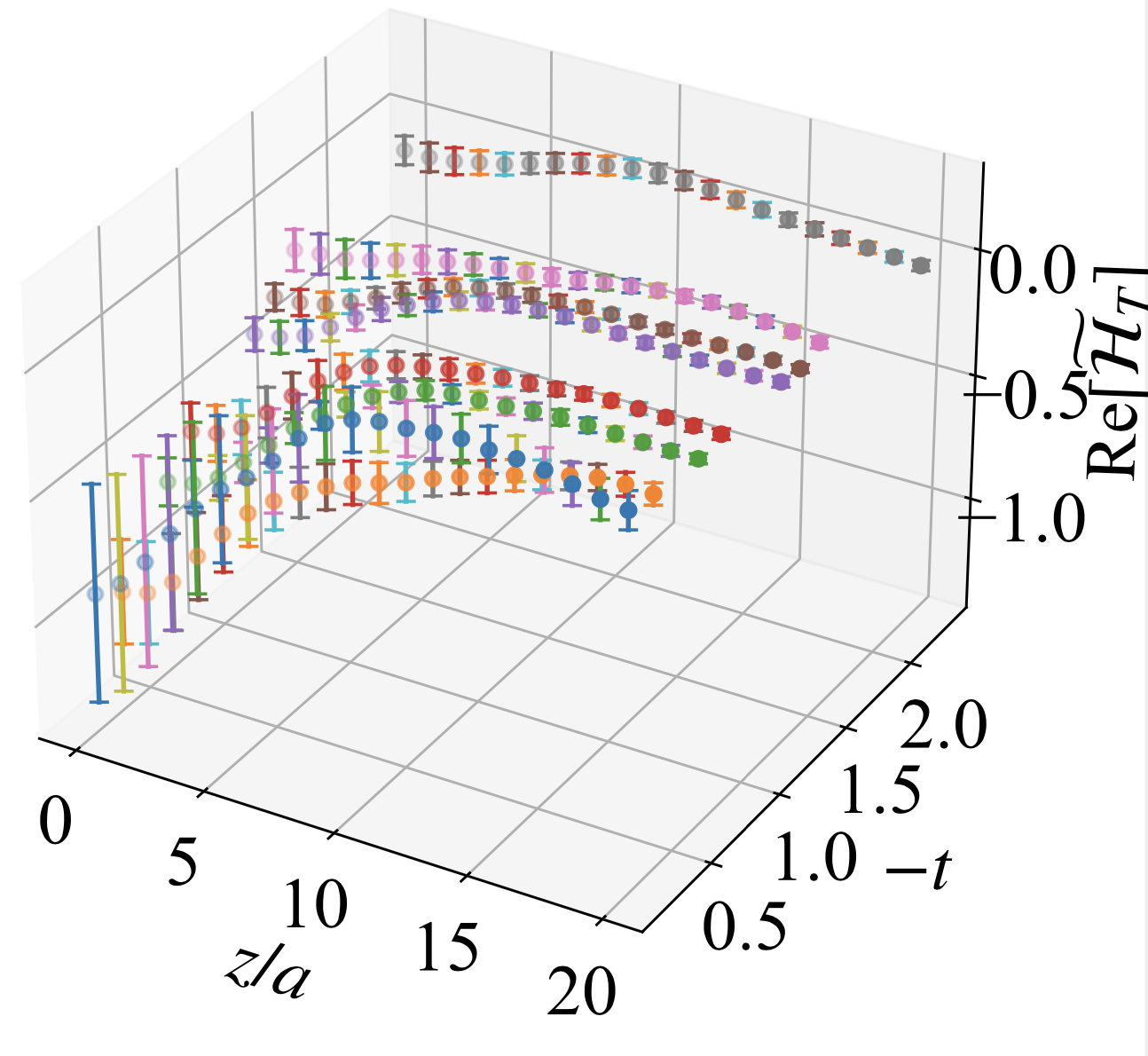} \hspace*{0.65cm}
    \includegraphics[scale=0.29]{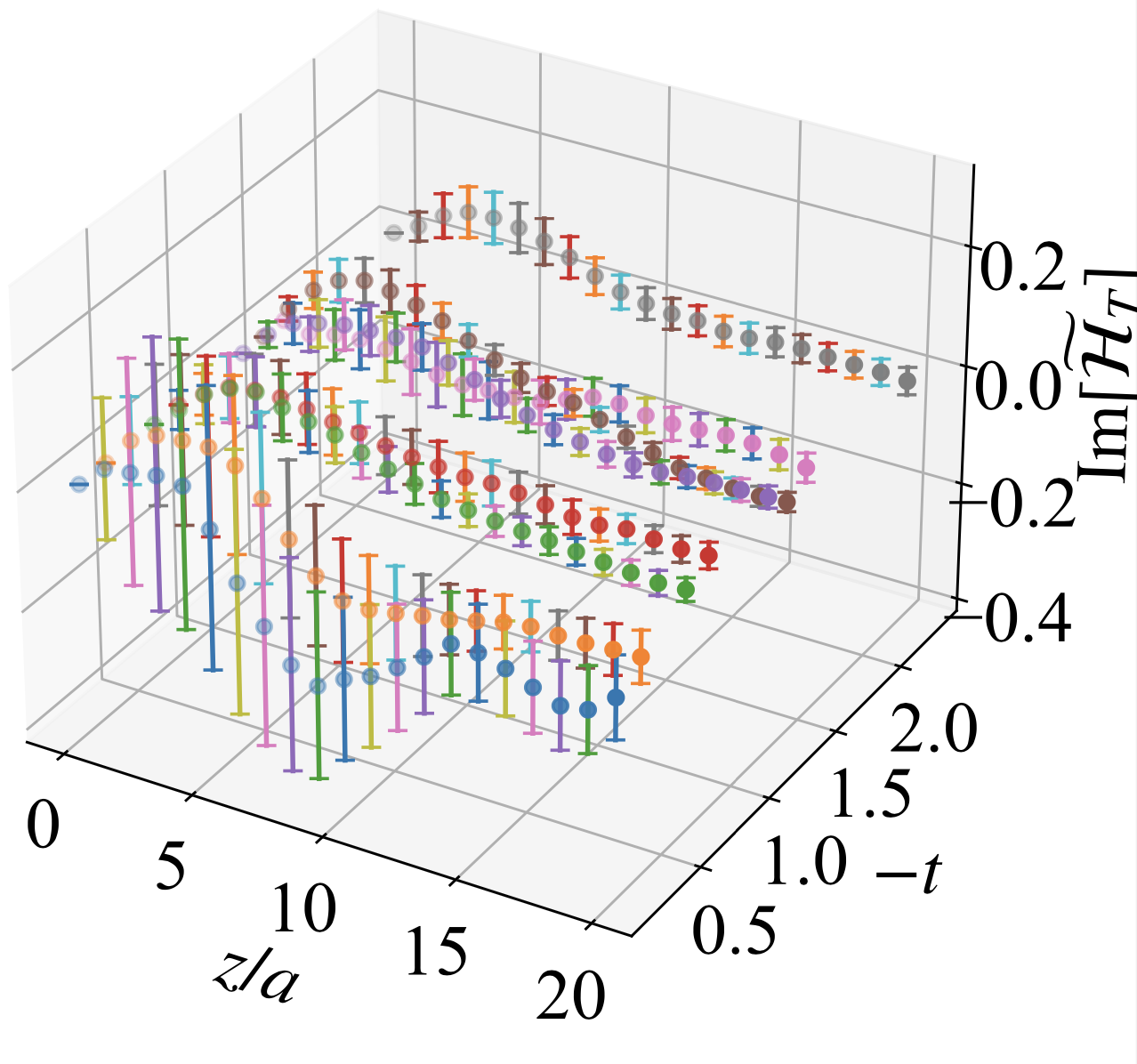}
    \vspace*{-0.1cm}
    \caption{\small{3D plot of the LI quasi-GPD $\widetilde{\mathcal{H}}_T$ as a function of $z/a$ and $-t$. The color coding for $-t$ follows that of Fig.~\ref{fig:HT_t}.}}
    \label{fig:3D_HTtilde_LI_position}
\end{figure}

\newpage
Finally, in Figs.~\ref{fig:3D_HT_position} - \ref{fig:3D_HTtilde_LI_position}, we present the results for the coordinate-space quasi-GPDs in three-dimensional plots as a function of $z/a$ and $-t$. These plots offer a little more clarity about the $-t$ dependence and emphasize the challenges in the signal for some of the momenta and some of the quasi-GPDs.
In terms of magnitude, $\mathcal{E}_T$ is the dominant quantity followed by $\mathcal{H}_T$, while $\mathcal{\widetilde{H}}_T$ has a clearly negative real part and becomes very small as $-t$ increases. The relative errors for $\mathcal{E}_T$ and $\mathcal{\widetilde{H}}_T$ are very large. Further discussion of the signal and magnitude is presented in Sec.~\ref{sec:light_cone_GPDs}.

\newpage
\subsection{Quasi-GPDs in momentum space}

Reconstructing the light-cone GPDs from lattice QCD data requires transforming the data into momentum space to obtain their $x$-dependence, for example
\begin{equation}
\label{eq:X2F}
{\cal H}_T(x,\xi,t,P_3,\mu_R) = \int_{-\infty}^\infty dz \, e^{-i x P_3 z} \, {\cal H}_T(z,\xi,t,P_3,\mu_R)\,.
\end{equation}
The same expressions hold for the other GPDs. 
As can be seen, this reconstruction involves integrating over a continuous range of $z$. 
However, lattice QCD calculations provide data only at discrete points, specifically for integer multiples of $z/a$, up to approximately half the lattice extent in the boosted direction, $L/2a$. 
This limitation is often referred to as the ``inverse problem''~\footnote{\textbf{We refer to Refs.~\cite{Karpie:2019eiq,Dutrieux:2025jed} for a comprehensive discussion on this issue in the context of parton distribution functions.}}, which predominantly affects the small-$x$ region. However, the moderate-to-large-$x$ domain remains relatively insensitive to this challenge, allowing for reliable predictions within that range.

\begin{figure}[h!]
    \centering
 \hspace*{-0.5cm}   \includegraphics[scale=0.32]{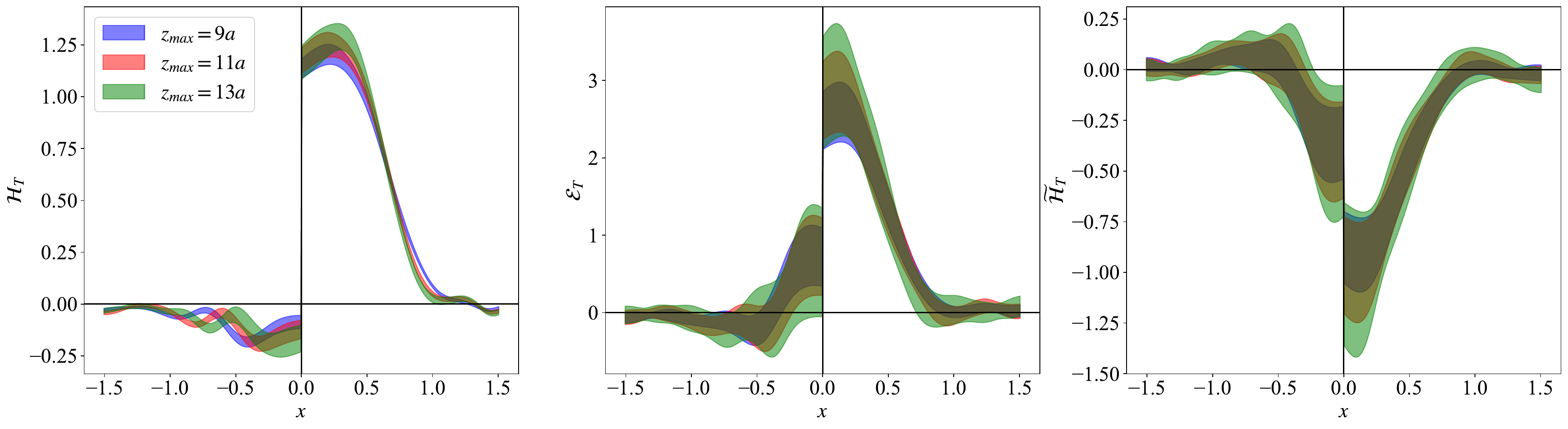}
    \vspace*{-0.4cm}
    \caption{\small{Dependence of momentum-space LI quasi-GPD on $z_{max}$ for $z_{max}=9a,~11a,~13a$. Results are shown at $-t^a=0.65~\mathrm{GeV}^2$ and $|P_3|=1.25$ GeV.}}
    \label{fig:GPD_BG_zmax}
\end{figure}

\begin{figure}[h!]
    \centering
 \hspace*{-0.5cm}   \includegraphics[scale=0.30]{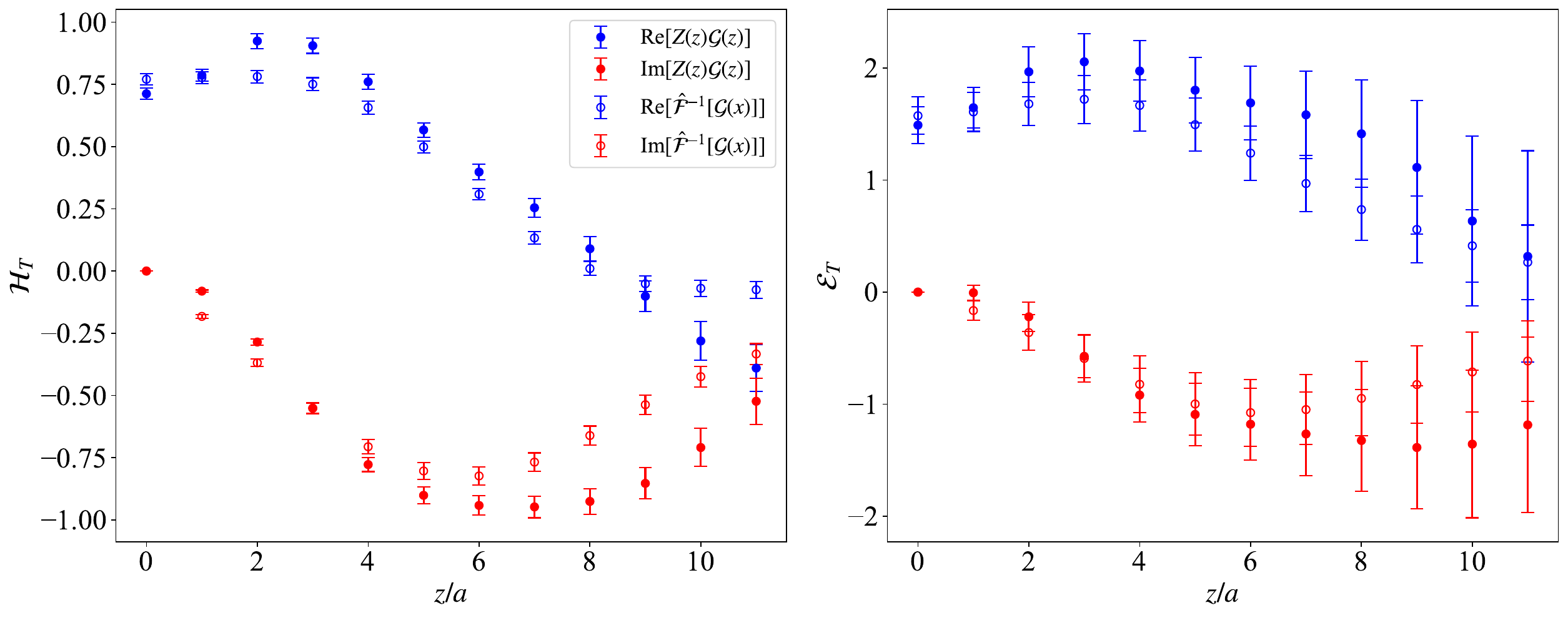}
 \includegraphics[scale=0.30]{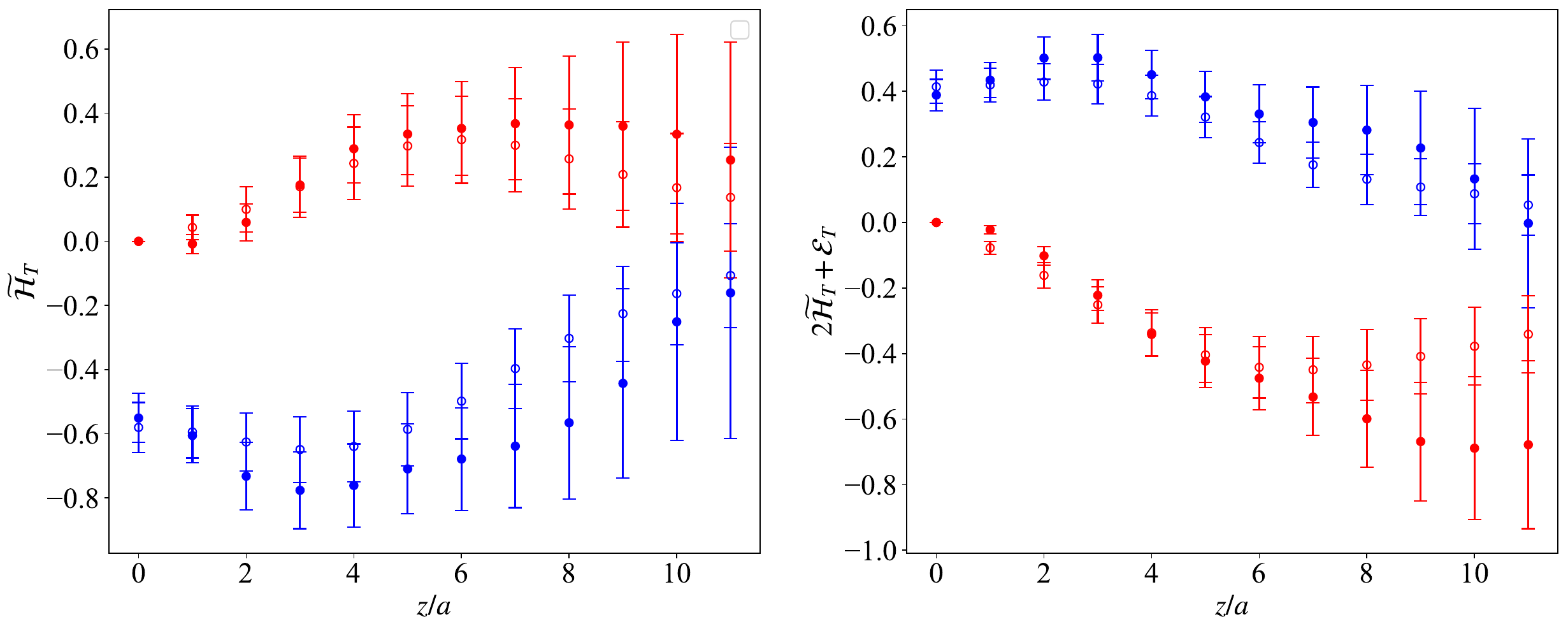}
    \vspace*{-0.4cm}
    \caption{\small{Consistency check of Backus-Gilbert reconstruction by comparing the real (blue) and imaginary (red) components of $Z(z)\mathcal{G}(z)$ (filled symbols) and the inverse Fourier Transform, $\hat{\mathcal{F}}^{-1}[\mathcal{G}(x)]$ (open symbols), for $\mathcal{G}\in \{\mathcal{H}_T,~\mathcal{E}_T,~\widetilde{\mathcal{H}}_T,~2\widetilde{\mathcal{H}}_T+\mathcal{E}_T\}$ at $|P_3|=1.25$ GeV and $-t=0.65~\mathrm{GeV}^2$.}}
    \label{fig:GPG_BG_check}
\end{figure}

In this analysis, we adopt the Backus-Gilbert (BG) reconstruction method~\cite{BackusGilbert}, a model-independent technique designed to address the inverse problem by selecting solutions that minimize statistical variance in the reconstructed distributions. 
The method works independently for each $x$ value, providing a stable momentum-space representation. It should be noted that, despite its advantages, the BG method is not without limitations, mainly due to the finite set of lattice data available. Members of our group are exploring alternative reconstruction methods~\cite{Chu:2025jsi}.
To partially explore these constraints, we examine reconstructions using varying sets of data, specifically $z_{\rm max}=9a,\,11a,\,13a$.
The comparison is illustrated in Fig.~\ref{fig:GPD_BG_zmax} for each quasi-GPD.
These plots can shed light on the influence of these choices on the GPDs: the reconstructed quasi-${\mathcal{\widetilde H}}_T$ for $z_{\rm max}=9a$ exhibits some differences from the highest choice of $z_{\rm max}=13a$ in the anti-quark region, as well as the small-to-intermediate quark region and in the region around $x=1$~\footnote{We remind the reader that the quasi-GPDs are defined in the infinite range of $x$.}. 
The other quasi-GPDs mostly exhibit agreement, an effect that is due to the large statistical uncertainties.
Based on this analysis, we choose $z_{\rm max}=11a$ when calculating the light-cone GPDs.
Given that the BG reconstruction carries inherent systematic uncertainties, a useful cross-check is to perform an inverse Fourier transform of the quasi-GPDs in momentum space and compare the result with the original input data. Any observed discrepancies can be traced to limitations of the BG method. This comparison, illustrated in Fig.~\ref{fig:GPG_BG_check}, reveals noticeable differences for the ${\cal H}_T$ GPD, while the other three distributions remain consistent. Such behavior is expected: the statistical uncertainties of ${\cal H}_T$ are significantly smaller than those of the other cases, making residual systematic effects more visible. As lattice calculations achieve higher precision, it becomes essential to explore alternative reconstruction strategies. An initial step in this direction by members of our group has been reported in Ref.~\cite{Chu:2025jsi}.

\subsection{Light-cone GPDs}
\label{sec:light_cone_GPDs}

Once the $x$-dependence of the quasi-GPDs is reconstructed, the next step involves deriving the light-cone quantities. This connection between the physical GPDs and quasi-GPDs is achieved through a perturbative matching procedure. As an example, we show the corresponding factorization formula for the ${\cal H}_T$ GPD,
\begin{equation}
\label{eq:matching}
{\cal H}_T(x,\xi,t,P_3,\mu^R,p_3^R) =\! \int_{-1}^1 \frac{dy}{|y|}\, C_G \left(\frac{x}{y},\frac{\xi}{y},\frac{y P_3}{\mu},\frac{y P_3}{p_3^R},\frac{(\mu^R)^2}{(p_3^R)^2}\right) H_T(y,t,\xi,\mu)+\mathcal{O}\left(\frac{m^2}{P_3^2},\frac{t}{P_3^2},\frac{\Lambda_{\rm QCD}^2}{x^2P_3^2},\frac{\Lambda_{\rm QCD}^2}{(1-x)^2P_3^2}\right).
\end{equation} 
The matching kernel, denoted as $C_G $, has been determined at one-loop in perturbation theory \cite{Liu:2019urm}. The relevant renormalization scales include: $\mu^R$, the RI renormalization scale; its $z$-component $p_3^R$ (where $p^2=(\mu^R)^2$), and $\mu$, the final $\overline{\rm MS}$ scale, which we set to $\mu = 2 \; {\rm GeV}$.
The matching coefficient for transversity quasi-GPDs are derived in Ref.~\cite{Xiong:2015nua} for the transverse momentum cutoff scheme. More recently, Ref.~\cite{Liu:2019urm} extended this work to all Dirac structures, deriving a matching formula that relates quasi-GPDs renormalized in a variant of the RI/MOM scheme to $\overline{ \rm MS}$ light-cone GPDs using the so-called minimal projectors. This is the matching formalism that we employ here. These studies revealed that the matching for GPDs at zero skewness aligns with the matching for PDFs. Additionally, it was established that, at the one-loop level, the matching formulas for $H$-type and $E$-type GPDs are identical. The explicit matching kernel for transversity GPDs can be found in the original references, as well as in Ref.~\cite{Alexandrou:2021bbo}.

\begin{figure}[h!]
    \centering
    \includegraphics[scale=0.48]{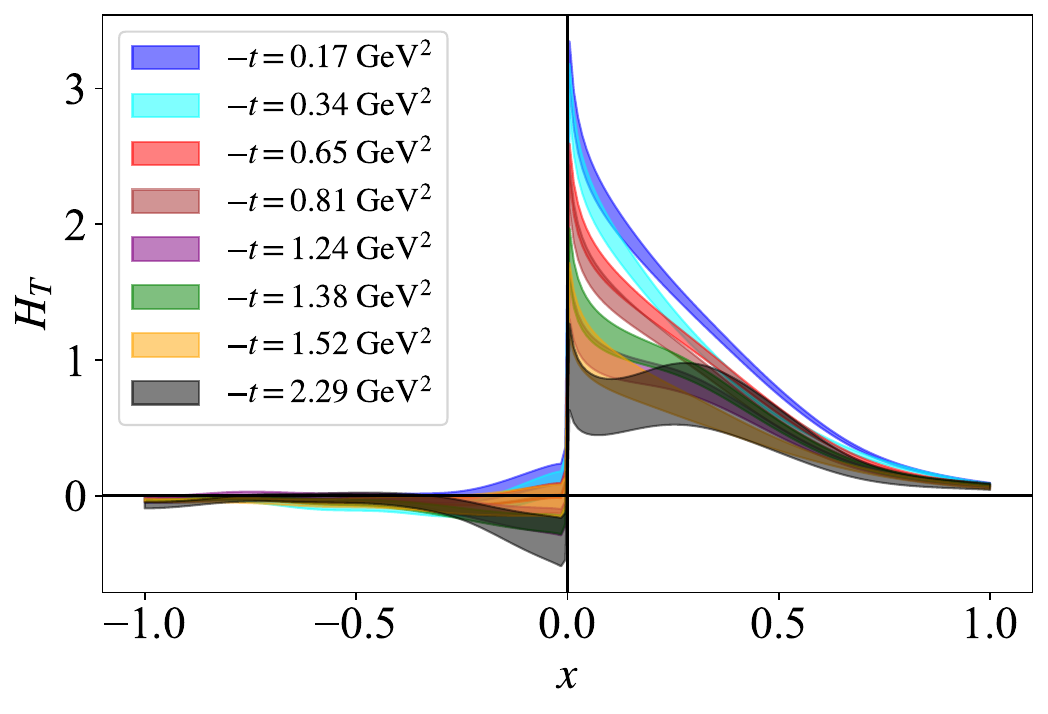}\,\,
    \includegraphics[scale=0.48]{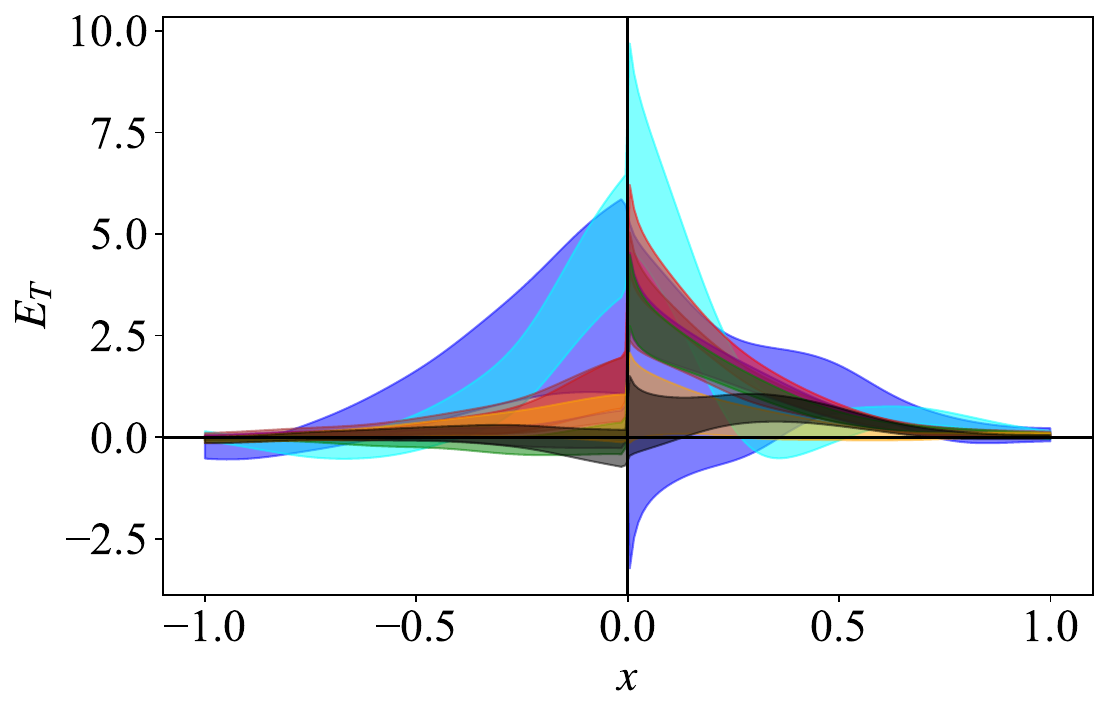}
    \vspace*{-0.4cm}
    \caption{\small{Light-cone GPD $H_T$ (left) and $E_T$ (right) in the $\overline{\mathrm{MS}}$ at 2 GeV.}}
    \label{fig:HT_GPD}
\end{figure}

\begin{figure}[h!]
    \centering
    \includegraphics[scale=0.48]{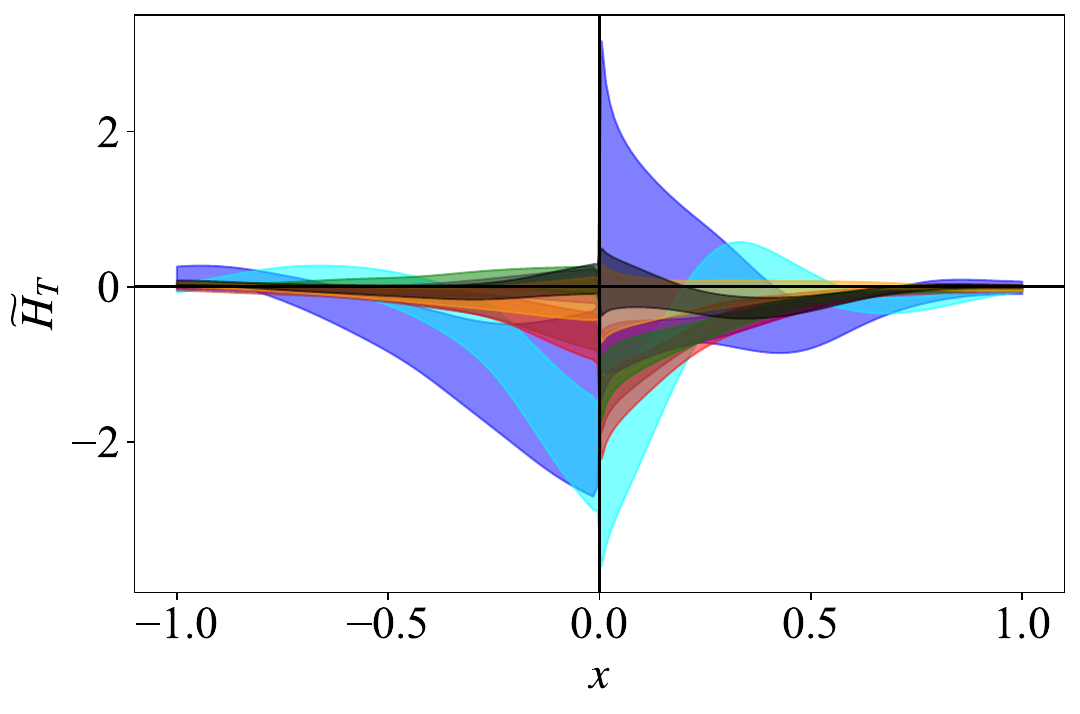}\,\,
        \includegraphics[scale=0.48]{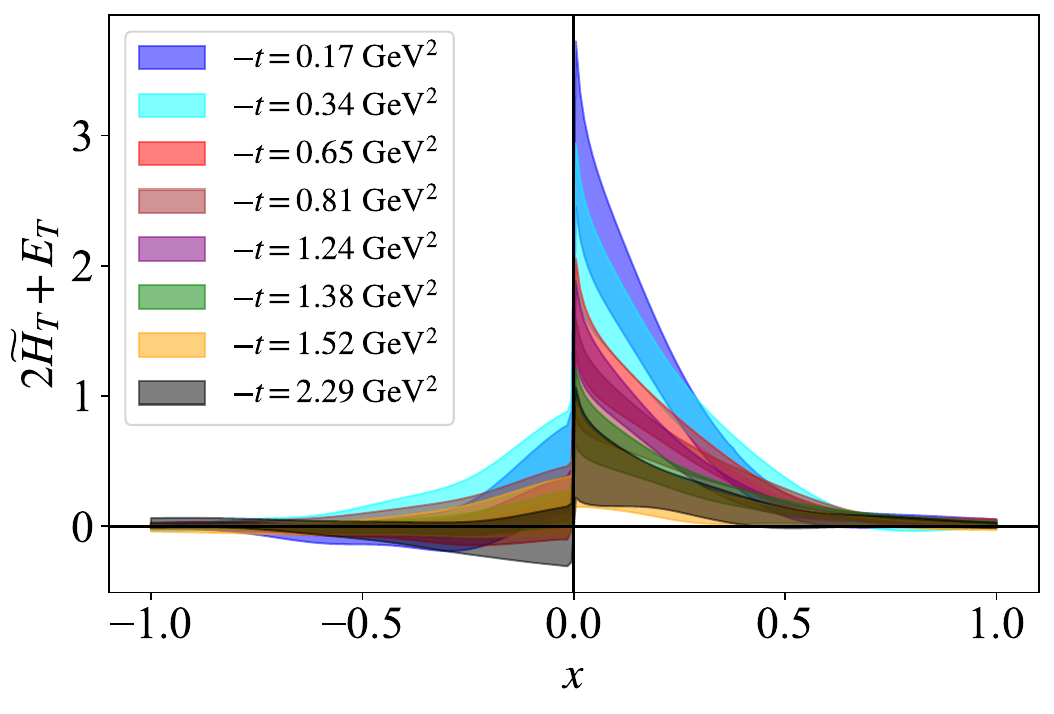}
    \vspace*{-0.4cm}
    \caption{\small{Light-cone GPD $\widetilde{H}_T$ (left) and the $2\widetilde{H}_T+E_T$ combination (right) in the $\overline{\mathrm{MS}}$ at 2 GeV.} }
    \label{fig:HTtilde_GPD}
\end{figure}

The final results for $H_T$, $E_T$, and $\widetilde{H}_T$ are shown in Figs.~\ref{fig:HT_GPD} - \ref{fig:HTtilde_GPD}. $E_T$ has the largest magnitude, followed by $H_T$. However, the latter has the most precise signal compared to $E_T$ and $\widetilde{H}_T$.
We also investigate the combination $E_T + 2 \widetilde{H}_T$, which plays a crucial role in understanding the transverse spin structure of the proton~\cite{Diehl:2005jf}. 
In impact parameter space, this quantity has a clear physical interpretation as it describes the deformation in the distribution of transversely polarized quarks within an unpolarized proton. 
As already mentioned in the Introduction, the quantity $\kappa_T = \int dx \big(E_T(x, 0, 0) + 2\widetilde{H}_T(x, 0, 0) \big) $ governs the size of the dipole moment given by this distribution~\cite{Burkardt:2005hp}.
In the right panel of Fig.~\ref{fig:HTtilde_GPD}, we present our results for $E_T + 2 \widetilde{H}_T$. 
It is interesting to observe that there is a good signal despite the involvement of the noisier $E_T$  and $\widetilde{H}_T$. Note that in the LI definition at zero skewness we obtain $\mathcal{E}_T + 2 \widetilde{\mathcal{H}}_T =  -A_{T4}$; the amplitude $A_{T2}$ cancels in this combination. In addition, we find numerically that this linear combination is as large as $H_T$.
On the other hand, the situation is more complicated for $E_T$ and $2 \widetilde{H}_T$ individually.  Specifically, $E_T$ by itself has no physical interpretation as a density, while  $\widetilde{H}_T$ has a connection with the quadrupole deformation of the distribution of transversely polarized quarks in a transversely polarized proton~\cite{Diehl:2005jf, Meissner:2007rx}.

\begin{figure}[h!]
    \centering
    \includegraphics[scale=0.33]{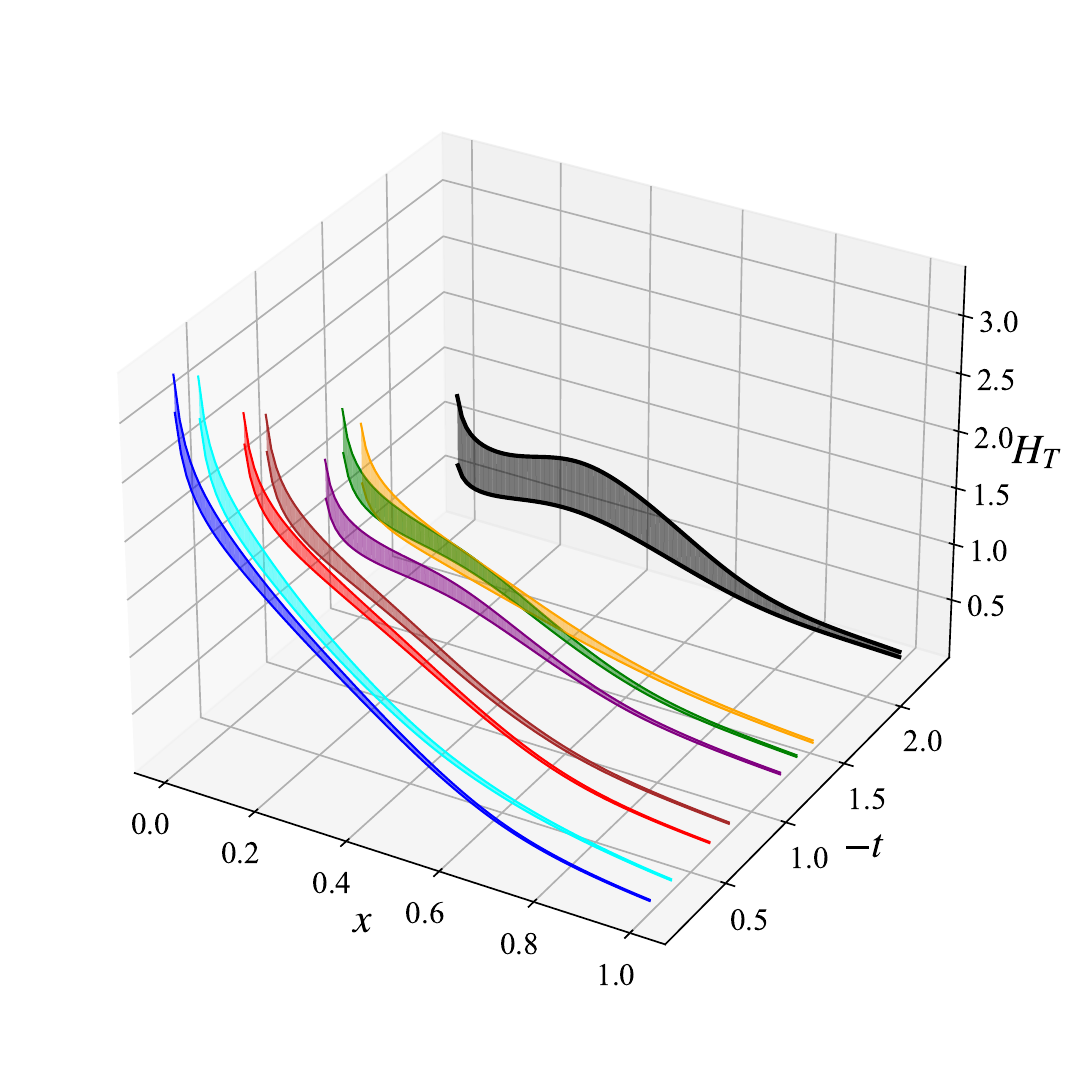} 
     \includegraphics[scale=0.33]{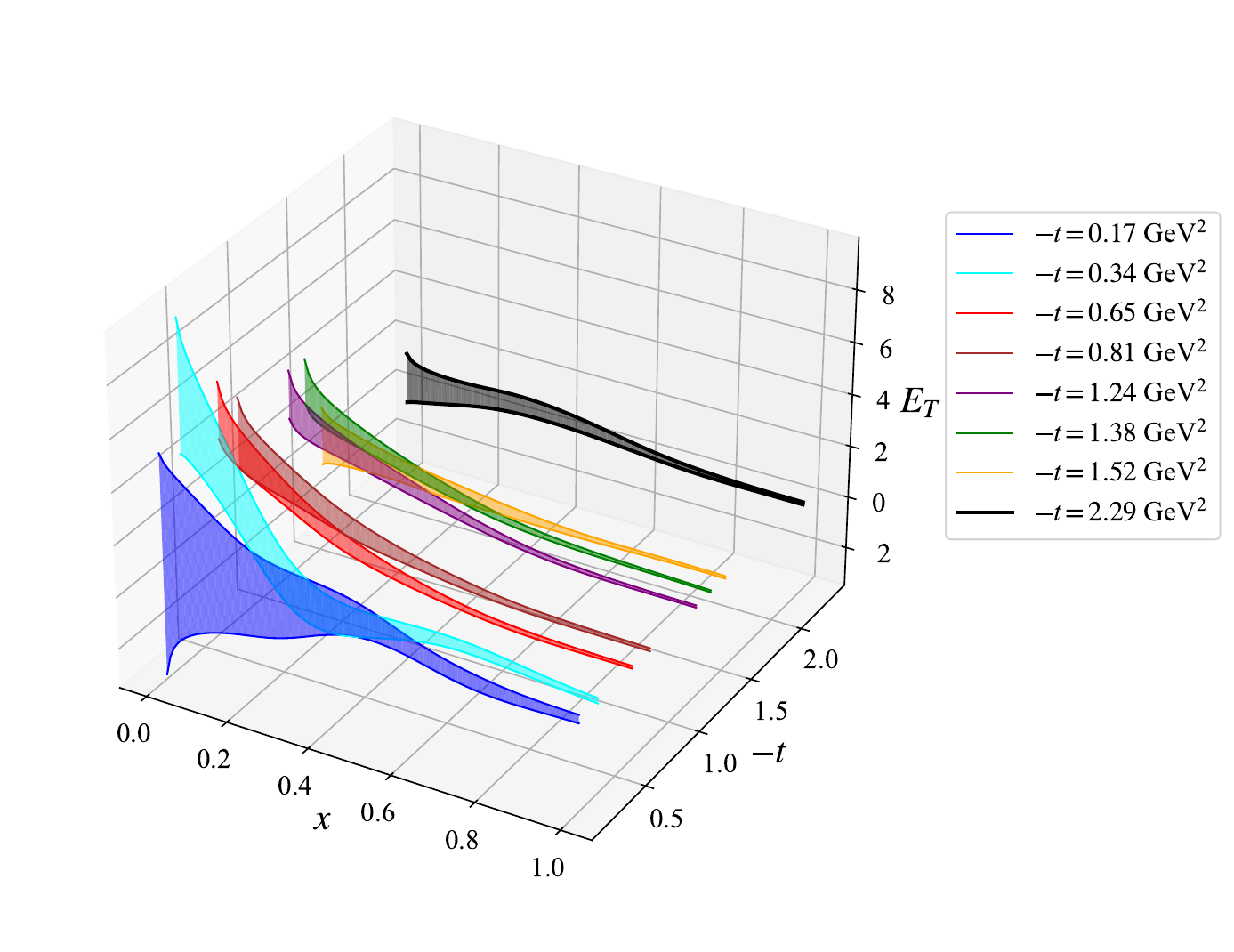}
    \vspace*{-0.1cm}
    \caption{\small{3D plot of the light-cone GPD ${H}_T$ (left) and  ${E}_T$ (right) as a function of $x$ and $-t$.}}
    \label{fig:3D_HT_LI_matched}
\end{figure}

\begin{figure}[h!]
    \centering
    \includegraphics[scale=0.33]{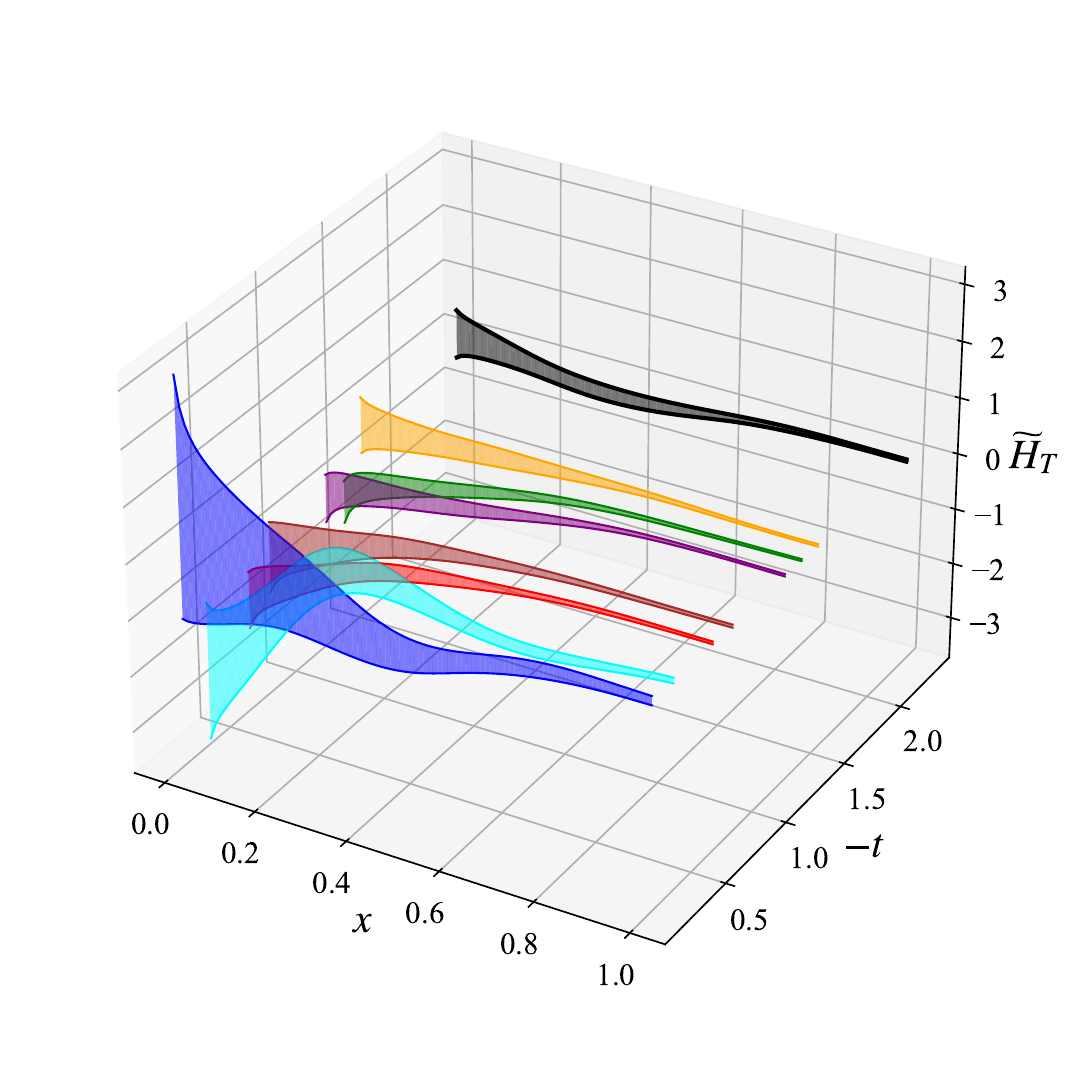}
    \includegraphics[scale=0.33]{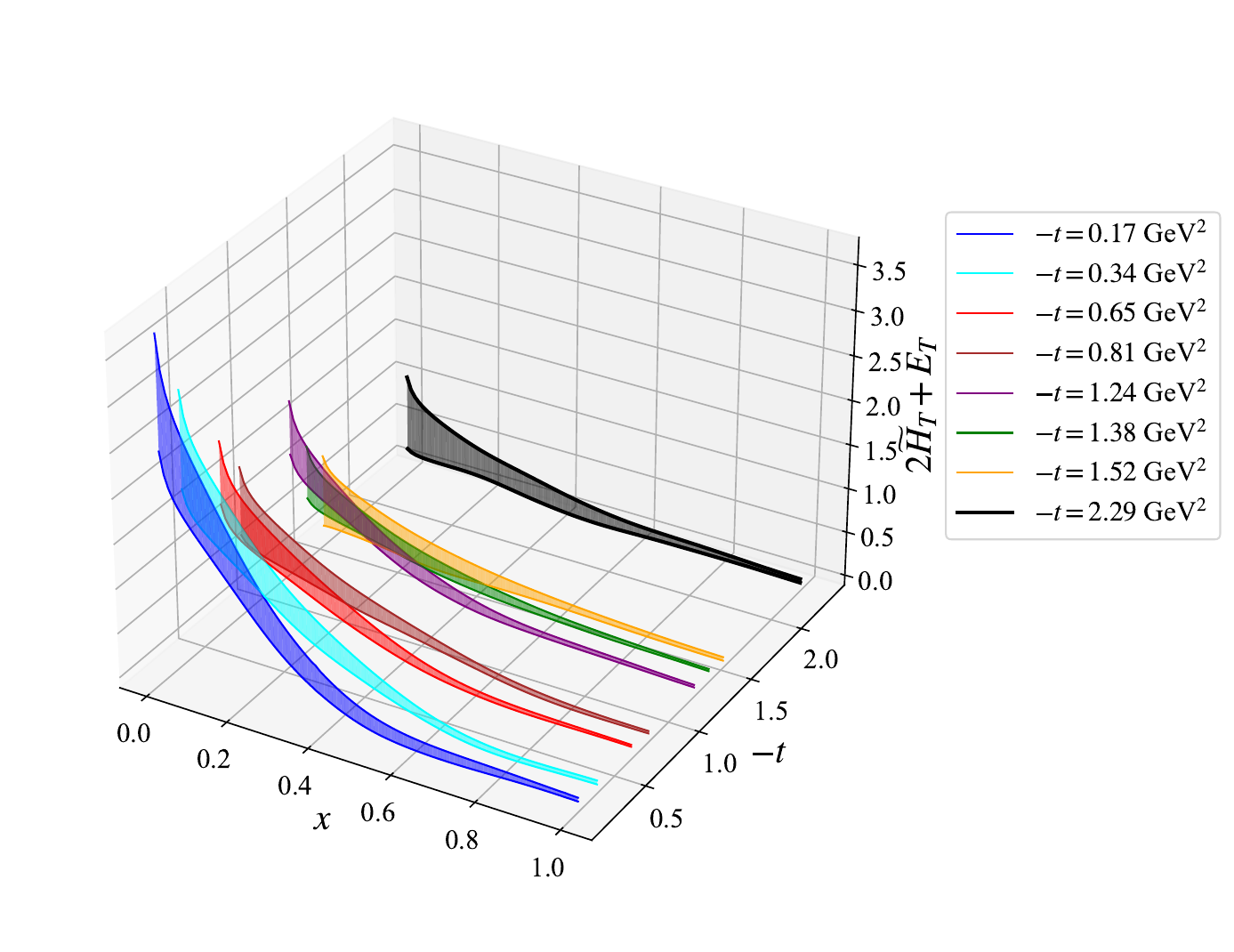}
    \vspace*{-0.1cm}
    \caption{\small{3D plot of the light-cone GPD ${\widetilde{H}}_T$ (left) and the combination $2\widetilde{H}_T + E_T$ (right) as a function of $x$ and $-t$.}}
    \label{fig:3D_HTtilde_LI_matched}
\end{figure}

\newpage
\noindent
Finally, as an alternative presentation, in Figs.~\ref{fig:3D_HT_LI_matched} - \ref{fig:3D_HTtilde_LI_matched}
we show the light-cone GPDs as a function of $x$ and $-t$.


\section{Summary and Outlook}
\label{sec:summary}
Recent advancements in lattice QCD have made it possible to calculate the $x$-dependence of GPDs for different kinematic frames for the unpolarized~\cite{Bhattacharya:2022aob} and helicity~\cite{Bhattacharya:2023jsc} cases of the proton. 
Leveraging these developments, this work focuses on developing a theoretical framework for evaluating transversity GPDs using Lorentz-invariant amplitudes. 
We find that the non-local tensor operator can be decomposed in terms of twelve basis structures, with each of them multiplied by a Lorentz-invariant amplitude.
In addition, we present the first lattice QCD calculation of the isovector flavor combination for the twist-2 quark transversity GPDs utilizing an asymmetric frame. As in our previous work, we explore two distinct definitions for quasi-GPDs, the first being the conventional $\sigma^{j0} \gamma_5$ definition (equivalently $\sigma^{3j}$) and the second being a Lorentz-invariant construction that mirrors the functional form of the light-cone GPDs. 
As a proof-of-concept, we illustrate the Lorentz invariance of the amplitudes $A_{Ti}$ by performing a calculation also in the symmetric frame for the smallest value of $-t$ that can match a value obtained from the asymmetric frames (up to about $6\%$ difference). 
Our analysis concentrates on the special yet very important case $\xi = 0$. We examine the Lorentz-invariant amplitudes $A_{Ti}$ and confirm that their numerical behavior aligns with the theoretical expectation of frame-independence.
At zero skewness, some $A_{Ti}$ as well as the GPD $\widetilde{E}_T$ are expected to vanish, as they are odd in $\xi$. 
In this work, we confirm numerically that these quantities are zero within errors.

A combination of operators, boosted sources and sinks, and parity projectors are employed to disentangle the four transversity GPDs: $H_T$, $E_T$, $\widetilde{H}_T$, and $\widetilde{E}_T$. 
The calculated GPDs span the $-t$ range $[0.17, 2.77]~\mathrm{GeV}^2$ at zero skewness, and most of the results exhibit a clear signal for all momenta. 
The lattice matrix elements are obtained at a momentum boost of about 1.25 GeV. 
As anticipated, statistical noise is a persistent challenge for off-forward matrix elements, particularly as momentum boost and momentum transfer increase. 
We note that the light-cone GPDs for values of $-t$ that are close to or larger than the boost $P_3$ may suffer from significant higher-twist contaminations and be unreliable.
The GPD $\widetilde{E}_T$ exhibits a very noisy signal that is zero within statistical uncertainties. As previously mentioned, this is expected theoretically.
On the other hand, we get a very good signal for $H_T$ and the linear combination $E_T + 2 \widetilde{H}_T$, both of which have a clear physical meaning. The former is the off-forward counterpart of the transversity PDF, while the latter governs the dipole-type deformation of the distribution of transversely polarized quarks inside an unpolarized proton. 
Future directions include addressing systematic uncertainties related to the lattice calculation of the matrix elements, such as the impact of excited-state contamination and pion mass dependence. Exploring nonzero skewness will further enhance the information extracted from lattice QCD calculations. Other investigations include alternative renormalization prescriptions, as well as reconstruction methods.

\begin{acknowledgements} 

We thank Xiang Gao, Swagato Mukherjee, and Yong Zhao for constructive discussions. The work of S.~B. has been supported by the Laboratory Directed Research and Development program of Los Alamos National Laboratory under project number 20240738PRD1.
S.~B. has also received support from the U.~S. Department of Energy through the Los Alamos National Laboratory. Los Alamos National Laboratory is operated by Triad National Security, LLC, for the National Nuclear Security Administration of U.~S. Department of Energy (Contract No. 89233218CNA000001). 
K.~C.\ is supported by the National Science Centre (Poland) grant OPUS No.\ 2021/43/B/ST2/00497. M.~C. and J. M. acknowledge financial support by the U.S. Department of Energy, Office of Nuclear Physics,  under Grant No.\ DE-SC0025218.
The work of A.~M. is supported by the National Science Foundation under grant number PHY-2110472 and PHY-2412792.
P.~P. was supported by
DOE under contract No. DE-SC0012704.
F.~S.\ was funded in part by the Deutsche Forschungsgemeinschaft (DFG, German Research Foundation) as part of the CRC 1639 NuMeriQS -- project no.\ 511713970.
The authors also acknowledge partial support by the U.S. Department of Energy, Office of Science, Office of Nuclear Physics under the umbrella of the Quark-Gluon Tomography (QGT) Topical Collaboration with Award DE-SC0023646.
Computations for this work were carried out in part on facilities of the USQCD Collaboration, which are funded by the Office of Science of the U.S. Department of Energy. 
This research used resources of the Oak Ridge Leadership Computing Facility, which is a DOE Office of Science User Facility supported under Contract DE-AC05-00OR22725.
This research was supported in part by PLGrid Infrastructure (Prometheus and Ares supercomputers at AGH Cyfronet in Cracow).
Computations were also partially performed at the Poznan Supercomputing and Networking Center (Eagle/Altair supercomputer), the Interdisciplinary Centre for Mathematical and Computational Modelling of the Warsaw University (Okeanos supercomputer), and at the Academic Computer Centre in Gda\'nsk (Tryton/Tryton Plus supercomputer).
The gauge configurations have been generated by the Extended Twisted Mass Collaboration on the KNL (A2) Partition of Marconi at CINECA, through the Prace project Pra13\_3304 ``SIMPHYS".
Inversions were performed using the DD-$\alpha$AMG solver~\cite{Frommer:2013fsa} with twisted mass support~\cite{Alexandrou:2016izb}. 

\end{acknowledgements}

\newpage
\appendix
\section{Symmetry property of the $A_{Ti}$ amplitudes}
\label{sec:symmetry}
In this appendix, we summarize the symmetry properties of the amplitudes as dictated by hermiticity and time-reversal transformation. Our derivation follows the steps outlined in our previous work~\cite{Bhattacharya:2022aob,Bhattacharya:2023jsc}.


\textbf{\textit{Symmetry of the $A_{Ti}$ using hermiticity}:} 
\begin{align}
A^*_{T1} (- z \cdot P, z \cdot \Delta, \Delta^2, z^2) & = A_{T1} (z \cdot P, z \cdot \Delta, \Delta^2, z^2) \, , \nonumber \\[0.2cm]
A^*_{T2} (- z \cdot P, z \cdot \Delta, \Delta^2, z^2) & = A_{T2} (z \cdot P, z \cdot \Delta, \Delta^2, z^2) \, , \nonumber \\[0.2cm]
- A^*_{T3} (- z \cdot P, z \cdot \Delta, \Delta^2, z^2) & = A_{T3} (z \cdot P, z \cdot \Delta, \Delta^2, z^2) \, , \nonumber \\[0.2cm]
A^*_{T4} (- z \cdot P, z \cdot \Delta, \Delta^2, z^2) & = A_{T4} (z \cdot P, z \cdot \Delta, \Delta^2, z^2) \, , \nonumber \\[0.2cm]
- A^*_{T5} (- z \cdot P, z \cdot \Delta, \Delta^2, z^2) & = A_{T5} (z \cdot P, z \cdot \Delta, \Delta^2, z^2) \, , \nonumber \\[0.2cm]
- A^*_{T6} (- z \cdot P, z \cdot \Delta, \Delta^2, z^2) & = A_{T6} (z \cdot P, z \cdot \Delta, \Delta^2, z^2) \, , \nonumber \\[0.2cm]
A^*_{T7} (- z \cdot P, z \cdot \Delta, \Delta^2, z^2) & = A_{T7} (z \cdot P, z \cdot \Delta, \Delta^2, z^2) \, , \nonumber \\[0.2cm]
A^*_{T8} (- z \cdot P, z \cdot \Delta, \Delta^2, z^2) & = A_{T8} (z \cdot P, z \cdot \Delta, \Delta^2, z^2) \, ,
\nonumber \\[0.2cm]
- A^*_{T9} (- z \cdot P, z \cdot \Delta, \Delta^2, z^2) & = A_{T9} (z \cdot P, z \cdot \Delta, \Delta^2, z^2) \, ,
\nonumber \\[0.2cm]
A^*_{T 10} (- z \cdot P, z \cdot \Delta, \Delta^2, z^2) & = A_{T 10} (z \cdot P, z \cdot \Delta, \Delta^2, z^2) \, ,
\nonumber \\[0.2cm]
A^*_{T 11} (- z \cdot P, z \cdot \Delta, \Delta^2, z^2) & = A_{T 11} (z \cdot P, z \cdot \Delta, \Delta^2, z^2) \, ,
\nonumber \\[0.2cm]
- A^*_{T 12} (- z \cdot P, z \cdot \Delta, \Delta^2, z^2) & = A_{T 12} (z \cdot P, z \cdot \Delta, \Delta^2, z^2) \, .
\label{e:hermicity_trans}
\end{align}

\textbf{\textit{Symmetry of the $A_{Ti}$ using Time-reversal}:}
\begin{align}
- A^*_{T1} (- \bar{z} \cdot \bar{P}, -\bar{z} \cdot \bar{\Delta}, \bar{\Delta}^2, \bar{z}^2) & = A_{T1} (z \cdot P, z \cdot \Delta, \Delta^2, z^2) \, , \nonumber \\[0.2cm]
A^*_{T2} (- \bar{z} \cdot \bar{P}, -\bar{z} \cdot \bar{\Delta}, \bar{\Delta}^2, \bar{z}^2) & = A_{T2} (z \cdot P, z \cdot \Delta, \Delta^2, z^2) \, , \nonumber \\[0.2cm]
- A^*_{T3} (- \bar{z} \cdot \bar{P}, -\bar{z} \cdot \bar{\Delta}, \bar{\Delta}^2, \bar{z}^2) & = A_{T3} (z \cdot P, z \cdot \Delta, \Delta^2, z^2) \, , \nonumber \\[0.2cm]
A^*_{T4} (- \bar{z} \cdot \bar{P}, -\bar{z} \cdot \bar{\Delta}, \bar{\Delta}^2, \bar{z}^2) & = A_{T4} (z \cdot P, z \cdot \Delta, \Delta^2, z^2) \, , \nonumber \\[0.2cm]
- A^*_{T5} (- \bar{z} \cdot \bar{P}, -\bar{z} \cdot \bar{\Delta}, \bar{\Delta}^2, \bar{z}^2) & = A_{T5} (z \cdot P, z \cdot \Delta, \Delta^2, z^2) \, , \nonumber \\[0.2cm]
A^*_{T6} (- \bar{z} \cdot \bar{P}, -\bar{z} \cdot \bar{\Delta}, \bar{\Delta}^2, \bar{z}^2) & = A_{T6} (z \cdot P, z \cdot \Delta, \Delta^2, z^2) \, , \nonumber \\[0.2cm]
A^*_{T7} (- \bar{z} \cdot \bar{P}, -\bar{z} \cdot \bar{\Delta}, \bar{\Delta}^2, \bar{z}^2) & = A_{T7} (z \cdot P, z \cdot \Delta, \Delta^2, z^2) \, , \nonumber \\[0.2cm]
- A^*_{T8} (- \bar{z} \cdot \bar{P}, -\bar{z} \cdot \bar{\Delta}, \bar{\Delta}^2, \bar{z}^2) & = A_{T8} (z \cdot P, z \cdot \Delta, \Delta^2, z^2) \, ,
\nonumber \\[0.2cm]
A^*_{T9} (- \bar{z} \cdot \bar{P}, -\bar{z} \cdot \bar{\Delta}, \bar{\Delta}^2, \bar{z}^2) & = A_{T9} (z \cdot P, z \cdot \Delta, \Delta^2, z^2) \, ,
\nonumber \\[0.2cm]
A^*_{T 10} (- \bar{z} \cdot \bar{P}, -\bar{z} \cdot \bar{\Delta}, \bar{\Delta}^2, \bar{z}^2) & = A_{T 10} (z \cdot P, z \cdot \Delta, \Delta^2, z^2) \, ,
\nonumber \\[0.2cm]
- A^*_{T 11} (- \bar{z} \cdot \bar{P}, -\bar{z} \cdot \bar{\Delta}, \bar{\Delta}^2, \bar{z}^2) & = A_{T 11} (z \cdot P, z \cdot \Delta, \Delta^2, z^2) \, ,
\nonumber \\[0.2cm]
- A^*_{T 12} (- \bar{z} \cdot \bar{P}, -\bar{z} \cdot \bar{\Delta}, \bar{\Delta}^2, \bar{z}^2) & = A_{T 12} (z \cdot P, z \cdot \Delta, \Delta^2, z^2) \, .
\label{e:time_trans}
\end{align}

\section{Consistency with the local current}
\label{sec:local}
It is important to investigate the consistency of our decomposition with the local tensor current at $z = 0$, which is parameterized in terms of three (real-valued) form factors. For $z=0$ our decomposition simplifies to the following expression:
\begin{align}
F^{[i\sigma^{\mu \nu}\gamma_5]} \big |_{z=0} & = \bar{u}(p_f,\lambda') \bigg [ \dfrac{P^{[\mu}\Delta^{\nu]}\gamma_5}{M^2} A_{T2} + \dfrac{\gamma^{[\mu}P^{\nu]}\gamma_5}{M} A_{T4} + \dfrac{\gamma^{[\mu}\Delta^{\nu]}\gamma_5}{M} A_{T6} + i \sigma^{\mu \nu} \gamma_5 A_{T10}  \bigg ] u(p_i, \lambda) \,.
\end{align}
Generally, the amplitudes $A_{Ti}$ are complex functions. However, in order to be consistent with the local tensor current, we need to demonstrate that the surviving $A_{Ti}$ for $z = 0$ are either purely real or imaginary. To do so, recall first that hermiticity leads to 
\begin{align}
A^*_{T2} = A_{T2} \, , \quad
A^*_{T4} = A_{T4} \, , \quad
- A^*_{T6}  = A_{T6} \, , \quad
A^*_{T10} = A_{T10} \, ,
\end{align}
which implies
\begin{align}
\textrm{Im.} (A_{T2})  = 0 \, , \qquad \, \textrm{Im.} (A_{T4})  = 0 \, , \qquad \textrm{Re.} (A_{T6})  = 0 \, , \qquad \textrm{Im.} (A_{T10})  = 0 \, .
\label{e:c1}
\end{align}
Second, we show that only three amplitudes are nonzero for $z=0$. The time-reversal transformation provides
\begin{align}
A^*_{T2} = A_{T2} \, , \quad
A^*_{T4} = A_{T4} \, , \quad
A^*_{T6}  = A_{T6} \, , \quad
A^*_{T10} = A_{T10} \,,
\end{align}
yielding
\begin{align}
\textrm{Im.} (A_{T2})  = 0 \, , \qquad \, \textrm{Im.} (A_{T4})  = 0 \, , \qquad \textrm{Im.} (A_{T6})  = 0 \, , \qquad \textrm{Im.} (A_{T10})  = 0 \, .
\label{e:c2}
\end{align}
By combining Eqs.~(\ref{e:c1}) and (\ref{e:c2}), we deduce
\begin{align}
\textrm{Re.} (A_{T6}) & = 0 \, , \qquad \textrm{Im.} (A_{T6}) = 0 \nonumber \\[0.2cm]
\therefore A_{T6} & = 0 \, .
\end{align}
Hence, Eqs.~(\ref{e:c1}) and (\ref{e:c2}) show that the only contributions at $z=0$ arise from
\begin{align}
\textrm{Re.} (A_{T2}) \neq  0 \, , \quad \textrm{Re.} (A_{T4})\neq 0 \, , \quad \textrm{Re.} (A_{T10})\neq 0 \, .
\end{align}
Therefore, our decomposition is fully consistent in that it reduces to three terms only in the local limit:
\begin{align}
F^{[i\sigma^{\mu \nu}\gamma_5]} \big |_{z=0} & = \bar{u}(p_f,\lambda') \bigg [ \dfrac{P^{[\mu}\Delta^{\nu]}\gamma_5}{M^2} A_{T2} + \dfrac{\gamma^{[\mu}P^{\nu]}\gamma_5}{M} A_{T4} + i \sigma^{\mu \nu} \gamma_5 A_{T10}  \bigg ] u(p_i, \lambda) \,.
\end{align}

\bibliography{references.bib}

\end{document}